\documentclass[11pt, draftclsnofoot, peerreview, letter, oneside, onecolumn]{IEEEtran}
\usepackage{graphicx}
\usepackage{amsmath}
\usepackage{amssymb}
\usepackage{mathrsfs}
\usepackage{stfloats}
\usepackage{dsfont}
\usepackage{setspace}
\newtheorem{corollary}{Corollary}
\newtheorem{proposition}{Proposition}
\newtheorem{lemma}{Lemma}
\begin{document}
%
%
%
%
%
%
\title{\Large \bf On the Diversity Order and Coding Gain of Multi--Source Multi--Relay Cooperative Wireless Networks with Binary Network Coding}
\author{{Marco~Di~Renzo,~\IEEEmembership{Member,~IEEE},
        Michela~Iezzi, and
        Fabio~Graziosi,~\IEEEmembership{Member,~IEEE}}
\thanks{\scriptsize Manuscript received April 23, 2012.
This paper was presented in part at the IEEE Global Commun. Conf., Houston, USA, Dec. 2011.}
\thanks{\scriptsize M. Di Renzo is with the Laboratory of Signals and Systems (LSS), French National Center for Scientific Research (CNRS) -- \'Ecole Sup\'erieure d'\'Electricit\'e (SUP\'ELEC) -- University of Paris--Sud 11, 3 rue Joliot--Curie, 91192 Gif--sur--Yvette (Paris), France, E--Mail:
marco.direnzo@lss.supelec.fr. M. Iezzi and F. Graziosi are with The University of L'Aquila, Department of Electrical and Information
Engineering (DIEI), Center of Excellence of Research DEWS, via G. Gronchi 18, 67100 L'Aquila, Italy. E--Mail:
\{michela.iezzi, fabio.graziosi\}@univaq.it.} \vspace{-0.5cm}}
%
%
%
%
%
%
\maketitle
\begin{abstract}
In this paper, a multi--source multi--relay cooperative wireless network with binary modulation and binary network coding is studied. The system model encompasses: i) a demodulate--and--forward protocol at the relays, where the received packets are forwarded regardless of their reliability; and ii) a
maximum--likelihood optimum demodulator at the destination, which accounts for possible demodulations errors at the relays. An asymptotically--tight and
closed--form expression of the end-to-end error probability is derived, which clearly showcases diversity order and coding gain of each source.
Unlike other papers available in the literature, the proposed framework has three main distinguishable features: i) it is useful for general
network topologies and arbitrary binary encoding vectors; ii) it shows how network code and two--hop forwarding protocol affect diversity order and
coding gain; and ii) it accounts for realistic fading channels and demodulation errors at the relays. The framework provides three main
conclusions: i) each source achieves a diversity order equal to the separation vector of the network code; ii) the coding gain of each source
decreases with the number of mixed packets at the relays; and iii) if the destination cannot take into account demodulation errors at the relays,
it loses approximately half of the diversity order.
\end{abstract}
%
%
%
%
%
\section{Introduction} \label{Introduction}
\PARstart{C}{ooperative} communications and Network Coding (NC) have recently emerged as strong candidate technologies for many future wireless
applications, such as relay--aided cellular networks \cite{Zorzi_CoopNC}, \cite{MitreCorp}. Since their inception in \cite{Laneman} and \cite{Ahlswede}, they have been extensively studied to improve performance and throughput of wireless networks, respectively. In particular, theory and experiments have shown that they can be extremely useful for wireless networks with disruptive channel and connectivity conditions \cite{Gerla}--\cite{Zorzi_Phoenix}.

However, similar to many other technologies, multi--hop/cooperative communications and NC are not without limitations \cite{Zorzi_CoopNC},
\cite{MDR_Springer2010}. Due to practical hardware limitations, \emph{e.g.}, the half--duplex constraint, relay transmissions consume extra
bandwidth, which implies that using cooperative diversity typically results in a loss of system throughput \cite{Thompson_Mag2011}.
On the other hand, NC is very susceptible to transmission errors caused by noise, fading, and interference. In fact, the algebraic operations
performed at the network nodes introduce some packet dependencies in a way that the injection of even a single erroneous packet has
the potential to corrupt every packet received at the destination \cite{Koetter}, \cite{Silva_PhD}. Due to their complementary merits
and limitations, it seems very natural to synergically exploit cooperation and NC to take advantage of their key benefits while
overcoming their limitations. For example, NC can be an effective enabler to recover the throughput loss experienced by multi--hop/cooperative
networking, while the redundancy inherently provided by cooperation might significantly help to alleviate the error propagation problem that
arises when mixing the packets \cite{Zorzi_CoopNC}.

In this context, multi--source multi--relay networks, which exploit cooperation and NC for performance and throughput improvement, are
receiving an always increasing interest for their inherent flexibility to achieving excellent performance and diversity/multiplexing tradeoffs
\cite{Shalvi2004}--\cite{XiaoJuly2011}. More specifically, considerable attention is currently devoted to understanding the achievable
performance of such networks when both cooperation and NC operations are pushed down to the physical layer, and their joint design and
optimization are closely tied to conventional physical layer functionalities, such as modulation, channel coding, and receiver design
\cite{Fasolo}, \cite{Laneman2010}. In particular, how to tackle the error propagation problem to guaranteeing a given quality--of--service
requirement, \emph{e.g.}, a distributed diversity order, plays a crucial role when these networks are deployed in error--prone environments,
\emph{e.g.}, in a wireless context. For example, simple case studies in \cite{Laneman2007}, \cite{Annavajjala}, \cite{XiaoDec2010}, and
\cite{Iezzi_ICC2011} have shown that a diversity loss occurs if cooperative protocols or detection algorithms are not adequately designed. To
counteract this issue, many solutions have been proposed in the literature, which can be divided in two main categories: i) adaptive (or
dynamic) solutions, \emph{e.g.}, \cite{PengAug2008}, \cite{XiaoDec2010}, \cite{Topakkaya}, \cite{Rebelatto}, and \cite{YuneKimIm}, which avoid
unnecessary error propagation that can be caused when encoding and forwarding erroneous data packets; and ii) non--adaptive solutions
\emph{e.g.}, \cite{Laneman2007}, \cite{GiannakisApr2008}, \cite{Hanzo2009}, \cite{Annavajjala}, and \cite{SchoberGLOBE2010}, which allow
erroneous packets to propagate through the network but exploit optimal detection mechanisms at the destination to counteract the error
propagation. Each category has its own merits and limitations.

Adaptive solutions rely, in general, on the following assumptions \cite{XiaoDec2010}, \cite{RayLiuGLOBE2010}, \cite{Topakkaya},
\cite{Rebelatto}: a) network code and cooperative strategy are adapted to the channel conditions and to the outcome/reliability of the
detection process at the relay nodes. This requires some overhead, since the network code must be communicated to the destination
for correct detection; b) powerful enough channel codes at the physical later are assumed to guaranteeing that the error performance is
dominated by outage events (according to the Shannon definition of outage capacity) \cite[Sec. II]{XiaoJuly2011}; and c) the adoption of ideal
Cyclic Redundancy Check (CRC) mechanisms for error detection, which guarantees that a packet is either dropped or injected into the network
without errors (\emph{i.e.}, erasure channel model). However, recent results have shown that, in addition to be highly spectral inefficient as
an entire packet is blocked if just one bit is in error, relaying based on CRC might not be very effective in block--fading channels
\cite{Nguyen2010}, \cite{AlHabian2011}. An interesting link--adaptive solution, which does not require CRC for error detection and
avoids full--CSI (Channel State Information) information at the relays, has been proposed in \cite{GiannakisEURASIP2008}. Therein, the achievable diversity (using the Singleton bound) is studied under the assumption that \emph{ad hoc} interleavers are used, while no analysis of the coding gain is conducted.

Non--adaptive solutions rely, in general, on the following assumptions \cite{Laneman2007}, \cite{GiannakisApr2008},
\cite{SchoberGLOBE2010}: a) neither error correction nor error detection mechanisms are needed at the physical layer, but the relays just
regenerate the incoming packets and forward them to the final destination (\emph{i.e.}, error channel model). This results in a simple
design of the relay nodes, as well as in a spectral efficient transmission scheme as the received packets are never blocked; and b) the
possibility to receive packets with errors needs powerful detection mechanisms at the destination, which require CSI of the whole network to
counteract the error propagation problem and to achieve full--diversity. Similar to adaptive solutions, this requires some overhead.

As far as adaptive solutions are concerned, \cite{XiaoDec2010}, \cite{Topakkaya}, \cite{Rebelatto} have recently provided a comprehensive study
of the diversity/multiplexing tradeoff for general multi--source multi--relay networks, and have shown that the design of diversity--achieving
network codes is equivalent to the design of systematic Maximum Distance Separable (MDS) codes for erasure channels. Thus, well--established
and general methods for the design of network codes exist, which can be borrowed from classical coding theory. On the other hand, as far as
non--adaptive solutions are concerned, theoretical analysis and guidelines for system optimization are available only for specific network
topologies and network codes. To the best of the authors knowledge, a general framework for performance analysis and code
design over fading channels is still missing. Motivated by these considerations, in this paper we focus our attention on non--adaptive solutions with a threefold
objective: i) to develop a general analytical framework to compute the Average Bit Error Probability (ABEP) of multi--source multi--relay
cooperative networks with arbitrary binary encoding vectors and realistic channel conditions over all the wireless links; ii) to provide
guidelines for network code design to achieve a given diversity and coding gain tradeoff; and iii) to understand the impact of the error
propagation problem and the role played by CSI at the destination on the achievable diversity order and coding gain.

More specifically, by carefully looking at recent literature related to the performance analysis and code design for non--adaptive solutions,
the following contributions are worth being mentioned: i) in \cite{Laneman2007}, the authors study a simple three--node network without
NC (a simple repetition code is considered) and they show that instantaneous CSI is needed at the destination to achieve full--diversity. No
closed--form expression of the coding gain is given; ii) in \cite{GiannakisApr2008} and \cite{GiannakisMay2011}, the authors introduce and
study Complex Field Network Coding (CFNC), which does not rely on Galois field operations and exploit interference and multi--user detection to
increase throughput and diversity. The analysis is valid for arbitrary network topologies. However, only the diversity order is computed analytically, while the coding gain is studied by simulation; iii) in \cite{Hanzo2009}, the authors study a simple three--node network with binary NC. Unlike other papers,
channel coding is considered in the analysis. However, the error performance is mainly estimated through Monte Carlo simulations; iv) in
\cite{Annavajjala}, the author considers multiple relay nodes but a simple repetition code is used (no NC). Main contribution is the study of the impact of channel estimation errors on the achievable diversity; v) in \cite{SchoberGLOBE2010}, the authors study a network topology with multiple sources but with just one relay. Also, a very specific network code is analyzed. This paper provides a simple and effective method to accurately computing the coding gain of error--prone cooperative networks with NC; vi) in \cite{XiaoICC2011}, the authors analyze generic multi--source multi--relay networks with binary NC, but error--free source--to--relay links are considered, and the performance (coding gain) is computed by using Monte Carlo simulations; vii) in \cite{Iezzi_ICC2011} and \cite{Iezzi_GLOBE2011}, we have studied the performance of network--coded cooperative networks with realistic
source--to--relay wireless channels. However, the analysis is useful only for two--source two--relay networks and for a very specific binary
network code; viii) in \cite{Ju2009}, a general framework to study the ABEP for arbitrary modulation schemes is provided, but a simple
three--node network without NC is considered; and ix) in \cite{Mallik2011}, the authors study a three--node network with a simple repetition
code. Exact results are provided for coding gain and diversity order. Finally, in \cite{XiaoNov2008} and \cite{XiaoMay2009}, NC with error--prone  source--to--relay links is studied, but the analysis is applicable only to noisy channels, while channel fading and distributed diversity issues are not investigated.

According to this up--to--date analysis of the state--of--the--art, it follows that no general framework for performance analysis and design of
non--adaptive solutions exists in the literature, which is useful for generic network topologies, for arbitrary encoding vectors, and which
provides an accurate characterization of diversity order and coding gain as a function of the CSI available at the destination. Motivated by these
considerations, in this paper we focus our attention on a general multi--source multi--relay network with realistic and error--prone channels
over all the wireless links. For analytical tractability (and to keep the implementation complexity of relays at a low level \cite{XiaoICC2011},
\cite{Fitzek_ICC2011}), we consider a binary network code, a Binary Phase Shift Keying (BPSK) modulation, and the
Demodulate--and--Forward (DemF) relay protocol. With these assumptions, the main contributions and outcomes of this paper are as follows: i) a Maximum--Likelihood (ML--) optimum demodulator is proposed, which allows the destination to exploit the distributed diversity inherently provided by cooperation and
NC. The demodulator takes into account demodulation errors that might occur at the relay nodes, as well as forwarding and NC operations. It is shown
that the demodulator resembles a Chase combiner \cite{ChaseCombining} with hard--decision decoding at the physical layer; ii) a simple but accurate
framework to compute the end--to--end ABEP of each source is proposed. The framework provides a closed--form expression of diversity order and coding
gain, and it clearly highlights the impact of error propagation and NC on the end--to--end performance; iii) it is proved that each source node
can achieve a diversity order that is equal to the separation vector \cite{Masnick_Oct1967}, \cite{Boyarinov_Marc1981} of the network code. In
particular, it is shown that the optimization of network codes is equivalent to the design of systematic linear block codes for
fully--interleaved fading channels, and that Equal and Unequal Error Protection (EEP/UEP) properties are preserved \cite{Masnick_Oct1967}; and
iv) the impact of CSI at the destination is studied, and it is shown that half of the diversity order is lost if the destination is
unable to account for possible demodulation errors at the relays.

The paper is organized as follows. In Section \ref{SystemModel}, network topology and system model are introduced.
In Section \ref{ReceiverDesign}, the ML--optimum demodulator that accounts for demodulation errors at the relays is proposed. In Section
\ref{FrameworkABEP}, a closed--form expression of the end--to--end ABEP is given. In Section
\ref{DiversityCodingGain}, diversity order and coding gain are studied for arbitrary binary network codes and network topologies. In Section
\ref{Results}, numerical results are presented to substantiate analysis and findings. Finally, Section \ref{Conclusion} concludes this
paper.
\section{System Model} \label{SystemModel}
We consider a generic multi--source multi--relay network with $N_S$ sources ($S_t$ for $t = 1,2, \ldots ,N_S$), $N_R$ relays ($R_q$ for $q =
1,2, \ldots ,N_R$), and, without loss of generality, a single destination $D$. We consider the baseline Time Division Multiple Access (TDMA)
protocol, where each transmission takes place in a different time--slot, and multiple--access interference can be neglected \cite{Laneman}. We
assume that direct links between sources and destination exist, and that the relays help the sources to deliver the information packets to the
final destination. The cooperative protocol is composed of two main phases: i) the broadcasting phase; and ii) the
relaying phase. During the first phase, the source $S_t$ transmits the information packet intended to the destination in time--slot $T_t$ for $t =
1,2, \ldots ,N_S$. These $N_S$ packets are overheard by the $N_R$ relays too, which store them in their buffers for further processing. This
phase lasts $N_S$ time--slots. During the second phase, the relay $R_q$ forwards a linear combination, \emph{i.e.}, NC is applied \cite{Ahlswede},
of some received packets to the destination in time--slot $T_{N_S+q}$ for $q = 1,2, \ldots ,N_R$. We consider a non--adaptive DemF relay protocol, which means that each relay demodulates the received packets, but perform NC and forward them regardless of their reliability. As a result, packets with erroneous bits can be injected into the network. However, these packets can be adequately used at the destination, by exploiting advanced detection and signal processing algorithms at the physical layer, to improve the system performance \cite[pp. 18--20]{Zorzi_CoopNC}. According to the working operation of the protocol, broadcasting and relaying phases last $N_S+N_R$ time--slots. Since $N_S$ information packets are transmitted by the sources, the protocol offers a fixed rate, $\mathcal{R}$, that is equal to $\mathcal{R} = {{N_S } \mathord{\left/ {\vphantom {{N_S } {\left( {N_S + N_R } \right)}}} \right. \kern-\nulldelimiterspace} {\left( {N_S  + N_R } \right)}}$. In this paper, we are interested in understanding how the operations, \emph{i.e.}, NC, performed at the relays affect the end--to--end performance for this given rate. Main objective is understanding the performance of cooperative networks with NC when physical layer terminologies are exploited to
counteract the error propagation problem \cite{Fasolo}, and, more specifically, when demodulation and network decoding are jointly performed at
the destination (\emph{i.e.}, cross--layer decoding). For analytical tractability and simplicity, we retain three main reasonable assumptions:
i) uncoded transmissions with no channel coding are considered. Accordingly, there is no loss of generality in considering symbol--by--symbol
transmission. Some preliminary results with channel coding are available in \cite{Poulliat}; ii) BPSK modulation is assumed to keep the
analytical complexity at a low level; and iii) binary NC at the relays is investigated. However, unlike many current papers in the literature,
\emph{e.g.}, \cite{SchoberGLOBE2010}, \cite{Iezzi_ICC2011}, \cite{Iezzi_GLOBE2011}, and references therein, no assumption about the encoding
vectors is made. These assumptions are widespread used in related literature \emph{e.g.}, \cite{Laneman2006},
\cite{Laneman2007}, \cite{Annavajjala}, and the references therein.
\subsection{Broadcasting and Relaying Phases}
According to the assumptions above, the generic source $S_t$ broadcasts, in time--slot $T_t$, a BPSK--modulated signal, $x_{S_t}$, with average
energy $E_m$, \emph{i.e.}, $x_{S_t } = \sqrt {E_m } \left( {1 - 2b_{S_t } } \right)$, where $b_{S_t }  \in \left\{ {0,1} \right\}$ is the bit
emitted by $S_t$. Then, the signals received at relays $R_q$ for $q = 1,2, \ldots ,N_R$ and destination $D$ are:
\begin{equation} \scriptsize
\label{Eq_1} \left\{ \begin{array}{l}
 y_{S_t R_q }  = h_{S_t R_q } x_{S_t }  + n_{S_t R_q }  \\
 y_{S_t D}  = h_{S_t D} x_{S_t }  + n_{S_t D}  \\
 \end{array} \right.
\end{equation}
\noindent where $h_{XY}$ is the fading coefficient from node $X$ to node $Y$, which is a circular symmetric complex Gaussian Random Variable
(RV) with zero mean and variance ${{\sigma _{XY}^2 } \mathord{\left/ {\vphantom {{\sigma _{XY}^2 } 2}} \right. \kern-\nulldelimiterspace} 2}$
per dimension (Rayleigh fading\footnote{\scriptsize The framework proposed in this paper is applicable to other fading distributions.
However, to keep the analytical development more concise and focused, we consider Rayleigh fading only. In Appendix \ref{Appendix_Lemmas}, we
provide some comments on how to extend the analysis to other fading distributions.}). Owing to the distributed nature of the network,
independent but non--identically identically distributed (i.n.i.d.) fading is considered. In particular, let
$d_{XY}$ be the distance between nodes $X$ and $Y$, and $\alpha$ be the path--loss exponent, we have $\sigma _{XY}^2
= d_{XY}^{ - \alpha }$ \cite{Proakis}, \cite{Simon}. Also, $n_{XY}$ is the complex Additive White Gaussian Noise (AWGN) at the input of
node $Y$ and related to the transmission from node $X$ to node $Y$. The AWGN in different time slots is independent and identically distributed
(i.i.d.) with zero mean and variance ${{N_0 } \mathord{\left/{\vphantom {{N_0 } 2}} \right. \kern-\nulldelimiterspace} 2}$ per dimension.

Upon reception of $y_{S_t R_q }$ and $y_{S_t D}$ in time--slot $T_t$, the relay $R_q$ for $q = 1,2, \ldots ,N_R$ and the destination $D$ demodulate
these received signals by using the ML--optimum criterion, as follows:
\begin{equation} \scriptsize
\label{Eq_2} \left\{ \begin{array}{l}
 \hat b_{S_t R_q }  = \mathop {\arg \min }\limits_{\tilde b_{S_t }  \in \left\{ {0,1} \right\}} \left\{ {\left| {y_{S_t R_q }  - \sqrt {E_m } h_{S_t R_q } \left( {1 - 2\tilde b_{S_t } } \right)} \right|^2 } \right\} \\
 \hat b_{S_t D}  = \mathop {\arg \min }\limits_{\tilde b_{S_t }  \in \left\{ {0,1} \right\}} \left\{ {\left| {y_{S_t D}  - \sqrt {E_m } h_{S_t D} \left( {1 - 2\tilde b_{S_t } } \right)} \right|^2 } \right\} \\
 \end{array} \right.
\end{equation}
\noindent where ${\hat{(\cdot)} }$ denotes the demodulated bit and ${\tilde {(\cdot)}}$ denotes the trial bit used in the hypothesis--testing
problem. More specifically, $\hat b_{S_t R_q }$ and $\hat b_{S_t D}$ are the estimates of $b_{S_t}$ at relay $R_q$ for $q = 1,2, \ldots ,N_R$,
and at destination $D$, respectively. We note that (\ref{Eq_2}) needs CSI about the source--to--relay and the relay--to--destination
channels at relay and destination nodes, respectively. In this paper, we assume that CSI is perfectly known at the receiver while it is not
known at the transmitter. This is obtained through adequate training \cite{Annavajjala}.

After estimating $\hat b_{S_t R_q }$ and $\hat b_{S_t D}$, the destination $D$ keeps the demodulated bit for further processing, as described in
Section \ref{ReceiverDesign}, while the relays initiate the relaying phase. More specifically, the generic relay, $R_q$, performs the following
three operations: i) it applies binary NC on the set of demodulated bits $\hat b_{S_t R_q }$ for $t = 1,2, \ldots ,N_S$; ii) it remodulates the
network--coded bit by using BPSK modulation; and iii) it transmits the modulated bit to the destination $D$ during time--slot $T_{N_S+q}$ for
$q = 1,2, \ldots ,N_R$. Once again, we emphasize that all the demodulated bits are considered in this phase, even though they are wrongly
detected, \emph{i.e.}, $b_{S_t } \ne \hat b_{S_t R_q }$. As far as NC is concerned, we denote the network--coded bit at relay $R_q$ by $b_{R_q
} = f_{R_q } \left( {\hat b_{S_1 R_q } ,\hat b_{S_2 R_q } , \ldots ,\hat b_{S_{N_S } R_q } } \right) = g_{S_1 R_q } \hat b_{S_1 R_q }  \oplus
g_{S_2 R_q } \hat b_{S_2 R_q }  \oplus  \ldots  \oplus g_{S_{N_S } R_q } \hat b_{S_{N_S } R_q }$, where: i) $f_{R_q } \left(  \cdot  \right)$
denotes the encoding function at relay $R_q$; ii) $\oplus$ denotes exclusive OR (XOR) operations; and iii) ${\bf{g}}_{R_q }  = \left[ {g_{S_1
R_q } ,g_{S_2 R_q } , \ldots ,g_{S_{N_S } R_q } } \right]^T$ is the binary encoding vector at relay $R_q$ \cite{Ahlswede}, where $g_{S_t R_q }
\in \left\{ {0,1} \right\}$ for $t = 1,2, \ldots ,N_S$. From this notation, it follows that only a sub--set of received bits are actually
network--coded at relay $R_q$, \emph{i.e.}, only those bits for which $g_{S_t R_q }  = 1$ for $t = 1,2, \ldots ,N_S$. Thus, our system setup is
very general: no assumptions are made on ${\bf{g}}_{R_q }$ for $q = 1,2, \ldots ,N_R$, and the encoding functions $f_{R_q }
\left( \cdot \right)$ can be different at each relay. The goal of this paper is to understand how a given choice of these functions affect the
end--to--end performance, as well as to provide guidelines for their design and optimization.

Thus, the signal received at destination $D$ in time--slot $T_{N_S+q}$ after NC and modulation is ($q = 1,2, \ldots ,N_R$):
\begin{equation} \scriptsize
\label{Eq_3} y_{R_q D}  = h_{R_q D} x_{R_q }  + n_{R_q D}
\end{equation}
\noindent where $x_{R_q }  = \sqrt {E_m } \left( {1 - 2b_{R_q } } \right)$. Let us note that the average transmit energy of each relay node is
the same as the average transmit energy of each source node, \emph{i.e.}, $E_m$. This uniform energy--allocation scheme stems from the
assumption of no CSI at the transmitter. The impact of optimal energy allocation is postponed to future research \cite{SchoberGLOBE2010}.
Thus, the total average transmit energy for broadcasting and relaying phases is $E_T  = E_m \left( {N_S + N_R }
\right)$, while the average transmit energy per network node is $E_A  = {{E_T } \mathord{\left/ {\vphantom {{E_T } {\left( {N_S + N_R }
\right)}}} \right. \kern-\nulldelimiterspace} {\left( {N_S  + N_R } \right)}} = E_m $.
\section{Receiver Design} \label{ReceiverDesign}
In this section, we develop a demodulator at the destination $D$ which is robust to the error propagation problem caused by forwarding
wrong detected bits from the relays. As explained in \cite[pp. 18--20]{Zorzi_CoopNC}, the main goal is to improve the end--to--end performance
by jointly performing demodulation and network decoding. To this end, we exploit the ML--optimum approach, which is composed of two main steps.
\paragraph{Step 1} Upon reception of $y_{R_q D}$ in time--slot $T_{N_S+q}$, the destination $D$ computes:
\begin{equation} \scriptsize
\label{Eq_4} \hat b_{R_q D}  = \mathop {\arg \min }\limits_{\tilde b_{R_q }  \in \left\{ {0,1} \right\}} \left\{ {\left| {y_{R_q D}  - \sqrt
{E_m } h_{R_q D} \left( {1 - 2\tilde b_{R_q } } \right)} \right|^2 } \right\}
\end{equation}
\noindent where $\hat b_{R_q D}$ is the estimate of $b_{R_q }$. Two important comments are worth being made. 1) At the end of broadcasting and relaying phases, the destination $D$ has $N_S+N_R$ estimated bits, \emph{i.e.}, $\hat b_{S_t D}$ for $t = 1,2, \ldots ,N_S$ from (\ref{Eq_2}) and $\hat b_{R_q D}$ for $q = 1,2, \ldots ,N_R$ from (\ref{Eq_4}), which can be seen as hard--decision estimates of all the bits transmitted in the network. These estimates are
exploited in \emph{Step 2}, as described below, to retrieve the information bits emitted by the sources and by taking into account NC
operations performed at the relays. 2) Hard--decision demodulation is performed before network decoding, but, as we will better show in
\emph{Step 2} below, the demodulator will take into account the reliability of these estimates when performing network decoding. Similar to
\cite[pp. 18--20]{Zorzi_CoopNC}, we will show that this is instrumental to achieve full--diversity.
\paragraph{Step 2} In this step we take advantage of physical layer methods to develop network demodulation schemes that are robust to the error propagation problem \cite{Zorzi_CoopNC}, \cite{MDR_Springer2010}, and, thus, to the injection into the network, according to (\ref{Eq_2}) and (\ref{Eq_3}), of wrong demodulated bits. This demodulator can be seen as a generalization of diversity--achieving demodulators for cooperative networks without NC \cite{Laneman2007}. The reader can notice that the proposed approach belong to the family of channel--aware detectors \cite{Varshney}, \cite{MDR_CR}. To the best of the authors knowledge, the only notable paper which has recently extended these decoders to cooperative networks with NC is \cite{SchoberGLOBE2010}. However, a single relay node and a fixed network code are considered in \cite{SchoberGLOBE2010}.

Using the ML criterion, $D$ demodulates the bits $b_{S_t }$ ($t = 1,2, \ldots ,N_S$) of the $N_S$ sources as \cite{Proakis}:
\begin{equation} \scriptsize
\label{Eq_5}
\begin{split}
 \left[ {\hat b_{S_1 } ,\hat b_{S_2 } , \ldots ,\hat b_{S_{N_S } } } \right] &= \mathop {\arg \max }\limits_{\tilde b_{S_1 }  \in \left\{ {0,1} \right\}, \ldots ,\tilde b_{S_{N_S } }  \in \left\{ {0,1} \right\}} \left\{ {\mathcal{P}\left( {\left. {\tilde b_{S_1 } ,\tilde b_{S_2 } , \ldots ,\tilde b_{S_{N_S } } } \right|\hat b_{S_1 D} , \ldots ,\hat b_{S_{N_S } D} ,\hat b_{R_1 D} , \ldots ,\hat b_{R_{N_R } D} } \right)} \right\} \\
  &\mathop \propto \limits^{\left( a \right)} \mathop {\arg \max }\limits_{\tilde b_{S_1 }  \in \left\{ {0,1} \right\}, \ldots ,\tilde b_{S_{N_S } }  \in \left\{ {0,1} \right\}} \left\{ {\left[ {\prod\limits_{t = 1}^{N_S } {\mathcal{P}\left( {\left. {\hat b_{S_t D} } \right|\tilde b_{S_t } } \right)} } \right]\left[ {\prod\limits_{q = 1}^{N_R } {\mathcal{P}\left( {\left. {\hat b_{R_q D} } \right|\tilde b_{S_1 } ,\tilde b_{S_2 } , \ldots ,\tilde b_{S_{N_S } } } \right)} } \right]} \right\} \\
  &\mathop \propto \limits^{\left( b \right)} \mathop {\arg \max }\limits_{\tilde b_{S_1 }  \in \left\{ {0,1} \right\}, \ldots ,\tilde b_{S_{N_S } }  \in \left\{ {0,1} \right\}} \left\{ { {\sum\limits_{t = 1}^{N_S } {\ln \left( {\mathcal{P}\left( {\left. {\hat b_{S_t D} } \right|\tilde b_{S_t } } \right)} \right)} } + {\sum\limits_{q = 1}^{N_R } {\ln \left( {\mathcal{P}\left( {\left. {\hat b_{R_q D} } \right|\tilde b_{S_1 } ,\tilde b_{S_2 } , \ldots ,\tilde b_{S_{N_S } } } \right)} \right)} } } \right\} \\
 \end{split}
\end{equation}
\noindent where: i) $\mathcal{P}\left( {\left. X \right|Y} \right)$ denotes the conditional Probability Density Function (PDF) of RV $X$ given
RV $Y$\footnote{\scriptsize Throughout this paper, the PDF of RV $X$ given RV $Y$ is denoted either by $\mathcal{P}\left( {\left. X \right|Y} \right)$ or
by $\mathcal{P}_X \left( {\left.  \cdot  \right|Y} \right)$.}; ii) $\propto$ denotes ``proportional to''; iii) $\mathop \propto \limits^{\left( a \right)}$ is obtained from the Bayes theorem, by exploiting the independence of the detection events in each time--slot, and by taking into account that the emitted bits are
equiprobable; and iv) $\mathop \propto \limits^{\left( b \right)}$ is obtained by moving to the logarithm domain, which preserves optimality.
Due to NC operations, in the second summation in the third row of (\ref{Eq_5}) each addend is conditioned upon all the bits
emitted from the source. In particular, from Section \ref{SystemModel}, we have: $\mathcal{P}\left( {\left. {\hat b_{R_q D} } \right|\tilde b_{S_1 } ,\tilde b_{S_2 } , \ldots ,\tilde b_{S_{N_S } } } \right) = \mathcal{P}\left( {\left. {\hat b_{R_q D} } \right|f_{R_q } \left( {\tilde b_{S_1 } ,\tilde b_{S_2 } , \ldots ,\tilde b_{S_{N_S } } } \right)} \right) = \mathcal{P}\left( {\left. {\hat b_{R_q D} } \right| \tilde b_{R_q } } \right)$ with $\tilde b_{R_q } = f_{R_q } \left(
{\tilde b_{S_1 } ,\tilde b_{S_2 } , \ldots ,\tilde b_{S_{N_S } } } \right)$.

The conditional probabilities in (\ref{Eq_5}) can be computed as follows. By direct inspection, it follows that ${\hat b_{S_t D} }$ for $t = 1,2,
\ldots ,N_S$ turns out to be the outcome of a Binary Symmetric Channel (BSC) with cross--over probability $P_{S_t D}  = \Pr \left\{ {\hat
b_{S_t D}  \ne b_{S_t } } \right\} = Q\left( {\sqrt {2\left( {{{E_m } \mathord{\left/ {\vphantom {{E_m } {N_0 }}} \right.
\kern-\nulldelimiterspace} {N_0 }}} \right)\left| {h_{S_t D} } \right|^2 } } \right)$, where $Q\left( x \right) = \left( {{1 \mathord{\left/
{\vphantom {1 {\sqrt {2\pi } }}} \right. \kern-\nulldelimiterspace} {\sqrt {2\pi } }}} \right)\int_x^{ + \infty } {\exp \left( { - {{t^2 }
\mathord{\left/ {\vphantom {{t^2 } 2}} \right. \kern-\nulldelimiterspace} 2}} \right)dt}$ is the Q--function, $\Pr \left\{  \cdot  \right\}$
denotes probability, and the last equality is due to using BPSK modulation. Accordingly,
${\mathcal{P}\left( {\left. {\hat b_{S_t D} } \right|\tilde b_{S_t } } \right)}$ follows a Bernoulli distribution, \emph{i.e.},
$\mathcal{P}\left( {\left. {\hat b_{S_t D} } \right|\tilde b_{S_t } } \right) = \left( {1 - P_{S_t D} } \right)^{1 - \left| {\hat b_{S_t D} -
\tilde b_{S_t } } \right|} P_{S_t D}^{\left| {\hat b_{S_t D}  - \tilde b_{S_t } } \right|}$. Similar arguments can be used to compute $\mathcal{P}\left( {\left. {\hat b_{R_q D} } \right| \tilde b_{R_q } } \right)$. In particular, for $q = 1,2, \ldots ,N_R$, we have $\mathcal{P}\left( {\left. {\hat b_{R_q D} } \right|\tilde b_{R_q } } \right) = \left( {1 - P_{S_{1:N_S } R_q D} }
\right)^{1 - \left| {\hat b_{R_q D}  - \tilde b_{R_q } } \right|} P_{S_{1:N_S } R_q D}^{\left| {\hat b_{R_q D}  - \tilde b_{R_q } } \right|}$.
However, in this case the cross--over probability $P_{S_{1:N_S } R_q D}  = \Pr \left\{ {\hat b_{R_q D}  \ne f_{R_q } \left( {b_{S_1 } ,b_{S_2 }
, \ldots ,b_{S_{N_S } } } \right)} \right\}$ is no longer related to a single--hop link, but it must be computed by taking into account: i)
dual--hop DemF protocol; and ii) NC operations performed at each relay node. To emphasize this fact, we use the subscript ${S_{1:N_S
} R_q D}$, where ${S_{1:N_S } }$ is a short--hand to denote the $N_S$ sources of the network. This probability is better defined and computed
in Section \ref{CrossoverProbability}.

By substituting ${\mathcal{P}\left( {\left. {\hat b_{S_t D} } \right|\tilde b_{S_t } } \right)}$ and $\mathcal{P}\left( {\left. {\hat
b_{R_q D} } \right| \tilde b_{R_q } } \right)$ in (\ref{Eq_5}), the ML--optimum demodulator simplifies, after some algebra
and by neglecting some terms that have no effect on the demodulation metric, as:
\begin{equation} \scriptsize
\label{Eq_6} \left[ {\hat b_{S_1 } ,\hat b_{S_2 } , \ldots ,\hat b_{S_{N_S } } } \right] \propto \mathop {\arg \min }\limits_{\tilde b_{S_1 }
\in \left\{ {0,1} \right\}, \ldots ,\tilde b_{S_{N_S } }  \in \left\{ {0,1} \right\}} \left\{ {\sum\limits_{t = 1}^{N_S } {\left( {w_{S_t D}
\left| {\hat b_{S_t D}  - \tilde b_{S_t } } \right|} \right)}  + \sum\limits_{q = 1}^{N_R } {\left( {w_{S_{1:N_S } R_q D} \left| {\hat b_{R_q
D}  - \tilde b_{R_q } } \right|} \right)} } \right\}
\end{equation}
\noindent where $w_{S_t D}  = \ln \left[ {{{\left( {1 - P_{S_t D} } \right)} \mathord{\left/ {\vphantom {{\left( {1 - P_{S_t D} } \right)}
{P_{S_t D} }}} \right. \kern-\nulldelimiterspace} {P_{S_t D} }}} \right]$ and $w_{S_{1:N_S } R_q D}  = \ln \left[ {{{\left( {1 - P_{S_{1:N_S }
R_q D} } \right)} \mathord{\left/ {\vphantom {{\left( {1 - P_{S_{1:N_S } R_q D} } \right)} {P_{S_{1:N_S } R_q D} }}} \right.
\kern-\nulldelimiterspace} {P_{S_{1:N_S } R_q D} }}} \right]$ for $t = 1,2, \ldots ,N_S$ and $q = 1,2, \ldots ,N_R$, respectively.

Three comments about (\ref{Eq_6}) are worth being made. 1) We can notice an evident resemblance with the well--known Chase combiner \cite[Eq. (13)]{ChaseCombining}. In spite of the similar structure, two fundamental differences exist between the original Chase combiner and (\ref{Eq_6}): i) the Chase combiner does not consider dual--hop networks, which means that all the packets reach the destination through direct links; and ii) the effect of error propagation caused by relaying and NC is not considered in the Chase combiner. These two differences are very important for two reasons: i) the detector in (\ref{Eq_6}) needs more CSI to work properly; and ii) the end--to--end performance of (\ref{Eq_6}) is affected by relaying and NC operations. Thus, the analysis of the performance of (\ref{Eq_6}) requires new analytical methodologies, as we will better describe in Section \ref{FrameworkABEP}. 2) For large
$N_S$ and $N_R$, the complexity of (\ref{Eq_6}) can be quite involving. As suggested in \cite[p. 19]{Zorzi_CoopNC}, this issue can be
mitigated by using near--optimum demodulation methods (\emph{e.g.}, sphere decoding \cite{Hassibi}), which attain ML optimality with an affordable complexity. 3) The demodulator in (\ref{Eq_6}) needs closed--form expressions of the cross--over probabilities $P_{S_{1:N_S } R_q
D}$, which, in Section \ref{CrossoverProbability}, is shown to depend on the CSI of the source--to--relay links, and on the NC
operations performed at the relay nodes. In general, the estimation of this CSI requires some overhead \cite{Annavajjala}. In Section
\ref{DiversityCodingGain}, we will analyze the impact of CSI on the achievable diversity order.
\subsection{Cross--Over Probabilities of DemF--based Dual--Hop Networks with Binary NC} \label{CrossoverProbability}
In this section, $P_{S_{1:N_S } R_q D}$ is computed in closed--form. \emph{Proposition \ref{CrossProb}} summarizes the main result.
\begin{proposition} \label{CrossProb}
Let us consider system model and notation in Section \ref{SystemModel} and Section \ref{ReceiverDesign}. The exact cross--over probability,
$P_{S_{1:N_S } R_q D}$, for arbitrary binary encoding vectors is $P_{S_{1:N_S } R_q D}  = P_{S_{1:N_S } R_q }  + P_{R_q D}  - 2P_{S_{1:N_S }
R_q } P_{R_q D}$, where $P_{R_q D}  = Q\left( {\sqrt {2\left( {{{E_m } \mathord{\left/ {\vphantom {{E_m } {N_0 }}} \right.
\kern-\nulldelimiterspace} {N_0 }}} \right)\left| {h_{R_q D} } \right|^2 } } \right)$, $P_{S_t R_q }  = Q\left( {\sqrt {2\left( {{{E_m }
\mathord{\left/ {\vphantom {{E_m } {N_0 }}} \right. \kern-\nulldelimiterspace} {N_0 }}} \right)\left| {h_{S_t R_q } } \right|^2 } } \right)$,
and:
\begin{equation} \scriptsize
\label{Eq_7} P_{S_{1:N_S } R_q }  = \sum\limits_{t = 1}^{N_S } {\left[ {g_{S_t R_q } P_{S_t R_q } \prod\limits_{r = t + 1}^{N_S } {\left( {1 -
2g_{S_r R_q } P_{S_r R_q } } \right)} } \right]}
\end{equation}

\smallskip \emph{Proof}: For the generic relay $R_q$, the end--to--end system can be seen as a dual--hop network where: i) the first hop is
given by an equivalent wireless link with $b_{R_q }^{\left( {{\rm{TX}}} \right)}  = f_{R_q } \left( {b_{S_1 } ,b_{S_2 } , \ldots ,b_{S_{N_S } }
} \right)$ at its input and $b_{R_q }  = f_{R_q } \left( {\hat b_{S_1 R_q } ,\hat b_{S_2 R_q } , \ldots ,\hat b_{S_{N_S } R_q } } \right)$ at
its output, respectively; ii) the second hop is given by the wireless link with $b_{R_q }  = f_{R_q } \left( {\hat b_{S_1 R_q } ,\hat b_{S_2
R_q } , \ldots ,\hat b_{S_{N_S } R_q } } \right)$ at its input and ${\hat b_{R_q D} }$ in (\ref{Eq_4}) at its output. Thus, $P_{S_{1:N_S } R_q
D}$ is given by $P_{S_{1:N_S } R_q D}  = \Pr \left\{ {\hat b_{R_q D}  \ne b_{R_q }^{\left( {{\rm{TX}}} \right)} } \right\}$, which, by using
\cite[Eq. 23]{Hasna2003}, is equal to:
\begin{equation} \scriptsize
\label{Eq_8} P_{S_{1:N_S } R_q D}  = \Pr \left\{ {b_{R_q }  \ne b_{R_q }^{\left( {{\rm{TX}}} \right)} } \right\} + \Pr \left\{ {\hat b_{R_q D}
\ne b_{R_q } } \right\} - 2\Pr \left\{ {b_{R_q }  \ne b_{R_q }^{\left( {{\rm{TX}}} \right)} } \right\}\Pr \left\{ {\hat b_{R_q D}  \ne b_{R_q }
} \right\}
\end{equation}

In (\ref{Eq_8}), $\Pr \left\{ {\hat b_{R_q D}  \ne b_{R_q } } \right\} = P_{R_q D}  = Q\left( {\sqrt {2\left( {{{E_m }
\mathord{\left/ {\vphantom {{E_m } {N_0 }}} \right. \kern-\nulldelimiterspace} {N_0 }}} \right)\left| {h_{R_q D} } \right|^2 } } \right)$, as it is the error probability of a single--hop link. On the other hand, $\Pr \left\{ {b_{R_q }  \ne b_{R_q }^{\left( {{\rm{TX}}} \right)} } \right\} = P_{S_{1:N_S } R_q }$ can be explicitly written as:
\begin{equation} \scriptsize
\label{Eq_9} P_{S_{1:N_S } R_q }  = \Pr \left\{ {g_{S_1 R_q } \hat b_{S_1 R_q }  \oplus g_{S_2 R_q } \hat b_{S_2 R_q }  \oplus  \ldots  \oplus
g_{S_{N_S } R_q } \hat b_{S_{N_S } R_q }  \ne g_{S_1 R_q } b_{S_1 R_q }  \oplus g_{S_2 R_q } b_{S_2 R_q }  \oplus  \ldots  \oplus g_{S_{N_S }
R_q } b_{S_{N_S } R_q } } \right\}
\end{equation}

Let us now introduce the notation ($t  = 1,2, \ldots ,N_S$):
\begin{equation} \scriptsize
\label{Eq_10} \left\{ \begin{array}{l}
 P_{S_{1:t} R_q }  = \Pr \left\{ {g_{S_1 R_q } \hat b_{S_1 R_q }  \oplus g_{S_2 R_q } \hat b_{S_2 R_q }  \oplus  \ldots  \oplus g_{S_t R_q } \hat b_{S_t R_q }  \ne g_{S_1 R_q } b_{S_1 R_q }  \oplus g_{S_2 R_q } b_{S_2 R_q }  \oplus  \ldots  \oplus g_{S_t R_q } b_{S_t R_q } } \right\} \\
 P_{S_t R_q }^{\left( {g_{S_t R_q } } \right)}  = \Pr \left\{ {g_{S_t R_q } \hat b_{S_t R_q }  \ne g_{S_t R_q } b_{S_t R_q } } \right\} = g_{S_t R_q } \Pr \left\{ {\hat b_{S_t R_q }  \ne b_{S_t R_q } } \right\} = g_{S_t R_q } P_{S_t R_q }  \\
 \end{array} \right.
\end{equation}
\noindent with $P_{S_{1:t} R_q }  = P_{S_t R_q }^{\left( {g_{S_t R_q } } \right)} = P_{S_1 R_q }^{\left( {g_{S_1 R_q } } \right)}$ if $t=1$.
Furthermore, similar to $P_{R_q D}$, $P_{S_t R_q } = \Pr \left\{ {\hat b_{S_t R_q }  \ne b_{S_t R_q } } \right\} = P_{S_t R_q }  = Q\left( {\sqrt {2\left( {{{E_m } \mathord{\left/ {\vphantom {{E_m } {N_0 }}} \right. \kern-\nulldelimiterspace} {N_0 }}} \right)\left| {h_{S_t R_q } } \right|^2 } } \right)$. By taking into account the properties of the XOR operator, (\ref{Eq_9}) can be computed by using the following chain of recurrence relations:
\begin{equation} \tiny
\label{Eq_11} \left\{ \begin{array}{l}
 P_{S_{1:N_S } R_q }  = P_{S_{1:N_S  - 1} R_q } \left( {1 - P_{S_{N_S } R_q }^{\left( {g_{S_{N_S } R_q } } \right)} } \right) + \left( {1 - P_{S_{1:N_S  - 1} R_q } } \right)P_{S_{N_S } R_q }^{\left( {g_{S_{N_S } R_q } } \right)}  \\
 P_{S_{1:N_S  - 1} R_q }  = P_{S_{1:N_S  - 2} R_q } \left( {1 - P_{S_{N_S  - 1} R_q }^{\left( {g_{S_{N_S  - 1} R_q } } \right)} } \right) + \left( {1 - P_{S_{1:N_S  - 2} R_q } } \right)P_{S_{N_S  - 1} R_q }^{\left( {g_{S_{N_S  - 1} R_q } } \right)}  \\
  \vdots  \\
 P_{S_{1:2} R_q }  = P_{S_{1:1} R_q } \left( {1 - P_{S_2 R_q }^{\left( {g_{S_2 R_q } } \right)} } \right) + \left( {1 - P_{S_{1:1} R_q } } \right)P_{S_2 R_q }^{\left( {g_{S_2 R_q } } \right)}  = P_{S_1 R_q }^{\left( {g_{S_1 R_q } } \right)} \left( {1 - P_{S_2 R_q }^{\left( {g_{S_2 R_q } } \right)} } \right) + \left( {1 - P_{S_1 R_q }^{\left( {g_{S_1 R_q } } \right)} } \right)P_{S_2 R_q }^{\left( {g_{S_2 R_q } } \right)}  \\
 \end{array} \right.
\end{equation}

A closed--form solution of a recurrence relation similar to (\ref{Eq_11}) has recently been given in \cite{Morgado} for multi--hop networks.
In particular, by using \cite[Eq. (9)]{Morgado}, (\ref{Eq_7}) can be obtained. This concludes the proof. \hfill $\Box$
\end{proposition}

\emph{Proposition \ref{CrossProb}} is instrumental for an efficient implementation of (\ref{Eq_6}). Furthermore, the proof sheds lights on the fundamental behavior of NC over fading channels. In fact, by comparing (\ref{Eq_7}) and \cite[Eq. (9)]{Morgado}, we notice that the cumulative error due to performing NC on wrong demodulated bits at the relay is equivalent to the error propagation problem in multi--hop networks. In other words, if the relay performs NC on the data received from $1 \le N_S^ *   \le N_S$ sources, then the error probability of the network--coded data is the same as a multi--hop network with $N_S^ *$ hops having fading channels given by the source--to--relay links. When adding the relay--to--destination link, the end--to--end network behaves like a $N_S^ * + 1$
multi--hop network. In other words, \emph{Proposition \ref{CrossProb}} clearly states that the larger the number of network--coded sources is
(\emph{i.e.}, the larger the number of non--zero elements of the encoding vector ${\bf{g}}_{R_q }$), the more important the error
propagation effect might be. In summary, \emph{Proposition \ref{CrossProb}} provides a simple, compact, and intuitive
characterization of the error propagation caused by DemF relaying and NC over fading channels.
\section{Performance Analysis -- ABEP} \label{FrameworkABEP}
In this section, we provide closed--form expressions of the ABEP for each source of the network. The framework takes into account the DemF
relay protocol and the characteristics of the network code. The departing point of our analysis consists in recognizing that, according to
Section \ref{SystemModel}, the network code can be seen as a $\left( {N_S  + N_R } \right)$--long distributed linear block code, whose first
$N_S$ bits can be seen as systematic information bits, and the last $N_R$ bits can be seen as the parity (redundant) bits. However, there are
two fundamental differences between the system model under analysis and classical linear block codes \cite{Proakis}, \cite{Knopp}: i) the
system model in Section \ref{SystemModel} encompasses a dual--hop network, while state--of--the--art analysis of classical codes usually
considers single--hop transmission; and ii) coding is not performed at the source nodes, but it is performed at the relay nodes. Due to the
distributed nature of the network code and the assumption of realistic fading channels, encoding operations at the relays are inherently
error--prone, as shown in \emph{Proposition \ref{CrossProb}}. Accordingly, new frameworks are needed to characterize the end--to--end
performance of dual--hop networks with NC, similar to the many frameworks that have been developed for cooperative/multi--hop
networks without NC \cite{Morgado}, \cite{MDR_TCOMSep2009}. Note that the frameworks proposed in \cite{XiaoNov2008} and \cite{XiaoMay2009} are not applicable to our setup since fading is neglected and no diversity--achieving demodulators are investigated.

Owing to the inherent similarly between the system model in Section \ref{SystemModel} and distributed linear block codes, we use union--bound
methods to compute the ABEP \cite[Eq. (12.44)]{Simon}. The main difference with respect to state--of--the--art frameworks is the computation of
each individual Average Pairwise Error Probability (APEP), which must account for the DemF protocol and for the error propagation introduced by
NC. Furthermore, in this paper we are interested in computing the ABEP of each source of the network instead of considering frame or codeword
error probabilities, as it is usually done for linear block codes \cite{Proakis}, \cite{Simon}. The reason is that in our distributed system each
source transmits independent information flows, and we are interested in characterizing the error performance of each of them.

Using the union--bound for equiprobable transmitted bits \cite[Eq. (12.44)]{Simon}, the ABEP of source $S_t$ is:
\begin{equation} \scriptsize
\label{Eq_12}
\begin{split}
 {\rm{ABEP}}_{S_t }  & \le \frac{1}{{2^{N_S } }}\sum\limits_{b_{S_1 }  = 0}^1 {\sum\limits_{b_{S_2 }  = 0}^1 { \cdots \sum\limits_{b_{S_{N_S } }  = 0}^1 {\sum\limits_{\bar b_{S_1 }  = 0}^1 {\sum\limits_{\bar b_{S_2 }  = 0}^1 { \cdots \sum\limits_{\bar b_{S_{N_S } }  = 0}^1 {\left[ {{\rm{APEP}}\left( {{\bf{c}} \to {\bf{\bar c}}} \right)\bar \Delta \left( {{\bf{c}}\left[ t \right],{\bf{\bar c}}\left[ t \right]} \right)} \right]} } } } } }  \\
  &\mathop  = \limits^{\left( a \right)} \frac{1}{{2^{N_S } }}\sum\limits_{\bf{b}} {\sum\limits_{{\bf{\bar b}}} {\left[ {{\rm{APEP}}\left( {{\bf{c}} \to {\bf{\bar c}}} \right)\bar \Delta \left( {{\bf{c}}\left[ t \right],{\bf{\bar c}}\left[ t \right]} \right)} \right]} }  \\
 \end{split}
\end{equation}
\noindent where: i) $\mathop  = \limits^{\left( a \right)}$ is a short--hand to avoid multi--fold summations; ii) $\left( \cdot \right)^T$
denotes transpose operations; iii) ${\bf{I}}_{n \times n}$ is an $n \times n$ identity matrix; iv) ${\bf{b}} = \left[ {b_{S_1 } ,b_{S_2 } ,
\ldots ,b_{S_{N_S } } } \right]^T$ and ${\bf{\bar b}} = \left[ {\bar b_{S_1 } ,\bar b_{S_2 } , \ldots ,\bar b_{S_{N_S } } } \right]^T$; v)
${\bf{c}} = {\bf{G}} \odot {\bf{b}}$ and ${\bf{\bar c}} = {\bf{G}} \odot {\bf{\bar b}}$, where ``$\odot$'' indicates that matrix operations
(additions and multiplications) are performed in the Galois field GF(2), ${\bf{\tilde G}} = \left[ {{\bf{g}}_{R_1 } ,{\bf{g}}_{R_2 } , \ldots
,{\bf{g}}_{R_{N_R } } } \right]^T$ is the $N_R \times N_S$ matrix containing the encoding vectors of all the relays, and ${\bf{G}} = \left[
{{\bf{I}}_{N_S \times N_S } |{\bf{\tilde G}}^T} \right]^T$ is the $\left( {N_S  + N_R } \right) \times N_S$ generator matrix of the whole
distributed network code; vi) ${{\bf{z}}\left[ m \right]}$ is the $m$--th entry of vector ${\bf{z}}$; vii) $\bar \Delta \left( {x,y} \right) =
1 - \Delta \left( {x,y} \right)$, where $\Delta \left( {\cdot,\cdot} \right)$ is the Kronecker delta function, \emph{i.e.}, $\Delta \left(
{x,y} \right) = 1$ if $x=y$ and $\Delta \left( {x,y} \right) = 0$ elsewhere; and viii) ${{\rm{APEP}}\left( {{\bf{c}} \to {\bf{\bar c}}}
\right)}$ is the probability, averaged over fading channel statistics, of detecting ${{\bf{\bar c}}}$ when, instead, ${\bf{c}}$ is actually
transmitted, and these are the only two codewords possibly being transmitted. The Kronecker delta function takes into account that a wrong demodulated codeword might not result in an error for the source, $S_t$, under analysis.

The next step is the computation of the APEP for a generic pair of distributed codewords. We proceed in two steps: i) the PEP conditioned
on fading channels is computed; and ii) the conditioning is removed.
\subsection{Computation of ${{\rm{PEP}}\left( {{\bf{c}} \to {\bf{\bar c}}} \right)}$} \label{PEP}
The decision metric in (\ref{Eq_6}) can be rewritten in a more compact form as follows:
\begin{equation} \scriptsize
\label{Eq_13} \sum\limits_{t = 1}^{N_S } {\left( {w_{S_t D} \left| {\hat b_{S_t D}  - \tilde b_{S_t } } \right|} \right)}  + \sum\limits_{q =
1}^{N_R } {\left( {w_{S_{1:N_S } R_q D} \left| {\hat b_{R_q D}  - \tilde b_{R_q } } \right|} \right)}  = \sum\limits_{m = 1}^{N_S  + N_R }
{\left\{ {{\bf{w}}\left[ m \right]\left| {{\bf{\hat c}}\left[ m \right] - {\bf{\tilde c}}\left[ m \right]} \right|} \right\}}  = \Lambda \left(
{{\bf{\tilde c}}} \right)
\end{equation}
\noindent where we have defined: ${\bf{w}} = \left[ {w_{S_1 D} ,w_{S_2 D} , \ldots ,w_{S_{N_S } D} ,w_{S_{1:N_S } R_1 D} ,w_{S_{1:N_S } R_2 D}
, \ldots ,w_{S_{1:N_S } R_{N_R } D} } \right]^T$, ${\bf{\hat b}} = \left[ {\hat b_{S_1 D} ,\hat b_{S_2 D} , \ldots ,\hat b_{S_{N_S } D} ,\hat
b_{R_1 D} ,\hat b_{R_2 D} , \ldots ,\hat b_{R_{N_R } D} } \right]^T$, ${\bf{\tilde b}} = \left[ {\tilde b_{S_1 D} ,\tilde b_{S_2 D} , \ldots
,\tilde b_{S_{N_S } D} ,\tilde b_{R_1 D} ,\tilde b_{R_2 D} , \ldots ,\tilde b_{R_{N_R } D} } \right]^T$, ${\bf{\hat c}} = {\bf{G\hat b}}$, and
${\bf{\tilde c}} = {\bf{G\tilde b}}$. From (\ref{Eq_13}), the PEP, \emph{i.e.}, ${\rm{PEP}}\left( {{\bf{c}} \to {\bf{\bar c}}} \right) = \Pr \left\{ {\Lambda \left( {\bf{c}} \right) > \Lambda \left( {{\bf{\bar c}}} \right)} \right\}$, is:
\begin{equation} \scriptsize
\label{Eq_14} \hspace{-0.25cm}
 {\rm{PEP}}\left( {{\bf{c}} \to {\bf{\bar c}}} \right) = \Pr \left\{ {\sum\limits_{m = 1}^{N_S  + N_R } {{\bf{w}}\left[ m \right]\left( {\left| {{\bf{\hat c}}\left[ m \right] - {\bf{c}}\left[ m \right]} \right| - \left| {{\bf{\hat c}}\left[ m \right] - {\bf{\bar c}}\left[ m \right]} \right|} \right)}  > 0} \right\}
  \mathop  = \limits^{\left( a \right)} \Pr \left\{ {\sum\limits_{m \in \Theta \left( {{\bf{c}},{\bf{\bar c}}} \right)} {{\bf{w}}\left[ m \right]\left( {\left| {{\bf{\hat c}}\left[ m \right] - {\bf{c}}\left[ m \right]} \right| - \left| {{\bf{\hat c}}\left[ m \right] - {\bf{\bar c}}\left[ m \right]} \right|} \right)}  > 0} \right\}
\end{equation}
\noindent where $\mathop  = \limits^{\left( a \right)}$ is obtained by taking into account that $D\left( {{\bf{c}}\left[ m \right],{\bf{\bar
c}}\left[ m \right]} \right) = {\bf{w}}\left[ m \right]\left( {\left| {{\bf{\hat c}}\left[ m \right] - {\bf{c}}\left[ m \right]} \right| -
\left| {{\bf{\hat c}}\left[ m \right] - {\bf{\bar c}}\left[ m \right]} \right|} \right)$ contributes to the summation if and only if
${\bf{c}}\left[ m \right] \ne {\bf{\bar c}}\left[ m \right]$, and, thus, the summation considers only the elements in the set $\Theta \left(
{{\bf{c}},{\bf{\bar c}}} \right) = \left\{ {m|{\bf{c}}\left[ m \right] \ne {\bf{\bar c}}\left[ m \right]} \right\}$. The cardinality, ${\rm{card}}\left\{
\cdot  \right\}$, of $\Theta \left( {{\bf{c}},{\bf{\bar c}}} \right)$ is given by the Hamming distance between ${\bf{c}}$ and ${{\bf{\bar c}}}$,
\emph{i.e.}, ${\rm{card}}\left\{ {\Theta \left( {{\bf{c}},{\bf{\bar c}}} \right)} \right\} = d_H \left( {{\bf{c}},{\bf{\bar c}}} \right) =
\sum\nolimits_{m = 1}^{N_S  + N_R } {\left| {{\bf{c}}\left[ m \right] - {\bf{\bar c}}\left[ m \right]} \right|}$.

By conditioning on ${m \in \Theta \left( {{\bf{c}},{\bf{\bar c}}} \right)}$, it can be shown, by direct inspection, that $D\left(
{{\bf{c}}\left[ m \right],{\bf{\bar c}}\left[ m \right]} \right)$ is a discrete RV which can only assume values ${{\bf{w}}\left[ m \right]}$
and $-{{\bf{w}}\left[ m \right]}$ with probability ${\bf{P}}\left[ m \right]$ and $1-{\bf{P}}\left[ m \right]$, respectively, where ${\bf{P}} =
\left[ {P_{S_1 D} ,P_{S_2 D} , \ldots ,P_{S_{N_S } D} ,P_{S_{1:N_S } R_1 D} ,P_{S_{1:N_S } R_2 D} , \ldots ,P_{S_{1:N_S } R_{N_R } D} }
\right]^T$ is the vector of cross--over probabilities computed in Section \ref{ReceiverDesign}. Accordingly, the conditional PDF of RV $D\left(
{{\bf{c}}\left[ m \right],{\bf{\bar c}}\left[ m \right]} \right)$ is:
\begin{equation} \scriptsize
\label{Eq_15} \mathcal{P}_{D\left( {{\bf{c}}\left[ m \right],{\bf{\bar c}}\left[ m \right]} \right)} \left( {\left. \xi  \right|m \in \Theta
\left( {{\bf{c}},{\bf{\bar c}}} \right)} \right) = {\bf{P}}\left[ m \right]\delta \left( {\xi  - {\bf{w}}\left[ m \right]} \right) + \left( {1
- {\bf{P}}\left[ m \right]} \right)\delta \left( {\xi  + {\bf{w}}\left[ m \right]} \right)
\end{equation}
\noindent where $\delta \left(  \cdot  \right)$ denotes the Dirac delta function. Since the RVs ${D\left( {{\bf{c}}\left[ m \right],{\bf{\bar c}}\left[ m \right]} \right)}$ are independent for ${m \in \Theta \left( {{\bf{c}},{\bf{\bar c}}} \right)}$, then $D\left( {{\bf{c}},{\bf{\bar c}}} \right) = \sum\nolimits_{m \in \Theta \left( {{\bf{c}},{\bf{\bar c}}} \right)} {D\left( {{\bf{c}}\left[ m \right],{\bf{\bar c}}\left[ m \right]} \right)}$ has a PDF given by the
convolution of the $d_H \left( {{\bf{c}},{\bf{\bar c}}} \right)$ PDFs of the individual RVs ${D\left( {{\bf{c}}\left[ m \right],{\bf{\bar
c}}\left[ m \right]} \right)}$. More specifically, let us denote by ${\bf{\bar m}}_{\left( {{\bf{c}},{\bf{\bar c}}} \right)}  = \left\{ {\bar
m_{\left( {{\bf{c}},{\bf{\bar c}}} \right)}^{\left( 1 \right)} ,\bar m_{\left( {{\bf{c}},{\bf{\bar c}}} \right)}^{\left( 2 \right)} , \ldots
,\bar m_{\left( {{\bf{c}},{\bf{\bar c}}} \right)}^{\left( {d_H \left( {{\bf{c}},{\bf{\bar c}}} \right)} \right)} } \right\}$ the specific set
of ${d_H \left( {{\bf{c}},{\bf{\bar c}}} \right)}$ indexes such that ${m \in \Theta \left( {{\bf{c}},{\bf{\bar c}}} \right)}$. Then, the PDF of
$D\left( {{\bf{c}},{\bf{\bar c}}} \right)$ can be written as:
\begin{equation} \scriptsize
\label{Eq_16} \mathcal{P}_{D\left( {{\bf{c}},{\bf{\bar c}}} \right)} \left( \xi  \right) = \left( {\mathcal{P}_{D\left( {{\bf{c}}\left[ {\bar
m_{\left( {{\bf{c}},{\bf{\bar c}}} \right)}^{\left( 1 \right)} } \right],{\bf{\bar c}}\left[ {\bar m_{\left( {{\bf{c}},{\bf{\bar c}}}
\right)}^{\left( 1 \right)} } \right]} \right)}  \otimes \mathcal{P}_{D\left( {{\bf{c}}\left[ {\bar m_{\left( {{\bf{c}},{\bf{\bar c}}}
\right)}^{\left( 2 \right)} } \right],{\bf{\bar c}}\left[ {\bar m_{\left( {{\bf{c}},{\bf{\bar c}}} \right)}^{\left( 2 \right)} } \right]}
\right)}  \otimes \ldots \otimes \mathcal{P}_{D\left( {{\bf{c}}\left[ {\bar m_{\left( {{\bf{c}},{\bf{\bar c}}} \right)}^{\left( {d_H \left(
{{\bf{c}},{\bf{\bar c}}} \right)} \right)} } \right],{\bf{\bar c}}\left[ {\bar m_{\left( {{\bf{c}},{\bf{\bar c}}} \right)}^{\left( {d_H \left(
{{\bf{c}},{\bf{\bar c}}} \right)} \right)} } \right]} \right)} } \right)\left( \xi  \right)
\end{equation}
\noindent where $\otimes$ denotes convolution operations. Thus, by definition, the PEP in (\ref{Eq_14}) can be computed as ${\rm{PEP}}\left( {{\bf{c}} \to {\bf{\bar c}}} \right) = \int\nolimits_0^{ + \infty } {\mathcal{P}_{D\left( {{\bf{c}},{\bf{\bar c}}} \right)} \left( \xi  \right)d\xi }$. A closed--form expression of this PEP is given in \emph{Proposition \ref{PEP}}.
\begin{proposition} \label{PEP}
Let $\mathcal{P}_{D\left( {{\bf{c}},{\bf{\bar c}}} \right)} \left( \cdot  \right)$ in (\ref{Eq_16}), for high--SNR (\emph{i.e.}, ${{E_m }
\mathord{\left/ {\vphantom {{E_m } {N_0 }}} \right. \kern-\nulldelimiterspace} {N_0 }} \to \infty$), ${\rm{PEP}}\left( {{\bf{c}} \to {\bf{\bar
c}}} \right) = \int\nolimits_0^{ + \infty } {\mathcal{P}_{D\left( {{\bf{c}},{\bf{\bar c}}} \right)} \left( \xi \right)d\xi }$ can be tightly
upper--bounded as follows:
\begin{equation} \tiny
\label{Eq_17}
\begin{split}
 {\rm{PEP}}\left( {{\bf{c}} \to {\bf{\bar c}}} \right) &\to \prod\limits_{k = 1}^{d_H \left( {{\bf{c}},{\bf{\bar c}}} \right)} {{\bf{P}}\left[ {\bar m_{\left( {{\bf{c}},{\bf{\bar c}}} \right)}^{\left( k \right)} } \right]}  + \sum\limits_{\scriptstyle k = 1 \atop
  \scriptstyle k \in \Psi _1 \left( {{\bf{c}},{\bf{\bar c}}} \right)}^{ \binom {d_H \left( {{\bf{c}},{\bf{\bar c}}} \right)} {1} } {\min \left\{ {{\bf{P}}\left[ {\bar m_{\left( {{\bf{c}},{\bf{\bar c}}} \right)}^{\left( k \right)} } \right],\prod\limits_{\scriptstyle h = 1 \hfill \atop
  \scriptstyle h \ne k \hfill}^{d_H \left( {{\bf{c}},{\bf{\bar c}}} \right)} {{\bf{P}}\left[ {\bar m_{\left( {{\bf{c}},{\bf{\bar c}}} \right)}^{\left( h \right)} } \right]} } \right\}}  \\
  &+ \sum\limits_{\scriptstyle k = 1 \atop
  {\scriptstyle k \in \Psi _2 \left( {{\bf{c}},{\bf{\bar c}}} \right) \atop
  \scriptstyle {\bf{v}}_k  \in \Phi _2 \left( {{\bf{c}},{\bf{\bar c}}} \right)}}^{ \binom {d_H \left( {{\bf{c}},{\bf{\bar c}}} \right)} {2} } {\min \left\{ {{\bf{P}}\left[ {\bar m_{\left( {{\bf{c}},{\bf{\bar c}}} \right)}^{\left( {{\bf{v}}_k \left[ 1 \right]} \right)} } \right]{\bf{P}}\left[ {\bar m_{\left( {{\bf{c}},{\bf{\bar c}}} \right)}^{\left( {{\bf{v}}_k \left[ 2 \right]} \right)} } \right],\prod\limits_{\scriptstyle h = 1 \atop
  \scriptstyle h \ne {\bf{v}}_k \left[ 1 \right],h \ne {\bf{v}}_k \left[ 2 \right]}^{d_H \left( {{\bf{c}},{\bf{\bar c}}} \right)} {{\bf{P}}\left[ {\bar m_{\left( {{\bf{c}},{\bf{\bar c}}} \right)}^{\left( h \right)} } \right]} } \right\}}  \\
  &+ \sum\limits_{\scriptstyle k = 1 \atop
  {\scriptstyle k \in \Psi _3 \left( {{\bf{c}},{\bf{\bar c}}} \right) \atop
  \scriptstyle {\bf{v}}_k  \in \Phi _3 \left( {{\bf{c}},{\bf{\bar c}}} \right)}}^{ \binom {d_H \left( {{\bf{c}},{\bf{\bar c}}} \right)} {3} } {\min \left\{ {{\bf{P}}\left[ {\bar m_{\left( {{\bf{c}},{\bf{\bar c}}} \right)}^{\left( {{\bf{v}}_k \left[ 1 \right]} \right)} } \right]{\bf{P}}\left[ {\bar m_{\left( {{\bf{c}},{\bf{\bar c}}} \right)}^{\left( {{\bf{v}}_k \left[ 2 \right]} \right)} } \right]{\bf{P}}\left[ {\bar m_{\left( {{\bf{c}},{\bf{\bar c}}} \right)}^{\left( {{\bf{v}}_k \left[ 3 \right]} \right)} } \right],\prod\limits_{\scriptstyle h = 1 \atop
  \scriptstyle h \ne {\bf{v}}_k \left[ 1 \right],h \ne {\bf{v}}_k \left[ 2 \right],h \ne {\bf{v}}_k \left[ 3 \right]}^{d_H \left( {{\bf{c}},{\bf{\bar c}}} \right)} {{\bf{P}}\left[ {\bar m_{\left( {{\bf{c}},{\bf{\bar c}}} \right)}^{\left( h \right)} } \right]} } \right\}}  \\
  &\vdots  \\
  &+ \sum\limits_{\scriptstyle k = 1 \atop
  {\scriptstyle k \in \Psi _{\left\lfloor {d_H \left( {{\bf{c}},{\bf{\bar c}}} \right)/2} \right\rfloor } \left( {{\bf{c}},{\bf{\bar c}}} \right) \atop
  \scriptstyle {\bf{v}}_k  \in \Phi _{\left\lfloor {d_H \left( {{\bf{c}},{\bf{\bar c}}} \right)/2} \right\rfloor } \left( {{\bf{c}},{\bf{\bar c}}} \right)}}^{ \binom{d_H \left( {{\bf{c}},{\bf{\bar c}}} \right)}{\left\lfloor {d_H \left( {{\bf{c}},{\bf{\bar c}}} \right)/2} \right\rfloor}    } {\min \left\{ {{\bf{P}}\left[ {\bar m_{\left( {{\bf{c}},{\bf{\bar c}}} \right)}^{\left( {{\bf{v}}_k \left[ 1 \right]} \right)} } \right] \cdots {\bf{P}}\left[ {\bar m_{\left( {{\bf{c}},{\bf{\bar c}}} \right)}^{\left( {{\bf{v}}_k \left[ {\left\lfloor {d_H \left( {{\bf{c}},{\bf{\bar c}}} \right)/2} \right\rfloor } \right]} \right)} } \right],\prod\limits_{\scriptstyle h = 1 \atop
  {\scriptstyle h \ne {\bf{v}}_k \left[ 1 \right],h \ne {\bf{v}}_k \left[ 2 \right] \atop
  \scriptstyle  \ldots ,h \ne {\bf{v}}_k \left[ {d_H \left( {{\bf{c}},{\bf{\bar c}}} \right)/2} \right]}}^{d_H \left( {{\bf{c}},{\bf{\bar c}}} \right)} {{\bf{P}}\left[ {\bar m_{\left( {{\bf{c}},{\bf{\bar c}}} \right)}^{\left( h \right)} } \right]} }
  \right\}} \\
  \end{split}
\end{equation}
\noindent where: i) $\binom{\cdot}{\cdot}$ is the binomial coefficient; ii) $\left\lfloor  \cdot  \right\rfloor$ is the floor integer part;
iii) $\Psi _n \left( {{\bf{c}},{\bf{\bar c}}} \right)$ is a set of indexes defined as $\Psi _n \left( {{\bf{c}},{\bf{\bar c}}} \right) =
\left\{ {\left. k \right|k + \sum\nolimits_{h = 1}^{n - 1} { \binom{{d_H \left( {{\bf{c}},{\bf{\bar c}}} \right)}}{h}  }  \le 2^{d_H \left(
{{\bf{c}},{\bf{\bar c}}} \right) - 1}  - 1} \right\}$; and iv) $\Phi _n \left( {{\bf{c}},{\bf{\bar c}}} \right)$ is the set of all possible
combinations of the indexes in ${\bf{\bar m}}_{\left( {{\bf{c}},{\bf{\bar c}}} \right)}$ taken in sets of $n$, and it is defined as $\Phi _n
\left( {{\bf{c}},{\bf{\bar c}}} \right) = \left\{ {\left. {{\bf{v}}_k } \right|{\bf{v}}_k  = \left\{ {{\bf{v}}_k \left[ 1 \right],{\bf{v}}_k
\left[ 2 \right], \ldots ,{\bf{v}}_k \left[ n \right]} \right\}} \right\}$, where ${\bf{v}}_k$ is its $k$--th element, \emph{i.e.}, the $k$--th
combination of the indexes in ${\bf{\bar m}}_{\left( {{\bf{c}},{\bf{\bar c}}} \right)}$. The cardinality of $\Phi _n \left( {{\bf{c}},{\bf{\bar
c}}} \right)$ is ${\rm{card}}\left\{ {\Phi _n \left( {{\bf{c}},{\bf{\bar c}}} \right)} \right\} = \binom{{d_H \left( {{\bf{c}},{\bf{\bar c}}}
\right)}}{n}$.

\smallskip \emph{Proof}: We proceed in two steps: i) first, we describe the step--by--step methodology to compute ${\rm{PEP}}\left( {{\bf{c}} \to {\bf{\bar c}}} \right)$ in (\ref{Eq_17}) for $d_H \left( {{\bf{c}},{\bf{\bar c}}} \right) = 3$; and ii) then, we describe how the approach can be generalized to generic $d_H \left( {{\bf{c}},{\bf{\bar c}}} \right)$. Let us start with $d_H \left( {{\bf{c}},{\bf{\bar c}}} \right) = 3$. In this case, we have ${\bf{\bar m}}_{\left( {{\bf{c}},{\bf{\bar c}}} \right)}  = \left\{ {\bar m_{\left( {{\bf{c}},{\bf{\bar c}}} \right)}^{\left( 1 \right)} ,\bar m_{\left( {{\bf{c}},{\bf{\bar c}}} \right)}^{\left( 2 \right)} ,\bar m_{\left( {{\bf{c}},{\bf{\bar c}}} \right)}^{\left( 3 \right)} } \right\}$, and $\mathcal{P}_{D\left( {{\bf{c}},{\bf{\bar c}}} \right)} \left(  \cdot  \right)$ in (\ref{Eq_16}) can be computed by using some properties of the Dirac delta function. By doing so, and substituting the obtained PDF in ${\rm{PEP}}\left( {{\bf{c}} \to {\bf{\bar c}}} \right) = \int\nolimits_0^{ + \infty } {\mathcal{P}_{D\left( {{\bf{c}},{\bf{\bar c}}} \right)} \left( \xi  \right)d\xi }$, we get:
\begin{equation} \tiny
\label{Eq_18}
\begin{split}
 {\rm{PEP}}\left( {{\bf{c}} \to {\bf{\bar c}}} \right) &= {\bf{P}}\left[ {\bar m_{\left( {{\bf{c}},{\bf{\bar c}}} \right)}^{\left( 1 \right)} } \right]{\bf{P}}\left[ {\bar m_{\left( {{\bf{c}},{\bf{\bar c}}} \right)}^{\left( 2 \right)} } \right]{\bf{P}}\left[ {\bar m_{\left( {{\bf{c}},{\bf{\bar c}}} \right)}^{\left( 3 \right)} } \right]\mathcal{H}\left( {{\bf{w}}\left[ {\bar m_{\left( {{\bf{c}},{\bf{\bar c}}} \right)}^{\left( 1 \right)} } \right] + {\bf{w}}\left[ {\bar m_{\left( {{\bf{c}},{\bf{\bar c}}} \right)}^{\left( 2 \right)} } \right] + {\bf{w}}\left[ {\bar m_{\left( {{\bf{c}},{\bf{\bar c}}} \right)}^{\left( 3 \right)} } \right]} \right) \\
  &+ \left( {1 - {\bf{P}}\left[ {\bar m_{\left( {{\bf{c}},{\bf{\bar c}}} \right)}^{\left( 1 \right)} } \right]} \right)\left( {1 - {\bf{P}}\left[ {\bar m_{\left( {{\bf{c}},{\bf{\bar c}}} \right)}^{\left( 2 \right)} } \right]} \right)\left( {1 - {\bf{P}}\left[ {\bar m_{\left( {{\bf{c}},{\bf{\bar c}}} \right)}^{\left( 3 \right)} } \right]} \right)\mathcal{H}\left( { - {\bf{w}}\left[ {\bar m_{\left( {{\bf{c}},{\bf{\bar c}}} \right)}^{\left( 1 \right)} } \right] - {\bf{w}}\left[ {\bar m_{\left( {{\bf{c}},{\bf{\bar c}}} \right)}^{\left( 2 \right)} } \right] - {\bf{w}}\left[ {\bar m_{\left( {{\bf{c}},{\bf{\bar c}}} \right)}^{\left( 3 \right)} } \right]} \right) \\
  &+ {\bf{P}}\left[ {\bar m_{\left( {{\bf{c}},{\bf{\bar c}}} \right)}^{\left( 1 \right)} } \right]\left( {1 - {\bf{P}}\left[ {\bar m_{\left( {{\bf{c}},{\bf{\bar c}}} \right)}^{\left( 2 \right)} } \right]} \right)\left( {1 - {\bf{P}}\left[ {\bar m_{\left( {{\bf{c}},{\bf{\bar c}}} \right)}^{\left( 3 \right)} } \right]} \right)\mathcal{H}\left( {{\bf{w}}\left[ {\bar m_{\left( {{\bf{c}},{\bf{\bar c}}} \right)}^{\left( 1 \right)} } \right] - {\bf{w}}\left[ {\bar m_{\left( {{\bf{c}},{\bf{\bar c}}} \right)}^{\left( 2 \right)} } \right] - {\bf{w}}\left[ {\bar m_{\left( {{\bf{c}},{\bf{\bar c}}} \right)}^{\left( 3 \right)} } \right]} \right) \\
  &+ {\bf{P}}\left[ {\bar m_{\left( {{\bf{c}},{\bf{\bar c}}} \right)}^{\left( 2 \right)} } \right]\left( {1 - {\bf{P}}\left[ {\bar m_{\left( {{\bf{c}},{\bf{\bar c}}} \right)}^{\left( 1 \right)} } \right]} \right)\left( {1 - {\bf{P}}\left[ {\bar m_{\left( {{\bf{c}},{\bf{\bar c}}} \right)}^{\left( 3 \right)} } \right]} \right)\mathcal{H}\left( { - {\bf{w}}\left[ {\bar m_{\left( {{\bf{c}},{\bf{\bar c}}} \right)}^{\left( 1 \right)} } \right] + {\bf{w}}\left[ {\bar m_{\left( {{\bf{c}},{\bf{\bar c}}} \right)}^{\left( 2 \right)} } \right] - {\bf{w}}\left[ {\bar m_{\left( {{\bf{c}},{\bf{\bar c}}} \right)}^{\left( 3 \right)} } \right]} \right) \\
  &+ {\bf{P}}\left[ {\bar m_{\left( {{\bf{c}},{\bf{\bar c}}} \right)}^{\left( 3 \right)} } \right]\left( {1 - {\bf{P}}\left[ {\bar m_{\left( {{\bf{c}},{\bf{\bar c}}} \right)}^{\left( 1 \right)} } \right]} \right)\left( {1 - {\bf{P}}\left[ {\bar m_{\left( {{\bf{c}},{\bf{\bar c}}} \right)}^{\left( 2 \right)} } \right]} \right)\mathcal{H}\left( { - {\bf{w}}\left[ {\bar m_{\left( {{\bf{c}},{\bf{\bar c}}} \right)}^{\left( 1 \right)} } \right] - {\bf{w}}\left[ {\bar m_{\left( {{\bf{c}},{\bf{\bar c}}} \right)}^{\left( 2 \right)} } \right] + {\bf{w}}\left[ {\bar m_{\left( {{\bf{c}},{\bf{\bar c}}} \right)}^{\left( 3 \right)} } \right]} \right) \\
  &+ {\bf{P}}\left[ {\bar m_{\left( {{\bf{c}},{\bf{\bar c}}} \right)}^{\left( 1 \right)} } \right]{\bf{P}}\left[ {\bar m_{\left( {{\bf{c}},{\bf{\bar c}}} \right)}^{\left( 2 \right)} } \right]\left( {1 - {\bf{P}}\left[ {\bar m_{\left( {{\bf{c}},{\bf{\bar c}}} \right)}^{\left( 3 \right)} } \right]} \right)\mathcal{H}\left( {{\bf{w}}\left[ {\bar m_{\left( {{\bf{c}},{\bf{\bar c}}} \right)}^{\left( 1 \right)} } \right] + {\bf{w}}\left[ {\bar m_{\left( {{\bf{c}},{\bf{\bar c}}} \right)}^{\left( 2 \right)} } \right] - {\bf{w}}\left[ {\bar m_{\left( {{\bf{c}},{\bf{\bar c}}} \right)}^{\left( 3 \right)} } \right]} \right) \\
  &+ {\bf{P}}\left[ {\bar m_{\left( {{\bf{c}},{\bf{\bar c}}} \right)}^{\left( 1 \right)} } \right]{\bf{P}}\left[ {\bar m_{\left( {{\bf{c}},{\bf{\bar c}}} \right)}^{\left( 3 \right)} } \right]\left( {1 - {\bf{P}}\left[ {\bar m_{\left( {{\bf{c}},{\bf{\bar c}}} \right)}^{\left( 2 \right)} } \right]} \right)\mathcal{H}\left( {{\bf{w}}\left[ {\bar m_{\left( {{\bf{c}},{\bf{\bar c}}} \right)}^{\left( 1 \right)} } \right] - {\bf{w}}\left[ {\bar m_{\left( {{\bf{c}},{\bf{\bar c}}} \right)}^{\left( 2 \right)} } \right] + {\bf{w}}\left[ {\bar m_{\left( {{\bf{c}},{\bf{\bar c}}} \right)}^{\left( 3 \right)} } \right]} \right) \\
  &+ {\bf{P}}\left[ {\bar m_{\left( {{\bf{c}},{\bf{\bar c}}} \right)}^{\left( 2 \right)} } \right]{\bf{P}}\left[ {\bar m_{\left( {{\bf{c}},{\bf{\bar c}}} \right)}^{\left( 3 \right)} } \right]\left( {1 - {\bf{P}}\left[ {\bar m_{\left( {{\bf{c}},{\bf{\bar c}}} \right)}^{\left( 1 \right)} } \right]} \right)\mathcal{H}\left( { - {\bf{w}}\left[ {\bar m_{\left( {{\bf{c}},{\bf{\bar c}}} \right)}^{\left( 1 \right)} } \right] + {\bf{w}}\left[ {\bar m_{\left( {{\bf{c}},{\bf{\bar c}}} \right)}^{\left( 2 \right)} } \right] + {\bf{w}}\left[ {\bar m_{\left( {{\bf{c}},{\bf{\bar c}}} \right)}^{\left( 3 \right)} } \right]} \right) \\
\end{split}
\end{equation}
\noindent where $\mathcal{H}\left( x \right) = \int\nolimits_0^{ + \infty } {\delta \left( {\xi  - x} \right)d\xi }$ is the Heaviside function: $\mathcal{H}\left( x \right) = 1$ if $x>0$ and $\mathcal{H}\left( x \right) = 0$ elsewhere.

The PEP in (\ref{Eq_18}) can be simplified and can be written in a form that is more useful to compute the average over fading statistics. The
main considerations to this end are as follows: i) since, by definition (see (\ref{Eq_6})), ${\bf{w}}\left[ m \right]
> 0$ for $m = 1,2, \ldots ,d_H \left( {{\bf{c}},{\bf{\bar c}}} \right)$, then $\mathcal{H}\left( {- \sum\nolimits_{k = 1}^{d_H \left(
{{\bf{c}},{\bf{\bar c}}} \right)} {{\bf{w}}\left[ {\bar m_{\left( {{\bf{c}},{\bf{\bar c}}} \right)}^{\left( k \right)} } \right]} } \right) =
0$ and $\mathcal{H}\left( { \sum\nolimits_{k = 1}^{d_H \left( {{\bf{c}},{\bf{\bar c}}} \right)} {{\bf{w}}\left[ {\bar m_{\left(
{{\bf{c}},{\bf{\bar c}}} \right)}^{\left( k \right)} } \right]} } \right) = 1$ for any $d_H \left( {{\bf{c}},{\bf{\bar c}}} \right)$ and for
any ${\bf{\bar m}}_{\left( {{\bf{c}},{\bf{\bar c}}} \right)}$; and ii) in the high--SNR regime, the BEP in (\ref{Eq_18}) can be tightly
upper--bounded by recognizing that $1 - {\bf{P}}\left[ m \right] \to 1$ for $m = 1,2, \ldots ,d_H \left( {{\bf{c}},{\bf{\bar c}}} \right)$.
Furthermore, by exploiting i) and ii), the resulting terms containing the Heaviside function can be grouped in three pairs of two addends each.
For example, a pair in (\ref{Eq_18}) is $Z = Z_1  + Z_2$ with:
\begin{equation} \scriptsize
\label{Eq_18bis} \left\{ \begin{array}{l}
 Z_1  = {\bf{P}}\left[ {\bar m_{\left( {{\bf{c}},{\bf{\bar c}}} \right)}^{\left( 1 \right)} } \right]{\bf{P}}\left[ {\bar m_{\left(
{{\bf{c}},{\bf{\bar c}}} \right)}^{\left( 2 \right)} } \right]\mathcal{H}\left( {{\bf{w}}\left[ {\bar m_{\left( {{\bf{c}},{\bf{\bar c}}}
\right)}^{\left( 1 \right)} } \right] + {\bf{w}}\left[ {\bar m_{\left( {{\bf{c}},{\bf{\bar c}}} \right)}^{\left( 2 \right)} } \right] -
{\bf{w}}\left[ {\bar m_{\left( {{\bf{c}},{\bf{\bar c}}} \right)}^{\left( 3 \right)} } \right]} \right)  \\
 Z_2  = {\bf{P}}\left[ {\bar
m_{\left( {{\bf{c}},{\bf{\bar c}}} \right)}^{\left( 3 \right)} } \right]\mathcal{H}\left( { - {\bf{w}}\left[ {\bar m_{\left(
{{\bf{c}},{\bf{\bar c}}} \right)}^{\left( 1 \right)} } \right] - {\bf{w}}\left[ {\bar m_{\left( {{\bf{c}},{\bf{\bar c}}} \right)}^{\left( 2
\right)} } \right] + {\bf{w}}\left[ {\bar m_{\left( {{\bf{c}},{\bf{\bar c}}} \right)}^{\left( 3 \right)} } \right]} \right)  \\
 \end{array} \right.
\end{equation}
\noindent while the other two pairs can be obtained by direct inspection of (\ref{Eq_18}) accordingly.

For generic $d_H \left( {{\bf{c}},{\bf{\bar c}}} \right)$, pairs as shown in (\ref{Eq_19}) can be obtained:
\begin{equation} \scriptsize
\label{Eq_19} \left\{ \begin{array}{l}
 Z_1  = \prod\limits_{k \in {\rm \mathcal{A}}} {\left\{ {{\bf{P}}\left[ {\bar m_{\left( {{\bf{c}},{\bf{\bar c}}} \right)}^{\left( k \right)} } \right]\mathcal{H}\left( {\sum\limits_{k \in {\rm \mathcal{A}}} {{\bf{w}}\left[ {\bar m_{\left( {{\bf{c}},{\bf{\bar c}}} \right)}^{\left( k \right)} } \right]}  - \sum\limits_{k \in \bar {\rm \mathcal{A}}} {{\bf{w}}\left[ {\bar m_{\left( {{\bf{c}},{\bf{\bar c}}} \right)}^{\left( k \right)} } \right]} } \right)} \right\}}  \\
 Z_2  = \prod\limits_{k \in \bar {\rm \mathcal{A}}} {\left\{ {{\bf{P}}\left[ {\bar m_{\left( {{\bf{c}},{\bf{\bar c}}} \right)}^{\left( k \right)} } \right]\mathcal{H}\left( { - \sum\limits_{k \in {\rm \mathcal{A}}} {{\bf{w}}\left[ {\bar m_{\left( {{\bf{c}},{\bf{\bar c}}} \right)}^{\left( k \right)} } \right]}  + \sum\limits_{k \in \bar {\rm \mathcal{A}}} {{\bf{w}}\left[ {\bar m_{\left( {{\bf{c}},{\bf{\bar c}}} \right)}^{\left( k \right)} } \right]} } \right)} \right\}}  \\
 \end{array} \right.
\end{equation}
\noindent where ${\rm \mathcal{A}}$ and ${\bar {\rm \mathcal{A}}}$ are two sets of indexes such that ${\bf{\bar m}}_{\left( {{\bf{c}},{\bf{\bar
c}}} \right)}  = {\rm \mathcal{A}} \cup \bar {\rm \mathcal{A}}$ and ${\rm \mathcal{A}} \cap \bar {\rm \mathcal{A}} = \emptyset$.

By taking into account that, for high--SNR, we have ${\bf{w}}\left[ m \right] = \ln \left[ {{{\left( {1 - {\bf{P}}\left[ m \right]} \right)}
\mathord{\left/ {\vphantom {{\left( {1 - {\bf{P}}\left[ m \right]} \right)} {{\bf{P}}\left[ m \right]}}} \right. \kern-\nulldelimiterspace}
{{\bf{P}}\left[ m \right]}}} \right] \to  - \ln \left( {{\bf{P}}\left[ m \right]} \right)$, and from the definition of Heaviside function,
$\mathcal{H}\left(  \cdot  \right)$, $Z_1$ and $Z_2$ in (\ref{Eq_19}) simplify as:
\begin{equation} \scriptsize
\label{Eq_20} \hspace{-0.5cm} \begin{array}{l} Z_1  \to {\begin{cases}
   \prod\limits_{k \in {\rm \mathcal{A}}} {{\bf{P}}\left[ {\bar m_{\left( {{\bf{c}},{\bf{\bar c}}} \right)}^{\left( k \right)} } \right]} & {\rm{if}} \prod\limits_{k \in \bar {\rm \mathcal{A}}} {{\bf{P}}\left[ {\bar m_{\left( {{\bf{c}},{\bf{\bar c}}} \right)}^{\left( k \right)} } \right]}  > \prod\limits_{k \in {\rm \mathcal{A}}} {{\bf{P}}\left[ {\bar m_{\left( {{\bf{c}},{\bf{\bar c}}} \right)}^{\left( k \right)} } \right]}  \hfill  \\
   0 & {\rm{elsewhere}}  \hfill  \\
\end{cases}};   \quad   Z_2  \to {\begin{cases}
   \prod\limits_{k \in \bar {\rm \mathcal{A}}} {{\bf{P}}\left[ {\bar m_{\left( {{\bf{c}},{\bf{\bar c}}} \right)}^{\left( k \right)} } \right]} & {\rm{if}} \prod\limits_{k \in \bar {\rm \mathcal{A}}} {{\bf{P}}\left[ {\bar m_{\left( {{\bf{c}},{\bf{\bar c}}} \right)}^{\left( k \right)} } \right]}  < \prod\limits_{k \in {\rm \mathcal{A}}} {{\bf{P}}\left[ {\bar m_{\left( {{\bf{c}},{\bf{\bar c}}} \right)}^{\left( k \right)} } \right]}  \hfill  \\
   0 & {\rm{elsewhere}}  \hfill  \\
\end{cases}} \end{array}
\end{equation}

Thus, for high--SNR, $Z=Z_1+Z_2$ can be re--written as follows:
\begin{equation} \scriptsize
\label{Eq_21} Z \to \left\{ {\begin{array}{*{20}c}
   {\prod\limits_{k \in {\rm \mathcal{A}}} {{\bf{P}}\left[ {\bar m_{\left( {{\bf{c}},{\bf{\bar c}}} \right)}^{\left( k \right)} } \right]} \quad \quad {\rm{if}}\quad \prod\limits_{k \in {\rm \mathcal{A}}} {{\bf{P}}\left[ {\bar m_{\left( {{\bf{c}},{\bf{\bar c}}} \right)}^{\left( k \right)} } \right]}  < \prod\limits_{k \in \bar {\rm \mathcal{A}}} {{\bf{P}}\left[ {\bar m_{\left( {{\bf{c}},{\bf{\bar c}}} \right)}^{\left( k \right)} } \right]} } \hfill  \\
   {\prod\limits_{k \in \bar {\rm \mathcal{A}}} {{\bf{P}}\left[ {\bar m_{\left( {{\bf{c}},{\bf{\bar c}}} \right)}^{\left( k \right)} } \right]} \quad \quad {\rm{if}}\quad \prod\limits_{k \in \bar {\rm \mathcal{A}}} {{\bf{P}}\left[ {\bar m_{\left( {{\bf{c}},{\bf{\bar c}}} \right)}^{\left( k \right)} } \right]}  < \prod\limits_{k \in {\rm \mathcal{A}}} {{\bf{P}}\left[ {\bar m_{\left( {{\bf{c}},{\bf{\bar c}}} \right)}^{\left( k \right)} } \right]} } \hfill  \\
\end{array}} \right. = \min \left\{ {\prod\limits_{k \in {\rm \mathcal{A}}} {{\bf{P}}\left[ {\bar m_{\left( {{\bf{c}},{\bf{\bar c}}} \right)}^{\left( k \right)} } \right]} ,\prod\limits_{k \in \bar {\rm \mathcal{A}}} {{\bf{P}}\left[ {\bar m_{\left( {{\bf{c}},{\bf{\bar c}}} \right)}^{\left( k \right)} } \right]} } \right\}
\end{equation}

In conclusion, by exploiting the properties of the Heaviside function, $\mathcal{H}\left(  \cdot  \right)$ and the high--SNR approximation in
(\ref{Eq_20}), the PEP in (\ref{Eq_18}) can be tightly upper--bounded as follows:
\begin{equation} \scriptsize
\label{Eq_22}
\begin{split}
 {\rm{PEP}}\left( {{\bf{c}} \to {\bf{\bar c}}} \right) &\to {\bf{P}}\left[ {\bar m_{\left( {{\bf{c}},{\bf{\bar c}}} \right)}^{\left( 1 \right)} } \right]{\bf{P}}\left[ {\bar m_{\left( {{\bf{c}},{\bf{\bar c}}} \right)}^{\left( 2 \right)} } \right]{\bf{P}}\left[ {\bar m_{\left( {{\bf{c}},{\bf{\bar c}}} \right)}^{\left( 3 \right)} } \right] + \min \left\{ {{\bf{P}}\left[ {\bar m_{\left( {{\bf{c}},{\bf{\bar c}}} \right)}^{\left( 1 \right)} } \right],{\bf{P}}\left[ {\bar m_{\left( {{\bf{c}},{\bf{\bar c}}} \right)}^{\left( 2 \right)} } \right]{\bf{P}}\left[ {\bar m_{\left( {{\bf{c}},{\bf{\bar c}}} \right)}^{\left( 3 \right)} } \right]} \right\} \\
  &+ \min \left\{ {{\bf{P}}\left[ {\bar m_{\left( {{\bf{c}},{\bf{\bar c}}} \right)}^{\left( 2 \right)} } \right],{\bf{P}}\left[ {\bar m_{\left( {{\bf{c}},{\bf{\bar c}}} \right)}^{\left( 1 \right)} } \right]{\bf{P}}\left[ {\bar m_{\left( {{\bf{c}},{\bf{\bar c}}} \right)}^{\left( 3 \right)} } \right]} \right\} + \min \left\{ {{\bf{P}}\left[ {\bar m_{\left( {{\bf{c}},{\bf{\bar c}}} \right)}^{\left( 3 \right)} } \right],{\bf{P}}\left[ {\bar m_{\left( {{\bf{c}},{\bf{\bar c}}} \right)}^{\left( 1 \right)} } \right]{\bf{P}}\left[ {\bar m_{\left( {{\bf{c}},{\bf{\bar c}}} \right)}^{\left( 2 \right)} } \right]} \right\} \\
 \end{split}
\end{equation}

The result in (\ref{Eq_22}) represents the first part of our proof, and allows us to explain two main aspects of (\ref{Eq_17}): i) its validity
and accuracy for high--SNRs only, as some approximations are used; and ii) the presence of the $\min \left\{ { \cdot , \cdot } \right\}$
function, which comes from grouping pairs of addends, and by exploiting definition and properties of the Heaviside function. The second step is
to provide a justification of (\ref{Eq_17}) for arbitrary $d_H \left( {{\bf{c}},{\bf{\bar c}}} \right)$. First, let us emphasize that, when
possible, the proof for $d_H \left( {{\bf{c}},{\bf{\bar c}}} \right) = 3$ has been given for arbitrary $d_H \left( {{\bf{c}},{\bf{\bar c}}}
\right)$, which provides a first sound proof of the generality of our approach. Second, we emphasize that the interested reader might repeat
the same steps as for the case study with $d_H \left( {{\bf{c}},{\bf{\bar c}}} \right) = 3$ for arbitrary $d_H \left( {{\bf{c}},{\bf{\bar c}}}
\right)$ and eventually lead to (\ref{Eq_17}). The only difficulty if the large number of terms arising when computing the convolution in
(\ref{Eq_16}). So, here we provide only some guidelines to understand (\ref{Eq_17}). The first thing to observe is that (\ref{Eq_22}) can be obtained from (\ref{Eq_17}), and, more specifically, it is given by the first two addends in the right--hand side of (\ref{Eq_17}). The other terms come from the fact that, for $d_H \left( {{\bf{c}},{\bf{\bar c}}} \right) >3$, in (\ref{Eq_21}) we have to consider all possible combinations of the indexes ${\bf{\bar m}}_{\left( {{\bf{c}},{\bf{\bar c}}} \right)}$ taken in sets of $1$, $2$, $3$, etc., since ${\rm \mathcal{A}}$ and ${\bar {\rm \mathcal{A}}}$ in (\ref{Eq_19}) are a partition of the $d_H
\left( {{\bf{c}},{\bf{\bar c}}} \right)$ indexes in ${\bf{\bar m}}_{\left( {{\bf{c}},{\bf{\bar c}}} \right)}$. This explains the presence of
all the other summations in (\ref{Eq_17}), along with the upper limit of each of them. The reason why the upper limit of the last summation is
$\binom{{d_H \left( {{\bf{c}},{\bf{\bar c}}} \right)}}{{\left\lfloor {d_H \left( {{\bf{c}},{\bf{\bar c}}} \right)/2} \right\rfloor }}$ is due
to the equality $\min \left\{ {\prod\nolimits_{k \in {\rm \mathcal{A}}} {{\bf{P}}\left[ {\bar m_{\left( {{\bf{c}},{\bf{\bar c}}}
\right)}^{\left( k \right)} } \right]} ,\prod\nolimits_{k \in \bar {\rm \mathcal{A}}} {{\bf{P}}\left[ {\bar m_{\left( {{\bf{c}},{\bf{\bar c}}}
\right)}^{\left( k \right)} } \right]} } \right\} = \min \left\{ {\prod\nolimits_{k \in \bar {\rm \mathcal{A}}} {{\bf{P}}\left[ {\bar m_{\left(
{{\bf{c}},{\bf{\bar c}}} \right)}^{\left( k \right)} } \right]} ,\prod\nolimits_{k \in {\rm \mathcal{A}}} {{\bf{P}}\left[ {\bar m_{\left(
{{\bf{c}},{\bf{\bar c}}} \right)}^{\left( k \right)} } \right]} } \right\}$, and because only one of these latter terms is explicitly present
in (\ref{Eq_16}). Furthermore, the need to compute all possible combinations of the indexes ${\bf{\bar m}}_{\left( {{\bf{c}},{\bf{\bar c}}}
\right)}$ clearly explains the definition of $\Phi _n \left( {{\bf{c}},{\bf{\bar c}}} \right)$ in (\ref{Eq_17}). The only thing left is to
understand why in each summation the index $k$ must belong to the set $\Psi _n \left( {{\bf{c}},{\bf{\bar c}}} \right)$. The motivation is as
follows. When computing the convolution in (\ref{Eq_16}), the total number of addends in the final result is $2^{d_H \left( {{\bf{c}},{\bf{\bar
c}}} \right)}$. In fact, the convolution of $2^{d_H \left( {{\bf{c}},{\bf{\bar c}}} \right)}$ PDFs is computed, each one given by the summation
of two terms. Among all these $2^{d_H \left( {{\bf{c}},{\bf{\bar c}}} \right)}$ terms, $\prod\nolimits_{k = 1}^{d_H \left( {{\bf{c}},{\bf{\bar
c}}} \right)} {{\bf{P}}\left[ {\bar m_{\left( {{\bf{c}},{\bf{\bar c}}} \right)}^{\left( k \right)} } \right]}$ and $\prod\nolimits_{k = 1}^{d_H
\left( {{\bf{c}},{\bf{\bar c}}} \right)} {\left( {1 - {\bf{P}}\left[ {\bar m_{\left( {{\bf{c}},{\bf{\bar c}}} \right)}^{\left( k \right)} }
\right]} \right)}$ are treated separately in (\ref{Eq_17}). More specifically, $\prod\nolimits_{k = 1}^{d_H \left( {{\bf{c}},{\bf{\bar c}}}
\right)} {{\bf{P}}\left[ {\bar m_{\left( {{\bf{c}},{\bf{\bar c}}} \right)}^{\left( k \right)} } \right]}$ is explicitly shown in (\ref{Eq_17})
as the first addend, while $\prod\nolimits_{k = 1}^{d_H \left( {{\bf{c}},{\bf{\bar c}}} \right)} {\left( {1 - {\bf{P}}\left[ {\bar m_{\left(
{{\bf{c}},{\bf{\bar c}}} \right)}^{\left( k \right)} } \right]} \right)}$ in zero because of the properties of the Heaviside function. The
remaining $2^{d_H \left( {{\bf{c}},{\bf{\bar c}}} \right)}  - 2$ are grouped in pairs of two addends, as shown in (\ref{Eq_19}). Furthermore,
each pair reduces to only one addend as shown in (\ref{Eq_21}). Accordingly, the number of terms in (\ref{Eq_17}) cannot be larger than
${{\left( {2^{d_H \left( {{\bf{c}},{\bf{\bar c}}} \right)}  - 2} \right)} \mathord{\left/ {\vphantom {{\left( {2^{d_H \left(
{{\bf{c}},{\bf{\bar c}}} \right)}  - 2} \right)} 2}} \right. \kern-\nulldelimiterspace} 2} = 2^{d_H \left( {{\bf{c}},{\bf{\bar c}}} \right) -
1}  - 1$. In other words, when the cumulative inequality in $\Psi _n \left( {{\bf{c}},{\bf{\bar c}}} \right)$ is no longer satisfied, we
can stop computing the summations in (\ref{Eq_17}). This concludes the proof. \hfill $\Box$
\end{proposition}

\emph{Proposition \ref{PEP}} is very general and can be applied to any ${d_H \left( {{\bf{c}},{\bf{\bar c}}} \right)}$.
However, it is not an exact result, as it holds for high--SNR only. For the special case ${d_H \left( {{\bf{c}},{\bf{\bar c}}} \right)} = 2$,
an exact expression of the PEP in (\ref{Eq_14}) can be obtained, which, in general, has to be preferred as it is accurate for any SNR. In
\emph{Corollary \ref{PEP_dH2}}, we provide the exact expression of the PEP in (\ref{Eq_14}) without any high--SNR approximations.
\begin{corollary} \label{PEP_dH2}
If  ${d_H \left( {{\bf{c}},{\bf{\bar c}}} \right)} = 2$, then ${\bf{\bar m}}_{\left( {{\bf{c}},{\bf{\bar c}}} \right)}  = \left\{ {\bar
m_{\left( {{\bf{c}},{\bf{\bar c}}} \right)}^{\left( 1 \right)} ,\bar m_{\left( {{\bf{c}},{\bf{\bar c}}} \right)}^{\left( 2 \right)} } \right\}$
and the PEP in (\ref{Eq_14}) is equal to:
\begin{equation} \scriptsize
\label{Eq_23} {\rm{PEP}}\left( {{\bf{c}} \to {\bf{\bar c}}} \right) = \min \left\{ {{\bf{P}}\left[ {\bar m_{\left( {{\bf{c}},{\bf{\bar c}}}
\right)}^{\left( 1 \right)} } \right],{\bf{P}}\left[ {\bar m_{\left( {{\bf{c}},{\bf{\bar c}}} \right)}^{\left( 2 \right)} } \right]} \right\}
\end{equation}

\smallskip \emph{Proof}: The proof follows from analytical steps similar to (\ref{Eq_18}) in \emph{Proposition \ref{PEP}}. In
particular, we have:
\begin{equation} \scriptsize
\label{Eq_24}
\begin{split}
 {\rm{PEP}}\left( {{\bf{c}} \to {\bf{\bar c}}} \right) &= {\bf{P}}\left[ {\bar m_{\left( {{\bf{c}},{\bf{\bar c}}} \right)}^{\left( 1 \right)} } \right]{\bf{P}}\left[ {\bar m_{\left( {{\bf{c}},{\bf{\bar c}}} \right)}^{\left( 2 \right)} } \right] + {\bf{P}}\left[ {\bar m_{\left( {{\bf{c}},{\bf{\bar c}}} \right)}^{\left( 1 \right)} } \right]\left( {1 - {\bf{P}}\left[ {\bar m_{\left( {{\bf{c}},{\bf{\bar c}}} \right)}^{\left( 2 \right)} } \right]} \right)\mathcal{H}\left( {{\bf{w}}\left[ {\bar m_{\left( {{\bf{c}},{\bf{\bar c}}} \right)}^{\left( 1 \right)} } \right] - {\bf{w}}\left[ {\bar m_{\left( {{\bf{c}},{\bf{\bar c}}} \right)}^{\left( 2 \right)} } \right]} \right) \\
  &+ {\bf{P}}\left[ {\bar m_{\left( {{\bf{c}},{\bf{\bar c}}} \right)}^{\left( 2 \right)} } \right]\left( {1 - {\bf{P}}\left[ {\bar m_{\left( {{\bf{c}},{\bf{\bar c}}} \right)}^{\left( 1 \right)} } \right]} \right)\mathcal{H}\left( { - {\bf{w}}\left[ {\bar m_{\left( {{\bf{c}},{\bf{\bar c}}} \right)}^{\left( 1 \right)} } \right] + {\bf{w}}\left[ {\bar m_{\left( {{\bf{c}},{\bf{\bar c}}} \right)}^{\left( 2 \right)} } \right]} \right) \\
 \end{split}
\end{equation}

Unlike \emph{Proposition \ref{PEP}}, there is no need to exploit the high--SNR approximation $1 - {\bf{P}}\left[ m \right] \to 1$. On the
contrary, by using the properties of the Heaviside function, (\ref{Eq_20}), and (\ref{Eq_21}), we get:
\begin{equation} \scriptsize
\label{Eq_25} {\rm{PEP}}\left( {{\bf{c}} \to {\bf{\bar c}}} \right) = \begin{cases}
   {\bf{P}}\left[ {\bar m_{\left( {{\bf{c}},{\bf{\bar c}}} \right)}^{\left( 1 \right)} } \right]{\bf{P}}\left[ {\bar m_{\left( {{\bf{c}},{\bf{\bar c}}} \right)}^{\left( 2 \right)} } \right] + \left( {1 - {\bf{P}}\left[ {\bar m_{\left( {{\bf{c}},{\bf{\bar c}}} \right)}^{\left( 1 \right)} } \right]} \right){\bf{P}}\left[ {\bar m_{\left( {{\bf{c}},{\bf{\bar c}}} \right)}^{\left( 2 \right)} } \right] = {\bf{P}}\left[ {\bar m_{\left( {{\bf{c}},{\bf{\bar c}}} \right)}^{\left( 2 \right)} } \right] & {\rm{if}} \; {\bf{P}}\left[ {\bar m_{\left( {{\bf{c}},{\bf{\bar c}}} \right)}^{\left( 2 \right)} } \right] < {\bf{P}}\left[ {\bar m_{\left( {{\bf{c}},{\bf{\bar c}}} \right)}^{\left( 1 \right)} } \right] \hfill  \\
   {\bf{P}}\left[ {\bar m_{\left( {{\bf{c}},{\bf{\bar c}}} \right)}^{\left( 1 \right)} } \right]{\bf{P}}\left[ {\bar m_{\left( {{\bf{c}},{\bf{\bar c}}} \right)}^{\left( 2 \right)} } \right] + \left( {1 - {\bf{P}}\left[ {\bar m_{\left( {{\bf{c}},{\bf{\bar c}}} \right)}^{\left( 2 \right)} } \right]} \right){\bf{P}}\left[ {\bar m_{\left( {{\bf{c}},{\bf{\bar c}}} \right)}^{\left( 1 \right)} } \right] = {\bf{P}}\left[ {\bar m_{\left( {{\bf{c}},{\bf{\bar c}}} \right)}^{\left( 1 \right)} } \right] & {\rm{if}} \; {\bf{P}}\left[ {\bar m_{\left( {{\bf{c}},{\bf{\bar c}}} \right)}^{\left( 1 \right)} } \right] < {\bf{P}}\left[ {\bar m_{\left( {{\bf{c}},{\bf{\bar c}}} \right)}^{\left( 2 \right)} } \right] \hfill  \\
\end{cases}
\end{equation}
\noindent which clearly leads to (\ref{Eq_23}). This concludes the proof. \hfill $\Box$
\end{corollary}

We note that the main difference between (\ref{Eq_17}) and (\ref{Eq_23}) is the absence in (\ref{Eq_23}) of the first addend in
(\ref{Eq_17}). In fact, this addend simplifies if the high--SNR approximation $1 - {\bf{P}}\left[ m \right] \to 1$ is not used in
(\ref{Eq_25}). This provides a better (and exact) estimate of the PEP. However, this procedure cannot be readily generalized to
network codes with ${d_H \left( {{\bf{c}},{\bf{\bar c}}} \right)} \ge 3$, without having a more complicated expression of the PEP, which is not useful for further analysis, and, more specifically, to remove the conditioning over fading statistics.
\subsection{Computation of ${{\rm{APEP}}\left( {{\bf{c}} \to {\bf{\bar c}}} \right)}$} \label{APEP}
The aim of this section is to provide a closed--form and insightful expression of the APEP, \emph{i.e.}, to average the PEP in (\ref{Eq_17})
over fading channel statistics. In spite of the apparent complexity of (\ref{Eq_17}), \emph{Proposition \ref{APEP}} shows that a surprisingly simple, compact, and insightful result can be obtained for i.n.i.d. fading.
\begin{proposition} \label{APEP}
Let us consider the Rayleigh fading channel model introduced in Section \ref{SystemModel}. The APEP, ${\rm{APEP}}\left( {{\bf{c}} \to
{\bf{\bar c}}} \right) = {\rm{E}}_{\bf{h}} \left\{ {{\rm{PEP}}\left( {{\bf{c}} \to {\bf{\bar c}}} \right)} \right\}$, is as follows:
\begin{equation} \scriptsize
\label{Eq_26}
\begin{split}
 {\rm{APEP}}\left( {{\bf{c}} \to {\bf{\bar c}}} \right) &\to \left( {4\frac{{E_m }}{{N_0 }}} \right)^{ - d_H \left( {{\bf{c}},{\bf{\bar c}}} \right)} \left[ {1 + 2\sqrt \pi  \Gamma \left( {d_H \left( {{\bf{c}},{\bf{\bar c}}} \right) + \frac{1}{2}} \right)\sum\limits_{d = 1}^{\left\lfloor {{{d_H \left( {{\bf{c}},{\bf{\bar c}}} \right)} \mathord{\left/
 {\vphantom {{d_H \left( {{\bf{c}},{\bf{\bar c}}} \right)} 2}} \right.
 \kern-\nulldelimiterspace} 2}} \right\rfloor } {\frac{{\mathcal{N}_d^{\left( {d_H \left( {{\bf{c}},{\bf{\bar c}}} \right)} \right)} }}{{\Gamma \left( {d + \frac{1}{2}} \right)\Gamma \left( {d_H \left( {{\bf{c}},{\bf{\bar c}}} \right) - d + \frac{1}{2}} \right)}}} } \right] \\
  &\times \prod\limits_{m = 1}^{N_S  + N_R } {\chi \left\{ {{\bf{\Delta }}_{{\bf{c}},{\bf{\bar c}}} \left[ m \right]{\bf{\bar \Sigma }}_{\rm{SRD}}^{\left( {\bf{G}} \right)} \left[ m \right]} \right\}}  \\
 \end{split}
\end{equation}
\noindent where:
\begin{equation} \scriptsize
\label{Eq_27} \mathcal{N}_d^{\left( {d_H \left( {{\bf{c}},{\bf{\bar c}}} \right)} \right)}  = \begin{cases}
 \binom{d_H \left( {{\bf{c}},{\bf{\bar c}}} \right)}{d} & {\rm{if}}\quad \sum\limits_{e = 1}^d {\binom {d_H \left( {{\bf{c}},{\bf{\bar c}}} \right)}{e}}  \le 2^{d_H \left( {{\bf{c}},{\bf{\bar c}}} \right) - 1}  - 1 \\
 2^{d_H \left( {{\bf{c}},{\bf{\bar c}}} \right) - 1}  - 1 - \sum\limits_{e = 1}^{d - 1} {\binom {d_H \left( {{\bf{c}},{\bf{\bar c}}} \right)}{e}} & {\rm{if}}\quad \sum\limits_{e = 1}^d {\binom {d_H \left( {{\bf{c}},{\bf{\bar c}}} \right)}{e}}  > 2^{d_H \left( {{\bf{c}},{\bf{\bar c}}} \right) - 1}  - 1 \\
 \end{cases}
\end{equation}
\noindent and: i) ${\rm{E}}_{\bf{h}} \left\{  \cdot  \right\}$ denotes the expectation operator computed over all fading gains of the network
model introduced in Section \ref{SystemModel}; ii) $\chi \left\{ \xi  \right\} = 1$ if $\xi=0$ and $\chi \left\{ \xi  \right\} = \xi$ if $\xi
\ne 0$; iii) ${\bf{\Delta }}_{{\bf{c}},{\bf{\bar c}}}  = {\bf{c}} \oplus {\bf{\bar c}} = \left({\bf{G}} \odot {\bf{b}}\right) \oplus
\left({\bf{G}} \odot {\bf{\bar b}}\right)$; iv) ${\bf{\bar \Sigma }}_{{\rm{SRD}}}^{\left( {\bf{G}} \right)}  = {\bf{\bar \Sigma }}_{{\rm{SD}}}
+ {\bf{\bar \Sigma }}_{{\rm{RD}}}  + {\bf{\bar \Sigma }}_{{\rm{SR}}}^{\left( {\bf{G}} \right)}$; v) ${\bf{\Sigma }}_{{\rm{SD}}}  = \left[ {{1
\mathord{\left/ {\vphantom {1 {\sigma _{S_1 D}^2 }}} \right. \kern-\nulldelimiterspace} {\sigma _{S_1 D}^2 }},{1 \mathord{\left/ {\vphantom {1
{\sigma _{S_2 D}^2 }}} \right. \kern-\nulldelimiterspace} {\sigma _{S_2 D}^2 }}, \ldots ,{1 \mathord{\left/ {\vphantom {1 {\sigma _{S_{N_S }
D}^2 }}} \right. \kern-\nulldelimiterspace} {\sigma _{S_{N_S } D}^2 }}} \right]^T$ and ${\bf{\bar \Sigma }}_{{\rm{SD}}}  = \left[ {{\bf{\Sigma
}}_{{\rm{SD}}}^T ,{\bf{0}}_{1 \times N_R } } \right]^T$, where ${{\bf{0}}_{1 \times n } }$ is a $1 \times n$ all--zero vector; vi) ${\bf{\Sigma
}}_{{\rm{SR}}_q }  = \left[ {{1 \mathord{\left/ {\vphantom {1 {\sigma _{S_1 R_q }^2 }}} \right. \kern-\nulldelimiterspace} {\sigma _{S_1 R_q
}^2 }},{1 \mathord{\left/ {\vphantom {1 {\sigma _{S_2 R_q }^2 }}} \right. \kern-\nulldelimiterspace} {\sigma _{S_2 R_q }^2 }}, \ldots ,{1
\mathord{\left/ {\vphantom {1 {\sigma _{S_{N_S } R_q }^2 }}} \right. \kern-\nulldelimiterspace} {\sigma _{S_{N_S } R_q }^2 }}} \right]^T$,
${\bf{\Sigma }}_{{\rm{SR}}}^{\left( {\bf{G}} \right)}  = \left[ {{\bf{g}}_{R_1 }^T {\bf{\Sigma }}_{{\rm{SR}}_{\rm{1}} } ,{\bf{g}}_{R_2 }^T
{\bf{\Sigma }}_{{\rm{SR}}_{\rm{2}} } , \ldots ,{\bf{g}}_{R_{N_R } }^T {\bf{\Sigma }}_{{\rm{SR}}_{N_R } } } \right]^T$, and ${\bf{\bar \Sigma
}}_{{\rm{SR}}}^{\left( {\bf{G}} \right)}  = \left[ {{\bf{0}}_{1 \times N_S } ,\left( {{\bf{\Sigma }}_{{\rm{SR}}}^{\left( {\bf{G}} \right)} }
\right)^T } \right]^T$; and vii) ${\bf{\Sigma }}_{{\rm{RD}}} = \left[ {{1 \mathord{\left/ {\vphantom {1 {\sigma _{R_1 D}^2 }}} \right.
\kern-\nulldelimiterspace} {\sigma _{R_1 D}^2 }},{1 \mathord{\left/ {\vphantom {1 {\sigma _{R_2 D}^2 }}} \right. \kern-\nulldelimiterspace}
{\sigma _{R_2 D}^2 }}, \ldots ,{1 \mathord{\left/ {\vphantom {1 {\sigma _{R_{N_R } D}^2 }}} \right. \kern-\nulldelimiterspace} {\sigma _{R_{N_R
} D}^2 }}} \right]^T$ and ${\bf{\bar \Sigma }}_{{\rm{RD}}}  = \left[ {{\bf{0}}_{1 \times N_S } ,{\bf{\Sigma }}_{{\rm{RD}}}^T } \right]^T$.
Finally, we emphasize that in ${\bf{\Sigma }}_{{\rm{SR}}}^{\left( {\bf{G}} \right)}$ usual matrix operations are used and arithmetic is not in
GF(2).

\smallskip \emph{Proof}: From the definition of APEP, \emph{i.e.}, ${\rm{APEP}}\left( {{\bf{c}} \to {\bf{\bar c}}} \right) = {\rm{E}}_{\bf{h}}
\left\{ {{\rm{PEP}}\left( {{\bf{c}} \to {\bf{\bar c}}} \right)} \right\}$ and the linearity property of the expectation operator, it follows
that two types of terms in (\ref{Eq_17}) have to be analyzed:
\begin{equation} \scriptsize
\label{Eq_28}
 T_1  = {\rm{E}}_{\bf{h}} \left\{ {\prod\limits_{k = 1}^{d_H \left( {{\bf{c}},{\bf{\bar c}}} \right)} {{\bf{P}}\left[ {\bar m_{\left( {{\bf{c}},{\bf{\bar c}}} \right)}^{\left( k \right)} } \right]} } \right\} \quad {\rm{and}} \quad
 T_2  = {\rm{E}}_{\bf{h}} \left\{ {\min \left\{ {\prod\limits_{k \in {\rm \mathcal{A}}} {{\bf{P}}\left[ {\bar m_{\left( {{\bf{c}},{\bf{\bar c}}} \right)}^{\left( k \right)} } \right]} ,\prod\limits_{k \in \bar {\rm \mathcal{A}}} {{\bf{P}}\left[ {\bar m_{\left( {{\bf{c}},{\bf{\bar c}}} \right)}^{\left( k \right)} } \right]} } \right\}} \right\}
\end{equation}

An asymptotically--tight (for ${{E_m } \mathord{\left/ {\vphantom {{E_m } {N_0  \to \infty }}} \right. \kern-\nulldelimiterspace}
{N_0  \to \infty }}$) approximation of $T_1$ and $T_2$ in (\ref{Eq_28}) can be obtained by using \emph{Lemma \ref{APEP_Lemma1}} and \emph{Lemma
\ref{APEP_Lemma2}} in Appendix \ref{Appendix_Lemmas}. In particular, for high--SNR, (\ref{Eq_28}) simplifies as follows:
\begin{equation} \scriptsize
\label{Eq_29} \left\{ \begin{array}{l}
 T_1  \to \left( {4\frac{{E_m }}{{N_0 }}} \right)^{ - d_H \left( {{\bf{c}},{\bf{\bar c}}} \right)} \prod\limits_{m = 1}^{N_S  + N_R } {\chi \left\{ {{\bf{\Delta }}_{{\bf{c}},{\bf{\bar c}}} \left[ m \right]{\bf{\bar \Sigma }}_{{\rm{SRD}}}^{\left( {\bf{G}} \right)} \left[ m \right]} \right\}}  \\
 T_2  \to \left( {4\frac{{E_m }}{{N_0 }}} \right)^{ - d_H \left( {{\bf{c}},{\bf{\bar c}}} \right)} \left[ {\frac{{2\sqrt \pi  \Gamma \left( {d_H \left( {{\bf{c}},{\bf{\bar c}}} \right) + \frac{1}{2}} \right)}}{{\Gamma \left( {d + \frac{1}{2}} \right)\Gamma \left( {d_H \left( {{\bf{c}},{\bf{\bar c}}} \right) - d + \frac{1}{2}} \right)}}} \right]\prod\limits_{m = 1}^{N_S  + N_R } {\chi \left\{ {{\bf{\Delta }}_{{\bf{c}},{\bf{\bar c}}} \left[ m \right]{\bf{\bar \Sigma }}_{{\rm{SRD}}}^{\left( {\bf{G}} \right)} \left[ m \right]} \right\}}  \\
 \end{array} \right.
\end{equation}
\noindent where $d = {\rm{card}}\left\{ {\rm{A}} \right\}$ denotes the cardinality of set $\mathcal{A}$.

From (\ref{Eq_29}), equation (\ref{Eq_26}) can be obtained from (\ref{Eq_17}) as follows: i) the first addend
in (\ref{Eq_17}) is $T_1$ in (\ref{Eq_28}) and, thus, it can directly be obtained from (\ref{Eq_29}); ii) each $\min \left\{ { \cdot , \cdot }
\right\}$ term in (\ref{Eq_17}) corresponds to $T_2$ in (\ref{Eq_28}) and, thus, it can directly be obtained from (\ref{Eq_29}); and iii) by
carefully studying $T_2$ in (\ref{Eq_29}), it can be noticed that it is independent of the particular sub--set of indexes in $\mathcal{A}$ and
$\mathcal{\bar A}$, as defined in (\ref{Eq_28}). The only thing which matters is the number of indexes in $\mathcal{A}$ and in
$\mathcal{\bar A}$, \emph{i.e.}, their cardinality ${\rm{card}}\left\{ {\rm{\mathcal{A}}} \right\} = d$ and
${\rm{card}}\left\{ {{\rm{\mathcal{\bar A}}}} \right\} = d_H \left( {{\bf{c}},{\bf{\bar c}}} \right) - d$, respectively. For example, if $d_H \left( {{\bf{c}},{\bf{\bar c}}} \right) = 3$ in (\ref{Eq_22}), then ${\rm{E}}_{\bf{h}} \left\{ {\min \left\{
{{\bf{P}}\left[ {\bar m_{\left( {{\bf{c}},{\bf{\bar c}}} \right)}^{\left( 1 \right)} } \right],{\bf{P}}\left[ {\bar m_{\left(
{{\bf{c}},{\bf{\bar c}}} \right)}^{\left( 2 \right)} } \right]{\bf{P}}\left[ {\bar m_{\left( {{\bf{c}},{\bf{\bar c}}} \right)}^{\left( 3
\right)} } \right]} \right\}} \right\} = {\rm{E}}_{\bf{h}} \left\{ {\min \left\{ {{\bf{P}}\left[ {\bar m_{\left( {{\bf{c}},{\bf{\bar c}}}
\right)}^{\left( 2 \right)} } \right],{\bf{P}}\left[ {\bar m_{\left( {{\bf{c}},{\bf{\bar c}}} \right)}^{\left( 1 \right)} }
\right]{\bf{P}}\left[ {\bar m_{\left( {{\bf{c}},{\bf{\bar c}}} \right)}^{\left( 3 \right)} } \right]} \right\}} \right\} = {\rm{E}}_{\bf{h}}
\left\{ {\min \left\{ {{\bf{P}}\left[ {\bar m_{\left( {{\bf{c}},{\bf{\bar c}}} \right)}^{\left( 3 \right)} } \right],{\bf{P}}\left[ {\bar
m_{\left( {{\bf{c}},{\bf{\bar c}}} \right)}^{\left( 1 \right)} } \right]{\bf{P}}\left[ {\bar m_{\left( {{\bf{c}},{\bf{\bar c}}}
\right)}^{\left( 2 \right)} } \right]} \right\}} \right\}$. This remark holds for generic i.n.i.d. channels, and it implies the identity (for $n = 1,2, \ldots ,\left\lfloor {d_H \left( {{\bf{c}},{\bf{\bar c}}} \right)/2} \right\rfloor$):
\begin{equation} \scriptsize
\label{Eq_30} {\rm{E}}_{\bf{h}} \left\{ {\sum\limits_{\scriptstyle k = 1 \atop
  {\scriptstyle k \in \Psi _n \left( {{\bf{c}},{\bf{\bar c}}} \right) \atop
  \scriptstyle {\bf{v}}_k  \in \Phi _n \left( {{\bf{c}},{\bf{\bar c}}} \right)}}^{\binom {d_H \left( {{\bf{c}},{\bf{\bar c}}} \right)} {n} } {\min \left\{ {\prod\limits_{h \in {\rm A}} {{\bf{P}}\left[ {\bar m_{\left( {{\bf{c}},{\bf{\bar c}}} \right)}^{\left( {{\bf{v}}_k \left[ h \right]} \right)} } \right]} ,\prod\limits_{h \in \bar {\rm A}} {{\bf{P}}\left[ {\bar m_{\left( {{\bf{c}},{\bf{\bar c}}} \right)}^{\left( {{\bf{v}}_k \left[ h \right]} \right)} } \right]} } \right\}} } \right\} = \mathcal{N}_d^{\left( {d_H \left( {{\bf{c}},{\bf{\bar c}}} \right)} \right)} T_2
\end{equation}
\noindent where $T_2$ is given in (\ref{Eq_29}), and $\mathcal{N}_d^{\left( {d_H \left( {{\bf{c}},{\bf{\bar c}}} \right)} \right)}$ is the
number of terms in (\ref{Eq_27}) that are actually summed in (\ref{Eq_30}).

By putting together these considerations, and by taking into account that there are $\left\lfloor {d_H \left( {{\bf{c}},{\bf{\bar c}}}
\right)/2} \right\rfloor$ summations with different ${\rm{card}}\left\{ {\rm{A}} \right\} = d$ in (\ref{Eq_17}), we obtain (\ref{Eq_26}). The only missing thing in our proof is to show that $\mathcal{N}_d^{\left( {d_H \left( {{\bf{c}},{\bf{\bar c}}} \right)} \right)}$ has the closed--form expression given in (\ref{Eq_27}). This result follows from the definition of $\Psi _n \left( {{\bf{c}},{\bf{\bar c}}} \right)$ for $n = 1,2, \ldots ,\left\lfloor {d_H \left( {{\bf{c}},{\bf{\bar c}}} \right)/2} \right\rfloor$ in (\ref{Eq_17}). In fact, since $\Psi _n \left( {{\bf{c}},{\bf{\bar c}}} \right) = \left\{
{\left. k \right|k + \sum\nolimits_{h = 1}^{n - 1} { \binom{{d_H \left( {{\bf{c}},{\bf{\bar c}}} \right)}}{h}    }  \le 2^{d_H \left(
{{\bf{c}},{\bf{\bar c}}} \right) - 1}  - 1} \right\}$, the number of elements in each summation in (\ref{Eq_30}) is: i) either $\binom{{d_H
\left( {{\bf{c}},{\bf{\bar c}}} \right)}}{d}$, if we have not reached the maximum number of indexes that can be summed, \emph{i.e.},
${2^{d_H \left( {{\bf{c}},{\bf{\bar c}}} \right) - 1}  - 1}$; ii) or, in the last summation, the remaining indexes if the cumulative
summation in $\Psi _n \left( {{\bf{c}},{\bf{\bar c}}} \right)$ exceeds this maximum number of indexes. Equation (\ref{Eq_27}) summarizes in
formulas these two cases. This concludes the proof. \hfill $\Box$
\end{proposition}

Similar to  \emph{Proposition \ref{PEP}}, the exact APEP can be obtained if $d_H \left( {{\bf{c}},{\bf{\bar c}}} \right) = 2$, as given in \emph{Corollary \ref{APEP_dH2}}.
\begin{corollary} \label{APEP_dH2}
Let us consider the Rayleigh fading channel model introduced in Section \ref{SystemModel}. Then, ${\rm{APEP}}\left( {{\bf{c}} \to {\bf{\bar
c}}} \right) = {\rm{E}}_{\bf{h}} \left\{ {{\rm{PEP}}\left( {{\bf{c}} \to {\bf{\bar c}}} \right)} \right\}$ with ${\rm{PEP}}\left( {{\bf{c}} \to
{\bf{\bar c}}} \right)$ given in (\ref{Eq_23}) for $d_H \left( {{\bf{c}},{\bf{\bar c}}} \right) = 2$ is as follows:
\begin{equation} \scriptsize
\label{Eq_31} {\rm{APEP}}\left( {{\bf{c}} \to {\bf{\bar c}}} \right) = \left( {\sqrt 2 \frac{{E_m }}{{N_0 }}} \right)^{ - 2} \prod\limits_{m =
1}^{N_S  + N_R } {\chi \left\{ {{\bf{\Delta }}_{{\bf{c}},{\bf{\bar c}}} \left[ m \right]{\bf{\bar \Sigma }}_{{\rm{SRD}}}^{\left( {\bf{G}}
\right)} \left[ m \right]} \right\}}
\end{equation}
\noindent where the same symbols and notation as in \emph{Proposition \ref{APEP}} are used.

\smallskip \emph{Proof}: It follows from (\ref{Eq_26}) with $d_H \left(
{{\bf{c}},{\bf{\bar c}}} \right) = 2$, by neglecting the ``1'' term as shown in \emph{Corollary \ref{PEP_dH2}}. \hfill $\Box$
\end{corollary}
\subsection{Particular Fading Channels} \label{ParticularChannels}
\emph{Proposition \ref{APEP}} is general and it can be applied to arbitrary i.n.i.d fading channels and network
topologies with generic binary NC. However, it is interesting to see what happens to the network performance for some
special channel models and operating conditions, which are often studied to shed lights on the fundamental behavior of complex systems. In this
section, we are interested in providing some simplified results for three notable scenarios of interest: i) i.i.d. fading, where we have
$\sigma _{XY}^2 = \sigma _0^2$ for every wireless link; ii) i.n.i.d. fading with high--reliable source--to--relay links, which is often
assumed to simplify the analysis, but, as described in \emph{Proposition \ref{CrossoverProbability}}, it does not account for the error
propagation effect due to NC; and iii) i.i.d. scenario with high--reliable source--to--relay links. The end--to--end APEP of these three
scenarios is summarized in \emph{Corollary \ref{APEP_IID}}, \emph{Corollary \ref{APEP_IdealSR}}, and \emph{Corollary \ref{APEP_IID_IdealSR}},
respectively.
\begin{corollary} \label{APEP_IID}
If the fading channels are i.i.d. with $\sigma _{XY}^2  = \sigma _0^2$, then the APEP in \emph{Proposition \ref{APEP}} and in \emph{Corollary
\ref{APEP_dH2}} can be simplified by taking into account the following identity:
\begin{equation} \scriptsize
\label{Eq_32} \prod\limits_{m = 1}^{N_S  + N_R } {\chi \left\{ {{\bf{\Delta }}_{{\bf{c}},{\bf{\bar c}}} \left[ m \right]{\bf{\bar \Sigma
}}_{{\rm{SRD}}}^{\left( {\bf{G}} \right)} \left[ m \right]} \right\}}  = \left( {\sigma _0^2 } \right)^{ - d_H \left( {{\bf{c}},{\bf{\bar c}}}
\right)} \prod\limits_{m = 1}^{N_S  + N_R } {\chi \left\{ {{\bf{\Delta }}_{{\bf{c}},{\bf{\bar c}}} \left[ m \right]{\bf{g}}^{\left( 0 \right)}
\left[ m \right]} \right\}}
\end{equation}
\noindent where: i) ${{\bf{1}}_{1 \times N_S } }$ is a $1 \times N_S$ all--one vector; ii) $g_{R_q }^{\left( 0 \right)}  = 1 + \sum\nolimits_{t
= 1}^{N_S } {g_{S_t R_q } } = 1 + N_S^{\left( {{\rm{eff}},R_q } \right)}$ for $q = 1,2, \ldots ,N_R$, where $N_S^{\left( {{\rm{eff}},R_q }
\right)}$ is the number of sources whose data is network--coded at relay node $R_q$; and iii) ${\bf{g}}^{\left( 0 \right)} = \left[
{{\bf{1}}_{1 \times N_S } ,g_{R_1 }^{\left( 0 \right)} ,g_{R_2 }^{\left( 0 \right)} , \ldots ,g_{R_{N_R } }^{\left( 0 \right)} } \right]^T$.

\smallskip \emph{Proof}: It follows from \emph{Proposition \ref{APEP}} with $\sigma _{XY}^2  = \sigma _0^2$. In particular,
${\bf{\Sigma }}_{{\rm{SD}}}$ and ${\bf{\Sigma }}_{{\rm{RD}}}$ simplify to all--one vectors multiplied by ${1
\mathord{\left/ {\vphantom {1 {\sigma _0^2 }}} \right. \kern-\nulldelimiterspace} {\sigma _0^2 }}$, and each entry of ${\bf{\Sigma
}}_{{\rm{SR}}}^{\left( {\bf{G}} \right)}$ reduces to the summation of the elements of the binary encoding vector used at each relay, which
is equal to the number of network--coded sources. \hfill $\Box$
\end{corollary}

The result in (\ref{Eq_32}) is very interesting as it clearly shows, through $N_S^{\left( {{\rm{eff}},R_q } \right)}$, that the larger the
number of network--coded sources is, the more pronounced the error propagation problem might be. Thus, depending on the quality of the fading
channels, it might be more or less convenient to mix at each relay the data packets transmitted from all the sources. Further
comments are postponed to Section \ref{Insights_APEP}.
\begin{corollary} \label{APEP_IdealSR}
Let us assume that the source--to--relay channels are very reliable, \emph{i.e.}, no demodulation errors at the relays. For example,
this can be achieved either by using very powerful error correction codes on the source--to--relay links, or when the relays are located very
close to the sources. Then, ${\bf{\bar \Sigma }}_{{\rm{SRD}}}^{\left( {\bf{G}} \right)}$ in \emph{Proposition \ref{APEP}} and \emph{Corollary
\ref{APEP_dH2}} simplifies to ${\bf{\bar \Sigma }}_{{\rm{SRD}}}^{\left( {\bf{G}} \right)}  = {\bf{\bar \Sigma }}_{{\rm{SD}}}  + {\bf{\bar
\Sigma }}_{{\rm{RD}}} = \left[ {{1 \mathord{\left/ {\vphantom {1 {\sigma _{S_1 D}^2 }}} \right. \kern-\nulldelimiterspace} {\sigma _{S_1 D}^2
}}, \ldots ,{1 \mathord{\left/ {\vphantom {1 {\sigma _{S_{N_S } D}^2 }}} \right. \kern-\nulldelimiterspace} {\sigma _{S_{N_S } D}^2 }},{1
\mathord{\left/ {\vphantom {1 {\sigma _{R_1 D}^2 }}} \right. \kern-\nulldelimiterspace} {\sigma _{R_1 D}^2 }}, \ldots ,{1 \mathord{\left/
{\vphantom {1 {\sigma _{R_{N_R } D}^2 }}} \right. \kern-\nulldelimiterspace} {\sigma _{R_{N_R } D}^2 }}} \right]^T$.

\smallskip \emph{Proof}: If the source--to--relay channels are very reliable, we have
$\sigma _{S_t R_q }^2  \to \infty$ for $t = 1,2, \ldots ,N_S$ and $q = 1,2, \ldots ,N_R$. Thus, by definition, ${\bf{\bar \Sigma
}}_{{\rm{SR}}}^{\left( {\bf{G}} \right)}  \to {\bf{0}}_{\left( {(N_S  + N_R)  \times 1} \right)}$. So, the simplified expression
of ${\bf{\bar \Sigma }}_{{\rm{SRD}}}^{\left( {\bf{G}} \right)}$ follows by taking into account the definition of ${\bf{\bar \Sigma
}}_{{\rm{SD}}}$ and ${\bf{\bar \Sigma }}_{{\rm{RD}}}$ as block matrices. This concludes the proof. \hfill $\Box$
\end{corollary}

Two important conclusions can be drawn from \emph{Corollary \ref{APEP_IdealSR}}. First, we notice that the APEP is affected by the encoding
operations performed at the relays only through the codeword's distance ${d_H \left( {{\bf{c}},{\bf{\bar c}}} \right)}$, which is the number of
distinct elements between $\bf{c}$ and $\bf{\bar c}$. This provides a very simple criterion to choose the network code for performance
optimization. Second, since $\left. {{\bf{\bar \Sigma }}_{{\rm{SRD}}}^{\left( {\bf{G}} \right)} \left[ m \right]} \right|_{\sigma _{S_t R_q }^2
\to \infty }  \le \left. {{\bf{\bar \Sigma }}_{{\rm{SRD}}}^{\left( {\bf{G}} \right)} \left[ m \right]} \right|_{\sigma _{S_t R_q }^2  < \infty
}$ for $m = 1,2, \ldots ,\left( {N_S  + N_R } \right)$, then $\left. {{\rm{APEP}}\left( {{\bf{c}} \to {\bf{\bar c}}} \right)} \right|_{\sigma
_{S_t R_q }^2  \to \infty }  \le \left. {{\rm{APEP}}\left( {{\bf{c}} \to {\bf{\bar c}}} \right)} \right|_{\sigma _{S_t R_q }^2  < \infty }$,
which is an expected result, and it confirms that, to limit the error propagation due to NC operations, the source--to--relay links should be
as reliable as possible. Further comments are postponed to Section \ref{Insights_APEP}.
\begin{corollary} \label{APEP_IID_IdealSR}
If the fading channels on the source--to--destination and relay--to--destination links are i.i.d. with $\sigma _{XY}^2  = \sigma _0^2$, and the
source--to--relay channels are very reliable with no decoding errors at the relays, then \emph{Proposition \ref{APEP}} and
\emph{Corollary \ref{APEP_dH2}} can be simplified by taking into account the identity:
\begin{equation} \scriptsize
\label{Eq_33} \prod\limits_{m = 1}^{N_S  + N_R } {\chi \left\{ {{\bf{\Delta }}_{{\bf{c}},{\bf{\bar c}}} \left[ m \right]{\bf{\bar \Sigma
}}_{{\rm{SRD}}}^{\left( {\bf{G}} \right)} \left[ m \right]} \right\}}  = \left( {\sigma _0^2 } \right)^{ - d_H \left( {{\bf{c}},{\bf{\bar c}}}
\right)}
\end{equation}

\smallskip \emph{Proof}: It follows from \emph{Corollary \ref{APEP_IdealSR}}, which for i.i.d. source--to--destination
and relay--to--destination links gives ${\bf{\bar \Sigma }}_{{\rm{SRD}}}^{\left( {\bf{G}} \right)}  = \left( {{1 \mathord{\left/ {\vphantom {1
{\sigma _0^2 }}} \right. \kern-\nulldelimiterspace} {\sigma _0^2 }}} \right){\bf{1}}_{\left( {(N_S  + N_R)  \times 1} \right)}$. Since there are ${d_H \left( {{\bf{c}},{\bf{\bar c}}} \right)}$ non--zero terms in ${{\bf{\Delta }}_{{\bf{c}},{\bf{\bar c}}} }$, we get (\ref{Eq_33}). \hfill $\Box$
\end{corollary}
\section{Analysis of Diversity Order and Coding Gain} \label{DiversityCodingGain}
To better understand the performance of the cooperative network under analysis, and to clearly showcase the impact of the distributed network
code on the end--to--end performance, in this section we study diversity order and coding gain according to the definition given in
\cite{Giannakis}. In particular, we are interested in re--writing the end--to--end ABEP in (\ref{Eq_12}) as ${\rm{ABEP}}_{S_t }  \to \left[
{\left( {{{E_m } \mathord{\left/ {\vphantom {{E_m } {N_0 }}} \right. \kern-\nulldelimiterspace} {N_0 }}} \right)G_c^{\left( {S_t } \right)} }
\right]^{ - G_d^{\left( {S_t } \right)} }$, where ${G_c^{\left( {S_t } \right)} }$ and ${G_d^{\left( {S_t } \right)} }$ are coding gain and
diversity order of $S_t$ for $t = 1,2, \ldots ,N_S$, respectively. This result is summarized in \emph{Proposition
\ref{CodingDiversityGain}}.
\begin{proposition} \label{CodingDiversityGain}
Given the ABEP in (\ref{Eq_12}) and the APEP in (\ref{Eq_26}), diversity order and coding gain of $S_t$ are:
\begin{equation} \scriptsize
\label{Eq_34} \left\{ \begin{array}{l}
 G_d^{\left( {S_t } \right)}  = {\rm{SV}}\left[ t \right] \\
 G_c^{\left( {S_t } \right)}  = 4\left\{ {\frac{1}{{2^{N_S } }}\sum\limits_{\scriptstyle {\bf{b}},{\bf{\bar b}} \atop
  \scriptstyle d_H \left( {{\bf{c}},\bar {\bf{c}}} \right) = {\rm{SV}}\left[ t \right]} {\left[ \begin{array}{l}
 \left( {1 + 2\sqrt \pi  \Gamma \left( {d_H \left( {{\bf{c}},{\bf{\bar c}}} \right) + \frac{1}{2}} \right)\sum\limits_{d = 1}^{\left\lfloor {{{d_H \left( {{\bf{c}},{\bf{\bar c}}} \right)} \mathord{\left/
 {\vphantom {{d_H \left( {{\bf{c}},{\bf{\bar c}}} \right)} 2}} \right.
 \kern-\nulldelimiterspace} 2}} \right\rfloor } {\frac{{\mathcal{N}_d^{\left( {d_H \left( {{\bf{c}},{\bf{\bar c}}} \right)} \right)} }}{{\Gamma \left( {d + \frac{1}{2}} \right)\Gamma \left( {d_H \left( {{\bf{c}},{\bf{\bar c}}} \right) - d + \frac{1}{2}} \right)}}} } \right) \\
  \times \left( {\prod\limits_{m = 1}^{N_S  + N_R } {\chi \left\{ {{\bf{\Delta }}_{{\bf{c}},{\bf{\bar c}}} \left[ m \right]{\bf{\bar \Sigma }}_{{\rm{SRD}}}^{\left( {\bf{G}} \right)} \left[ m \right]} \right\}} } \right)\bar \Delta \left( {{\bf{c}}\left[ t \right],{\bf{\bar c}}\left[ t \right]} \right) \\
 \end{array} \right]} } \right\}^{ - \frac{1}{{G_d^{\left( {S_t } \right)} }}}  \\
 \end{array} \right.
\end{equation}
\noindent where ${\rm{SV}}$ is known, in coding theory, as ``Separation Vector'' (SV) \cite[Def. 1]{Dunning}, and, for a given codebook
$\mathcal{C} = \left\{ {\left. {\bf{c}} \right|{\bf{c}} = {\bf{Gb}},\,\,\forall {\bf{b}}} \right\}$, its $t$--th entry, \emph{i.e.},
${\rm{SV}}\left[ t \right]$, is defined as the minimum Hamming distance between any pair of codewords ${\bf{c}} = {\bf{Gb}}  \in \mathcal{C}$
and ${\bf{\bar c}} = {\bf{G \bar b}} \in \mathcal{C}$ with different $t$--th bit, \emph{i.e.}, with ${\bf{b}}\left[ t \right] \ne {\bf{\bar
b}}\left[ t \right]$.

\smallskip \emph{Proof}: First of all, let us study $G_d^{\left( {S_t } \right)}$. From (\ref{Eq_26}) in \emph{Proposition \ref{APEP}} we notice
that the APEP has diversity order $d_H \left( {{\bf{c}},{\bf{\bar c}}} \right)$ \cite{Giannakis}, which is the Hamming distance between the
pair of codewords ${\bf{c}}$ and ${{\bf{\bar c}}}$. Furthermore, from (\ref{Eq_12}) we know that this APEP contributes to the ABEP of source
$S_t$ if and only if the $t$--th bits of ${\bf{c}}$ and ${{\bf{\bar c}}}$ are different, \emph{i.e.}, if and only if ${\bf{c}}\left[ t \right]
\ne {\bf{\bar c}}\left[ t \right]$. Since the network codes studied in this paper can be seen as systematic linear block codes, as explained in
Section \ref{SystemModel}, the latter condition implies ${\bf{b}}\left[ t \right] \ne {\bf{\bar b}}\left[ t \right]$. Accordingly, in
(\ref{Eq_26}) only the APEPs having a diversity order, \emph{i.e.}, a Hamming distance, equal to:
\begin{equation} \scriptsize
\label{Eq_35} d_H^{\left( {\min } \right)} \left( t \right) = \left\{ {\left. {d_H \left( {{\bf{c}},{\bf{\bar c}}} \right)} \right|d_H \left(
{{\bf{c}},{\bf{\bar c}}} \right) \le d_H \left( {{\bf{c}}^{'} ,{\bf{\bar c}}^{'} } \right)\;\;\forall {\bf{c}},{\bf{\bar c}},{\bf{c}}^{'}
,{\bf{\bar c}}^{'}  \in \mathcal{C}\;{\rm{with}}\;{\bf{c}}\left[ t \right] \ne {\bf{\bar c}}\left[ t \right]\;{\rm{and}}\;{\bf{c}}^{'} \left[ t
\right] \ne {\bf{\bar c}}^{'} \left[ t \right]} \right\}
\end{equation}
\noindent will dominate the performance for high--SNR. In fact, all the other APEPs will decay much faster with the SNR, thus providing a
negligible contribution. In formulas, the ABEP in (\ref{Eq_12}) can be re--written as:
\begin{equation} \scriptsize
\label{Eq_36} {\rm{ABEP}}_{S_t }  \le \frac{1}{{2^{N_S } }}\sum\limits_{\bf{b}} {\sum\limits_{{\bf{\bar b}}} {\left[ {{\rm{APEP}}\left(
{{\bf{c}} \to {\bf{\bar c}}} \right)\bar \Delta \left( {{\bf{c}}\left[ t \right],{\bf{\bar c}}\left[ t \right]} \right)} \right]} }  \to
\frac{1}{{2^{N_S } }}\sum\limits_{\scriptstyle {\bf{b}},{\bf{\bar b}} \atop
  \scriptstyle d_H \left( {{\bf{c}},{\bf{\bar c}}} \right) = d_H^{\left( {\min } \right)} \left( t \right)} {\left[ {{\rm{APEP}}\left( {{\bf{c}} \to {\bf{\bar c}}} \right)\bar \Delta \left( {{\bf{c}}\left[ t \right],{\bf{\bar c}}\left[ t \right]} \right)} \right]}
\end{equation}

From (\ref{Eq_35}) and (\ref{Eq_36}), by definition \cite[Def. 1]{Dunning}, $d_H^{\left( {\min }
\right)} \left( t \right)$ is exactly the $t$--th entry of SV, \emph{i.e.}, $d_H^{\left( {\min } \right)} \left( t \right) = {\rm{SV}}\left[ t
\right]$. Thus, we have proved that the end--to--end diversity order of source $S_t$ is equal to its SV. This result showcases
that, depending on the used network code, different sources in the network have, in general, different diversity orders. This observation has
important applications, as described in Section \ref{Insights_APEP}. Finally, the coding gain, $G_c^{\left( {S_t } \right)}$, can be obtained through algebraic manipulations by substituting (\ref{Eq_26}) in (\ref{Eq_36}), and equating the resulting expression to ${\rm{ABEP}}_{S_t }  \to \left[ {\left( {{{E_m } \mathord{\left/ {\vphantom {{E_m } {N_0 }}} \right. \kern-\nulldelimiterspace} {N_0 }}} \right)G_c^{\left( {S_t } \right)} } \right]^{ - G_d^{\left( {S_t }
\right)} }$. This concludes the proof. \hfill $\Box$
\end{proposition}
\subsection{Insights from the Analytical Framework} \label{Insights_APEP}
Even though the overall analytical derivation and proof to get (\ref{Eq_26}) in \emph{Proposition \ref{APEP}} are quite analytically involving,
the final expression of the APEP turns out to be very compact, elegant, and simple to compute. In Section \ref{Results}, via Monte Carlo simulations, we will substantiate its accuracy for high--SNR. In addition, the framework is very insightful, as it provides, via direct inspection, important considerations on how the network code affects the performance of the cooperative network, as well as how it can be optimized to improve the end--to--end performance. Important insights from the analytical framework are as follows. \vspace{-0.05cm}
\begin{list}{$\bullet$}{\leftmargin=0em \itemindent=1.5em}
\item \emph{End--to--end diversity order}. As far as diversity is concerned, in \emph{Proposition \ref{CodingDiversityGain}} we have proved that each
source can achieve a diversity order that is equal to the separation vector of the network code. This is a very important result as it shows
that even though a dual--hop network is considered, which is prone to error propagation due to relaying and to demodulation errors
that might happen at each relay node, the distance properties of the network code are still preserved as far as the end--to--end performance is
concerned. This result allows us to conclude that, if we want to guarantee a given diversity order for a given source, we can use conventional
linear block codes as network codes, and be sure that the end--to--end diversity order (and, thus, the error correction capabilities
\cite{Masnick_Oct1967}, \cite{Boyarinov_Marc1981}, \cite{Dunning}) of these codes is preserved even in the presence of error propagation due to
relaying and NC operations. This result and its proof is, to the best of the authors knowledge, new, as it is often assumed a priori that the
presence of error propagation does not affect the diversity properties of the network code \cite{XiaoICC2011}. On a practical point view, this
result suggests that, as far as only the diversity order is concerned, the network codes can be designed by using the same optimization criteria
as for single--hop networks. Finally, we note that the result obtained in this paper is more general than \cite{GiannakisEURASIP2008}, as our
proof is not based on the Singleton bound, and, more important, no \emph{ad hoc} interleavers are needed to achieve a distributed diversity equal to the SV.
\item \emph{Comparison with single--hop network and classical coding theory}. It is interesting to compare the result about the achievable
diversity order in \emph{Proposition \ref{CodingDiversityGain}} with the diversity order that is achievable in single--hop networks. From
\cite[Sec. 14--6--1)]{Proakis}, \cite[Ch. 12]{Simon}, and \cite[Sec. II]{Knopp}, we know that single--hop networks operating in
fully--interleaved fading channels and using soft--decision decoding have a diversity order that is equal to the minimum distance of the linear
code. The result in \emph{Proposition \ref{CodingDiversityGain}} can be seen as a generalization of the analysis of single--hop networks in
\cite{Proakis}, \cite{Simon}, \cite{Knopp} to dual--hop networks with NC. It is important to emphasize that in our analysis we have taken into
account realistic communication and channel conditions, which include demodulation errors at the relays and practical forwarding mechanisms. Also,
our results are in agreement with \cite{Zhang}, where the error correction properties of network codes for the single--source scenario have
been studied, and a strong connection with classical coding theory has been established. Our analysis extends the analysis to multi--source
networks, provides closed--form expressions of important performance metrics, and accounts for practical communication constraints. Finally, we
note that even though relays and destination compute hard--decision estimates of the incoming signals and send them to the network--layer to
exploit the redundancy introduced by cooperation and NC, the diversity order is the same as in single--hop networks with soft--decision decoding.
The reason is that at the network--layer we take into account the reliability of each bit through a demodulator that resembles the Chase combiner
\cite{ChaseCombining} (see also Section \ref{MSD_Decoder}).
\item \emph{Comparison with adaptive NC solutions}. In Section \ref{Introduction}, we have mentioned that another class of network code designs
aims at guaranteeing a given end--to--end diversity order without injecting erroneous packets into the network. In these solutions, the network
code changes according to the detection outcome at each relay node. Results and analysis in \cite{XiaoDec2010}, \cite{Topakkaya}, and
\cite{Rebelatto} have established a strong connection between the design of diversity--achieving network codes and linear block codes for
erasure channels. More specifically, \cite{XiaoDec2010}, \cite{Topakkaya}, and \cite{Rebelatto} have shown that MDS codes can be used as
network codes to achieve distributed diversity for erasure channels. The analysis conducted in the present paper complements design and
optimization of network codes for \emph{erasure channels} to the performance analysis and design of such codes for \emph{error channels}, where
all the bits are forwarded to the destination regardless of their reliability.
\item \emph{End--to--end coding gain}. As far as the coding gain in \emph{Proposition \ref{CodingDiversityGain}} is concerned, and unlike the
analysis of the diversity order, there are differences between single-- and dual--hop networks with and without NC. In fact, in \emph{Corollary
\ref{APEP_IdealSR}}, we have shown that both demodulation errors at the relays and dual--hop relaying introduce a coding gain loss if
compared to single--hop transmissions. Thus, even though NC and relaying, via a proper receiver design, do not reduce the diversity order
inherently provided by the distributed network code, they do reduce the coding gain, which results in a performance degradation that depends on
the quality of the source--to--relay channels. However, this performance degradation might be reduced, and even completely compensated, through
adequate network code optimization and design. In fact, \emph{Proposition \ref{CodingDiversityGain}} and \emph{Corollary
\ref{APEP_IdealSR}} provide closed--form expressions of the coding gain for both scenarios where we account for realistic and ideal
source--to--relay channels. A good criterion to design the network code, \emph{i.e.}, to exploit the inherent redundancy introduced by NC,
might be to choose the generator matrix of the network code such that the following condition, for each source node, is satisfied:
\begin{equation} \scriptsize
\label{Eq_37} {\bf{G}}_{{\rm{opt}}}  = \left\{ {\left. {\bf{G}} \right|\left| {\left. {\left. {G_c^{\left( {S_t } \right)} } \right|_{\sigma
_{S_t R_q }^2  < \infty }  - G_c^{\left( {S_t } \right)} } \right|_{\sigma _{S_t R_q }^2  \to \infty } } \right| \to 0\quad \forall t = 1,2,
\ldots ,N_S \quad {\rm{and}}\quad \forall q = 1,2, \ldots ,N_R } \right\}
\end{equation}

It is worth being mentioned that, in general, the most important criterion to satisfy is the diversity order requirement, as it has a more
pronounced effect on the system performance. The optimization condition in (\ref{Eq_37}) can be taken into consideration if there is no
reduction on the achievable diversity order for a given rate. Finally, we emphasize that both diversity order and coding gain can be adjusted by
adding or removing relay nodes from the network, which, however, has an effect on the achievable rate as shown in Section \ref{SystemModel}.
The framework proposed in this paper can be exploited for many network optimizations, such as: i) designing the network code to achieve the
best diversity order and coding gain for a given number of sources and relays (\emph{i.e.}, for a given rate); or ii) designing the network code to
have the minimum number of source and relay nodes (\emph{i.e.}, to maximize the rate), for a given diversity order and coding gain.
\item \emph{EEP/UEP Capabilities}. The diversity analysis in \emph{Proposition \ref{CodingDiversityGain}} has pointed out that each source of the
network can achieve a diversity order that is given by the separation vector of the network code. In other words, each source can achieve a
different diversity order. In coding theory, this class of codes is known as UEP codes \cite{Masnick_Oct1967}, and it can be very useful when
different sources have to transmit data with a different quality--of--service requirement or priority. In other words, the network code might be designed
to take into account the individual requirement of each source, instead of being designed by looking at the worst--case scenario only. For
example, let us consider a network with three sources, with one of them having data to be transmitted with very low ABEP. Looking at the worst
case scenario, we should optimize the system, and, thus, the network code as well, such that this source has, for a fixed transmit--power, a
very high diversity order. If we cannot tune the diversity order of each source individually, we are forced to adopt a network code that provides
the same high diversity order for all the sources of the network, which might have an impact on the achievable rate (see Section
\ref{SystemModel}). Our analysis showcases that UEP codes usually exploited in classical coding theory could be used to find the best
trade--off between the diversity order achieved by each source and the rate of the network. In our opinion, this provides design
flexibility, and introduces a finer level of granularity for system optimization, which has not been investigated yet for adaptive NC schemes.
In fact, in general, network codes are designed such that all the sources have the same diversity order \cite{XiaoDec2010}, \cite{Topakkaya}, \cite{Rebelatto}. Our framework provides a systematic way to guarantee unequal diversity orders for each source.

Another interesting application that might benefit from UEP capabilities provided by NC can be found in \cite{RayLiu_Loc2009} and
\cite[Ch 3 and Ch. 5]{HQ_Lai_Thesis2011}. More specifically, in \cite{RayLiu_Loc2009}, UEP capability is called ``incremental diversity''. The idea is that energy consumption can be reduced if source nodes located farther from the destination can transmit with the same power as
closer source nodes, and exploit UEP properties to achieve the same end--to--end performance. In other words, the incremental diversity offered
by UEP network codes might be used to have even energy consumption among the nodes of the network with important implications for green
applications \cite{MDR_GREENET}. Another application for energy saving is the exploitation of the proposed framework as an utility
function for energy efficient network formation through coalition formation games \cite[Ch. 5]{HQ_Lai_Thesis2011}.
\item \emph{Generalization of the performance analysis of dual--hop cooperative protocols}. The framework proposed in this paper for can be thought as a generalization of the many results available in the literature for cooperative networks without NC. Among the many papers described in Section \ref{Introduction}, let us consider, as an example,
\cite{Laneman2007}. In \cite{Laneman2007}, it is shown that a dual--hop three--node network using the DemF protocol can achieve full--diversity
equal to $2$ if the receiver has a reliable estimate of the instantaneous error probability at the relay. This result is included, as a
byproduct, in our analysis, which is more general as it accounts for arbitrary sources, relays, and binary encoding vectors at each relay. In
fact, under the classical coding theory framework, the distributed code used in \cite{Laneman2007} can be seen a repetition code with Hamming
distance equal to $2$ for the single source of the network. Accordingly, from \emph{Proposition \ref{CodingDiversityGain}} we know that the diversity order is equal to $2$, which confirms the analysis in \cite{Laneman2007} under a much broader perspective. In summary, the proposed framework
can be used to study the end--to--end performance of dual--hop cooperative networks without NC, since a repetition code is a special network
code.
\end{list} \vspace{-0.05cm}
\subsection{Impact of Receiver (Network) CSI on the Achievable Diversity} \label{MSD_Decoder}
In this section, we are interested in analyzing the importance of CSI at the receiver to achieve the full--diversity inherently available in the
structure of the network code, which is given by its SV. In fact, it is important to emphasize that the conclusions drawn in Section
\ref{Insights_APEP} hold if the receiver has perfect knowledge of the cross--over probabilities computed in Section \ref{CrossoverProbability}.
This implies that the receiver knows the encoding vectors used at each relay node, along with the CSI of all the wireless links
of the network. In general, the network code can be agreed during the initialization of the network or transmitted by each relay node over the
control plane (at the cost of some overhead). On the other hand, CSI must be estimated at the receiver. In this section, we are aimed at showing the
importance, to achieve full--diversity, of the knowledge of these cross--over probabilities. To this end, we assume that each
receive node, including the destination, has access to the CSI of the wireless links that are directly connected to it (single--hop). In other
words, the destination knows only the fading gains over the source--to--destination and relay--to--destination links, while it is not aware of
the fading gains over the source--to--relay links. On the other hand, we assume that the destination is aware of the network code used at the
relays. This is a requirement for any NC design.

With these assumptions, the destination is unable to compute the cross--over probabilities in (\ref{Eq_7}), and, thus, the received bits cannot
be properly weighted according to their reliability. In such a worst--case scenario, the destination can only assign the same reliability to
each received bit. This corresponds to set all the weights in (\ref{Eq_6}) equal to 1, \emph{i.e.}, ${\bf{w}}\left[ m \right] = 1$ for $m =
1,2, \ldots ,\left( {N_S  + N_R } \right)$. Accordingly, the demodulator in (\ref{Eq_6}) is no longer ML--optimum, and it simplifies to:
\begin{equation} \scriptsize
\label{Eq_38} \left[ {\hat b_{S_1 } ,\hat b_{S_2 } , \ldots ,\hat b_{S_{N_S } } } \right] \propto \mathop {\arg \min }\limits_{\tilde b_{S_1 }
\in \left\{ {0,1} \right\}, \ldots ,\tilde b_{S_{N_S } }  \in \left\{ {0,1} \right\}} \left\{ {\sum\limits_{t = 1}^{N_S } {\left| {\hat b_{S_t
D}  - \tilde b_{S_t } } \right|}  + \sum\limits_{q = 1}^{N_R } {\left| {\hat b_{R_q D}  - \tilde b_{R_q } } \right|} } \right\}
\end{equation}

By using the connection between network code design and classical coding theory described in Section \ref{Insights_APEP}, the decoder in
(\ref{Eq_38}) can be interpreted as a distributed Minimum Distance Decoder (MDD) applied to the overall network code \cite{Proakis}. The
fundamental difference with classical coding theory is that, even though the receiver is not aware of CSI on the source--to--relay links,
demodulation errors at the relay always take place and propagate through the network because of NC and forwarding operations. The demodulator in (\ref{Eq_38}) simply cannot counteract these effects. Of course, this is a worst--case scenario as the destination has no estimates, even imperfect, of this CSI.
The goal here is to understand the diversity order of this low--complexity but sub--optimal demodulator. \emph{Proposition \ref{nonML_Decoder}} provides
an answer to this question.
\begin{proposition} \label{nonML_Decoder}
Given the network model described in Section \ref{SystemModel}, the demodulator in (\ref{Eq_38}) provides an end--to--end diversity order equal to
($t=1,2,\ldots, N_S$):
\begin{equation} \scriptsize
\label{Eq_39} G_d^{\left( {S_t } \right)}  = {\rm{SV}}\left[ t \right] - \left\lfloor {\frac{{{\rm{SV}}\left[ t \right]}}{2}} \right\rfloor
\end{equation}

\smallskip \emph{Proof}: It follows by using the same steps as in Section \ref{FrameworkABEP} by setting
${\bf{w}}\left[ m \right] = 1$ for $m = 1,2, \ldots ,\left( {N_S  + N_R } \right)$. Due to space limitations, we describe only the main modifications of the proof that lead to (\ref{Eq_39}). In particular, when ${\bf{w}}\left[ m \right] = 1$ for $m = 1,2, \ldots ,\left( {N_S  + N_R } \right)$, (\ref{Eq_21}), (\ref{Eq_28}), and (\ref{Eq_29}) simplify as follows:
\begin{equation} \scriptsize
\label{Eq_40}
 Z \to \prod\limits_{k \in \bar {\rm \mathcal{A}}} {{\bf{P}}\left[ {\bar m_{\left( {{\bf{c}},{\bf{\bar c}}} \right)}^{\left( k \right)} } \right]} \quad {\rm{and}} \quad
 T_2  \to {\rm{E}}_{\bf{h}} \left\{ {\prod\limits_{k \in \bar {\rm \mathcal{A}}} {{\bf{P}}\left[ {\bar m_{\left( {{\bf{c}},{\bf{\bar c}}} \right)}^{\left( k \right)} } \right]} } \right\}\mathop  \to \limits^{\left( a \right)} \left( {4\frac{{E_m }}{{N_0 }}} \right)^{ - \left( {d_H \left( {{\bf{c}},{\bf{\bar c}}} \right) - d} \right)} \tilde G_d^{\left( {T_2 } \right)}
\end{equation}
\noindent where: i) the large--SNR approximation in $\mathop  \to \limits^{\left( a \right)}$ is obtained by using the same development as in
Appendix \ref{Lemma2}. More specifically, in (\ref{Eq_4__Lemma2}) we have proved that $Z$ can be seen as the error probability of a Maximum
Ratio Combining (MRC) scheme with ${\rm{card}}\left\{ {\bar {\rm \mathcal{A}}} \right\} = d_H \left( {{\bf{c}},{\bf{\bar c}}} \right) - d$
branches, where $d = {\rm{card}}\left\{ {\rm \mathcal{A}} \right\}$; and ii) $\tilde G_d^{\left( {T_2 } \right)}$ it related to the coding gain
of $T_2$, which is not shown here due to space limitations. By comparing (\ref{Eq_1__Lemma2}) and (\ref{Eq_40}), it follows that $T_2$ undergoes a reduction of the diversity order from ${d_H \left( {{\bf{c}},{\bf{\bar c}}} \right)}$ to ${\rm{card}}\left\{ {\bar {\rm
\mathcal{A}}} \right\} = d_H \left( {{\bf{c}},{\bf{\bar c}}} \right) - d$. From (\ref{Eq_26}), because of the
summation over $d$, each term of the APEP has no longer the same diversity order equal to ${d_H \left( {{\bf{c}},{\bf{\bar c}}} \right)}$, but the
allowed diversity orders fall in the range $\left[ {d_H \left( {{\bf{c}},{\bf{\bar c}}} \right) - \left\lfloor {{{d_H \left(
{{\bf{c}},{\bf{\bar c}}} \right)} \mathord{\left/ {\vphantom {{d_H \left( {{\bf{c}},{\bf{\bar c}}} \right)} 2}} \right.
\kern-\nulldelimiterspace} 2}} \right\rfloor ,d_H \left( {{\bf{c}},{\bf{\bar c}}} \right)} \right]$. Since end--to--end diversity is
given by the addend having the smallest diversity order, we conclude that ${\rm{DIV}}_{{\rm{APEP}}}  = d_H \left( {{\bf{c}},{\bf{\bar c}}}
\right) - \max \left\{ d \right\} = d_H \left( {{\bf{c}},{\bf{\bar c}}} \right) - \left\lfloor {{{d_H \left( {{\bf{c}},{\bf{\bar c}}} \right)}
\mathord{\left/ {\vphantom {{d_H \left( {{\bf{c}},{\bf{\bar c}}} \right)} 2}} \right. \kern-\nulldelimiterspace} 2}} \right\rfloor$. Finally,
by taking into account the relation between Hamming distance and SV given in Section \ref{DiversityCodingGain}, (\ref{Eq_39})
is obtained. This concludes the proof. \hfill $\Box$
\end{proposition}

\emph{Proposition \ref{nonML_Decoder}} brings to our attention the importance of the CSI of the source--to--relay links. In fact, the demodulator in (\ref{Eq_38}) loses approximately half of the potential diversity order inherently available in the network code. This result is in agreement with some studies available in the literature for simple cooperative networks
without NC, such as \cite{Laneman2006}, \cite{Laneman2007}, and \cite{Annavajjala}, where a similar diversity loss due to either non--coherent
demodulation or imperfect CSI has been observed. Furthermore, this result seems to agree with the diversity that can be achieved by linear
block codes over single--hop networks with hard--decision decoding \cite[Sec. 14--6--2]{Proakis}. However, it should be emphasized that in our
case the diversity loss is not due to hard--decision demodulation at the physical layer, which is actually used for both demodulators in (\ref{Eq_6})
and (\ref{Eq_38}), but it originates from the distributed nature of the network code, from demodulation errors at the relays, and from the demodulator
that does not adapt itself to the reliability of the source--to--relay links.
\section{Numerical and Simulation Results} \label{Results}
\begin{figure}
\centering
\includegraphics [width=\columnwidth] {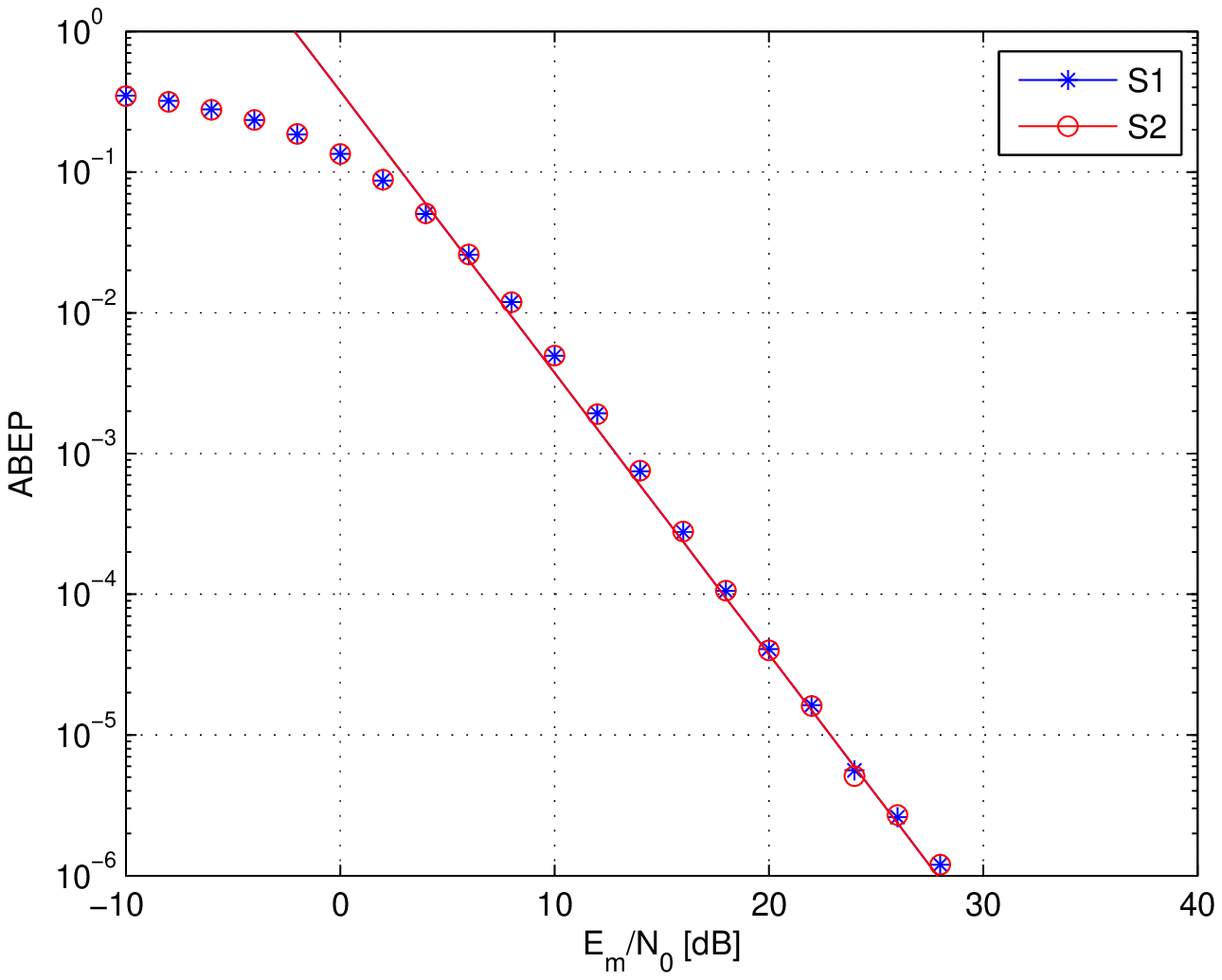}
\vspace{-0.75cm} \caption{\scriptsize ABEP of a 2--source 2--relay network. Markers show Monte Carlo simulations and solid lines show the analytical framework. Setup: i) i.i.d. fading with $\sigma _0^2  = 1$; and ii) $b_{R_1 }  = \hat b_{S_1 R_1 }  \oplus \hat b_{S_2 R_1 }$, $b_{R_2 }  = \hat b_{S_1 R_2 } \oplus
\hat b_{S_2 R_2 }$. The Separation Vector is ${\rm{SV}} = \left[ {{\rm{2}}{\rm{,2}}} \right]$.} \label{Fig1__2S_2R__AllNC}
\end{figure}

\begin{figure}
\centering
\includegraphics [width=\columnwidth] {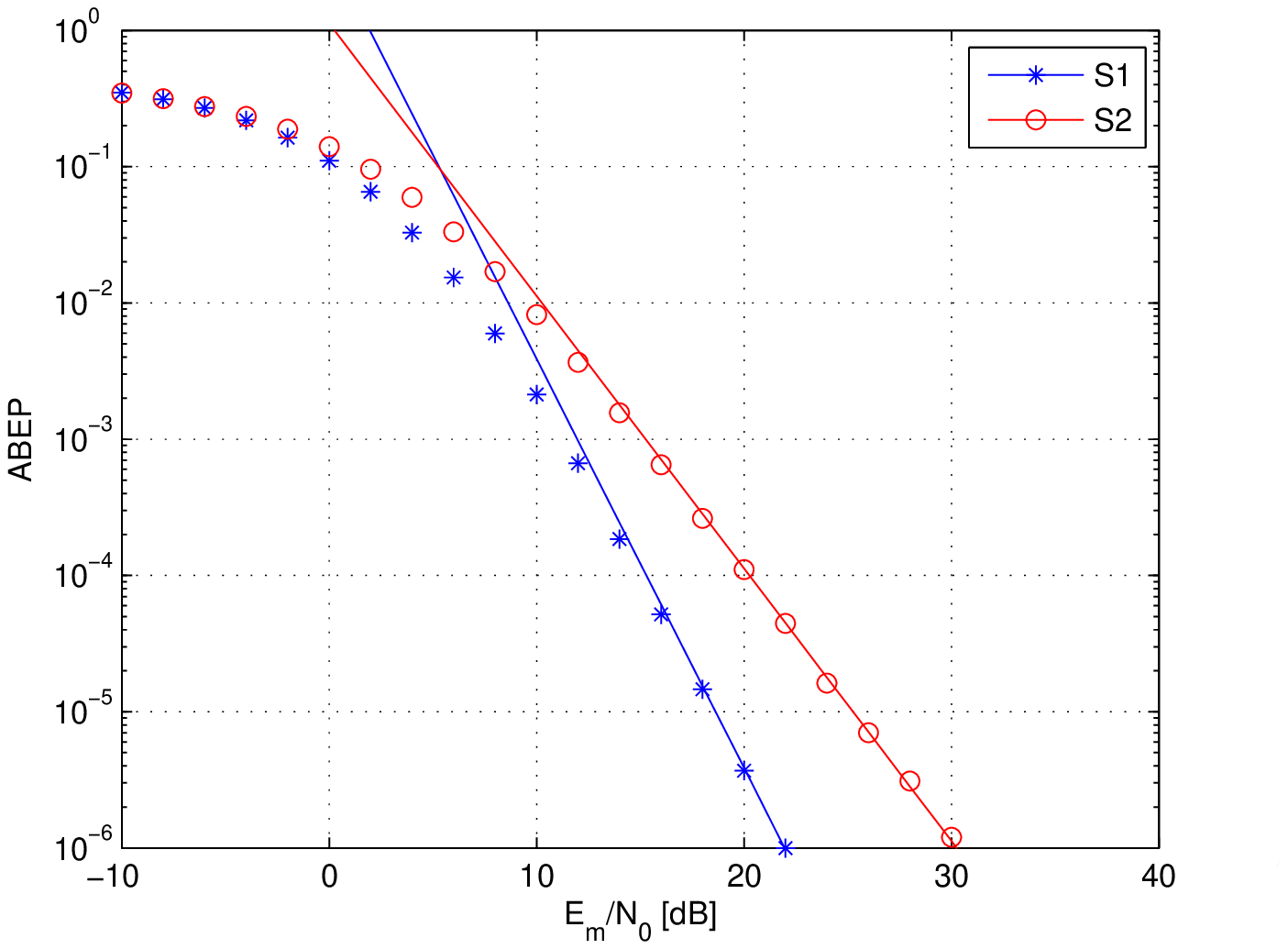}
\vspace{-0.75cm} \caption{\scriptsize ABEP of a 2--source 2--relay network. Markers show Monte Carlo simulations and solid lines show the analytical framework. Setup: i) i.i.d. fading with $\sigma _0^2  = 1$; and ii) $b_{R_1 }  = \hat b_{S_1 R_1 }$, $b_{R_2 }  = \hat b_{S_1 R_2 }  \oplus \hat b_{S_2 R_2 }$. The
Separation Vector is ${\rm{SV}} = \left[ {{\rm{3}}{\rm{,2}}} \right]$.} \label{Fig2__2S_2R__MixS1}
\end{figure}

\begin{figure}
\centering
\includegraphics [width=\columnwidth] {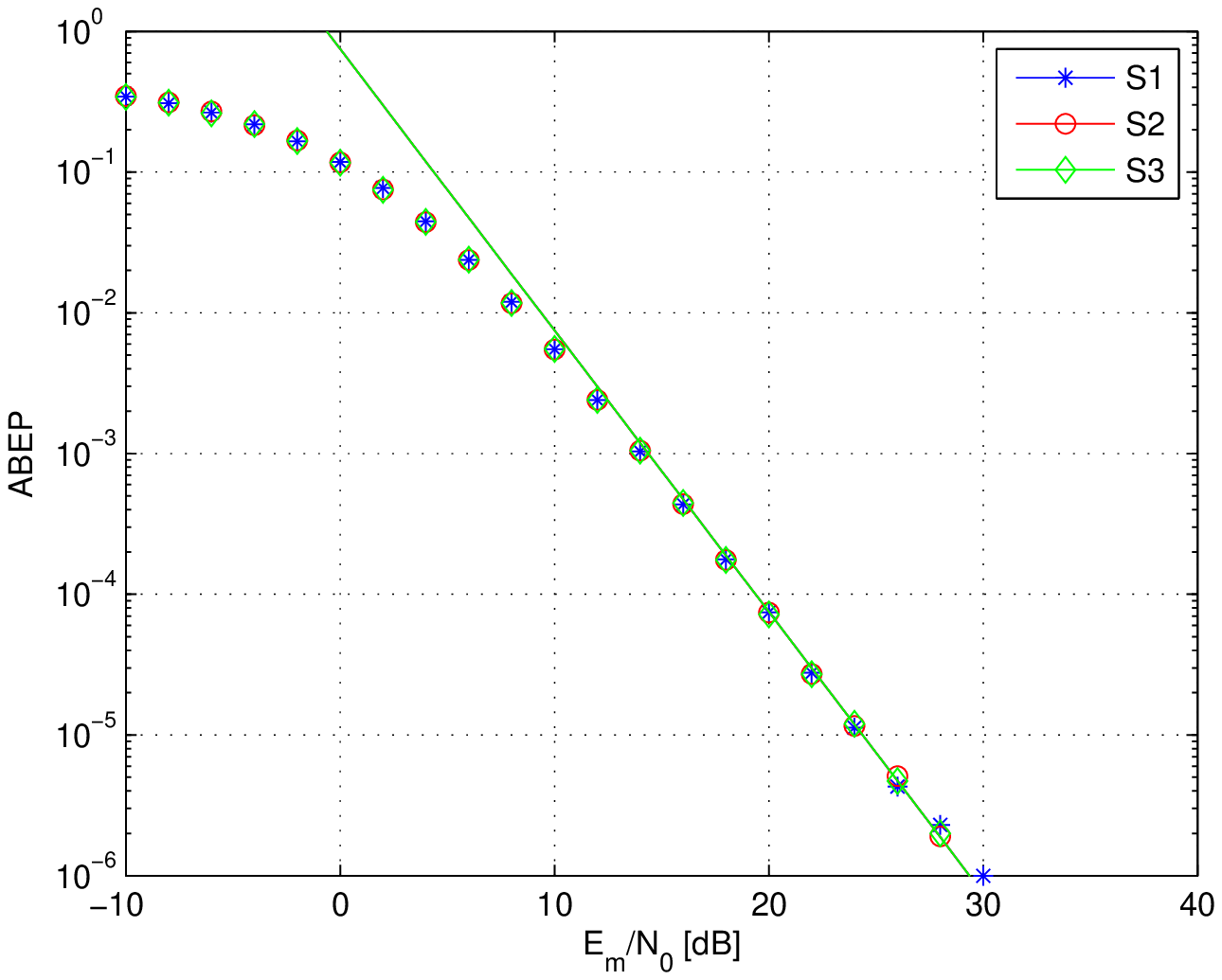}
\vspace{-0.75cm} \caption{\scriptsize ABEP of a 3--source 3--relay network. Markers show Monte Carlo simulations and solid lines show the analytical framework. Setup: i) i.i.d. fading with $\sigma _0^2  = 1$; and ii) $b_{R_1 }  = \hat b_{S_1 R_1 }$, $b_{R_2 }  = \hat b_{S_2 R_2 }$, $b_{R_3 }  = \hat b_{S_3 R_3
}$. The Separation Vector is ${\rm{SV}} = \left[ {{\rm{2}}{\rm{,2}}{\rm{,2}}} \right]$.} \label{Fig3__3S_3R__AllFW}
\end{figure}

\begin{figure}
\centering
\includegraphics [width=\columnwidth] {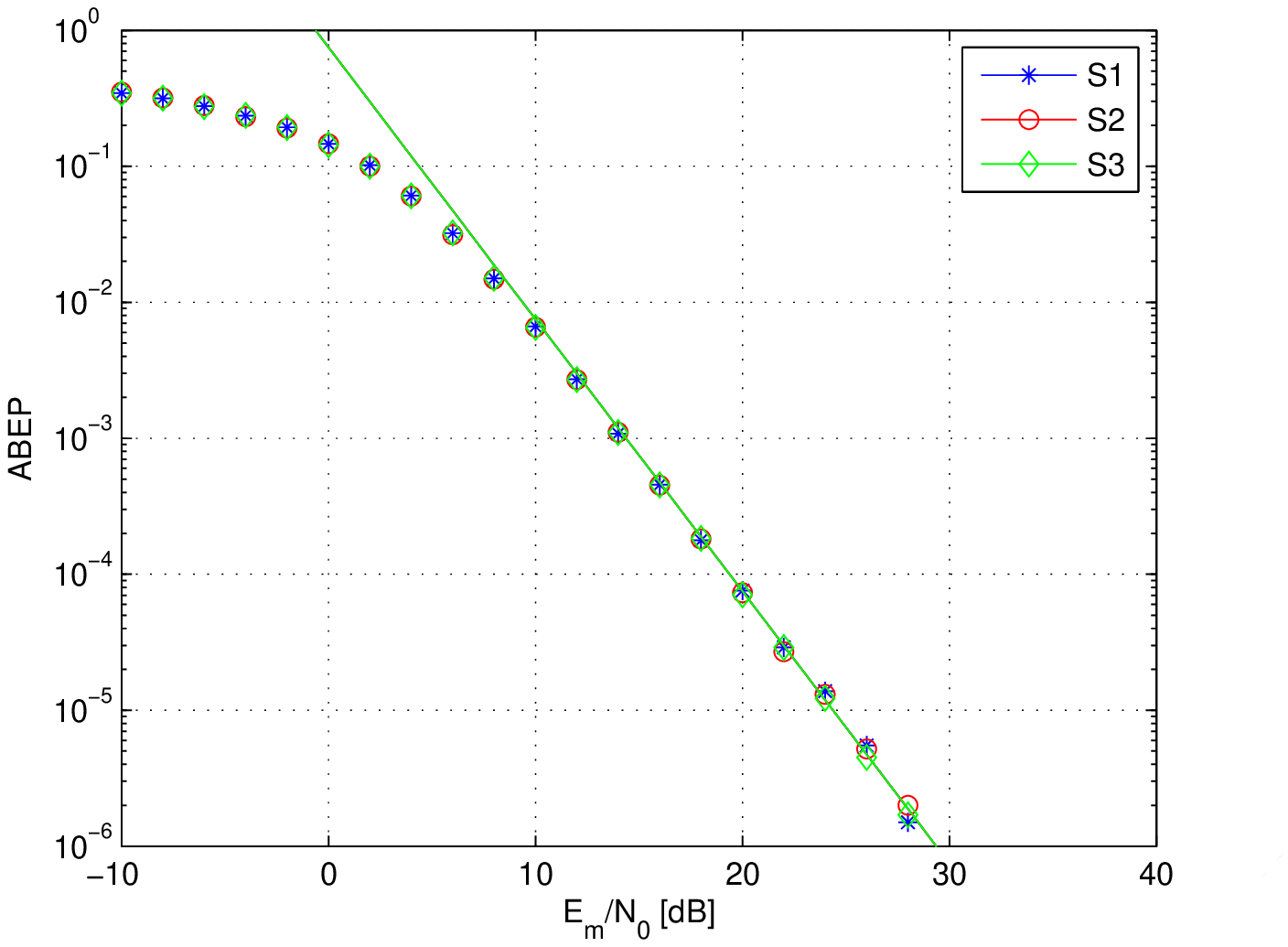}
\vspace{-0.75cm} \caption{\scriptsize ABEP of a 3--source 3--relay network. Markers show Monte Carlo simulations and solid lines show the analytical framework. Setup: i) i.i.d. fading with $\sigma _0^2  = 1$; and ii) $b_{R_1 }  = \hat b_{S_1 R_1 } \oplus \hat b_{S_2 R_1 } \oplus \hat b_{S_3 R_1 }$, $b_{R_2 } =
\hat b_{S_1 R_2 } \oplus \hat b_{S_2 R_2 } \oplus \hat b_{S_3 R_2 }$, $b_{R_3 }  = \hat b_{S_1 R_3 } \oplus \hat b_{S_2 R_3 } \oplus \hat
b_{S_3 R_3 }$. The Separation Vector is ${\rm{SV}} = \left[ {{\rm{2}}{\rm{,2}}{\rm{,2}}} \right]$.} \label{Fig4__3S_3R__AllNC}
\end{figure}

\begin{figure}
\centering
\includegraphics [width=\columnwidth] {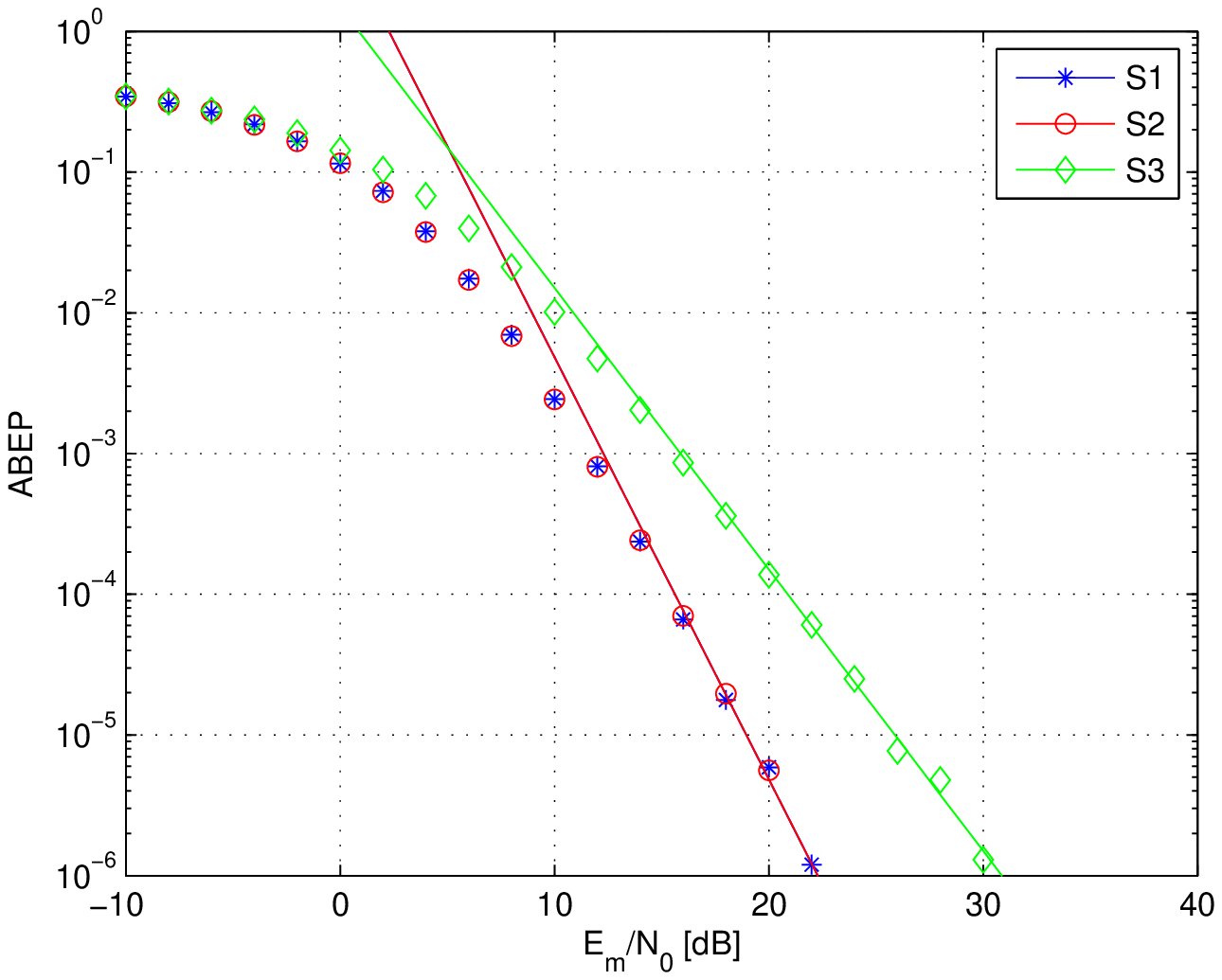}
\vspace{-0.75cm} \caption{\scriptsize ABEP of a 3--source 3--relay network. Markers show Monte Carlo simulations and solid lines show the analytical framework. Setup: i) i.i.d. fading with $\sigma _0^2  = 1$; and ii) $b_{R_1 }  = \hat b_{S_1 R_1 }$, $b_{R_2 }  = \hat b_{S_2 R_2 }$, $b_{R_3 } = \hat b_{S_1 R_3 }
\oplus \hat b_{S_2 R_3 } \oplus \hat b_{S_3 R_3 }$. The Separation Vector is ${\rm{SV}} = \left[ {{\rm{3}}{\rm{,3}}{\rm{,2}}} \right]$.}
\label{Fig5__3S_3R__MixS1S2}
\end{figure}

\begin{figure}
\centering
\includegraphics [width=\columnwidth] {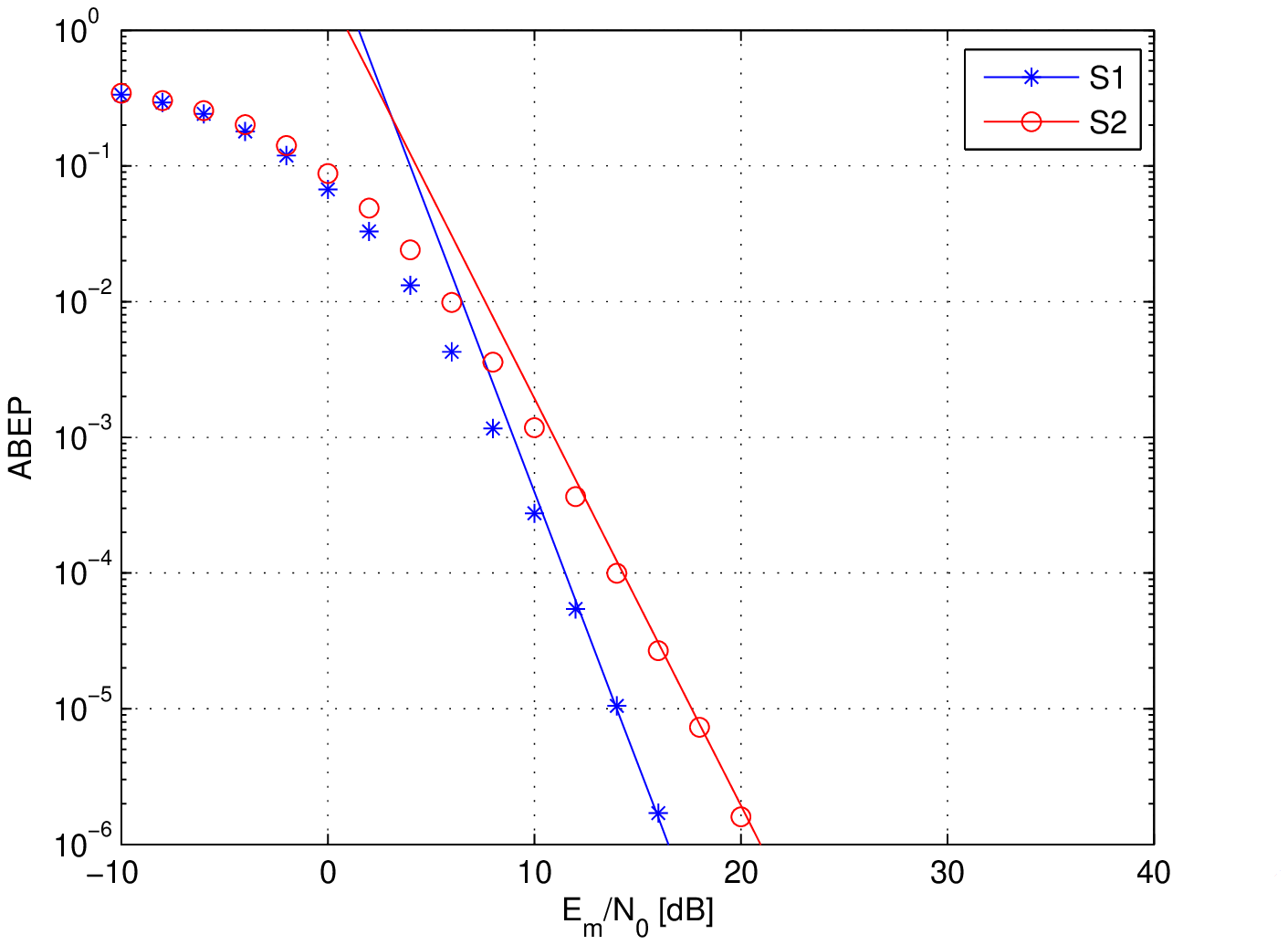}
\vspace{-0.75cm} \caption{\scriptsize ABEP of a 2--source 5--relay network. Markers show Monte Carlo simulations and solid lines show the analytical framework. Setup: i) i.i.d. fading with $\sigma _0^2  = 1$; and ii) $b_{R_1 }  = \hat b_{S_1 R_1 }$, $b_{R_2 }  = \hat b_{S_1 R_2 }$, $b_{R_3 }  = \hat b_{S_1 R_3
}$, $\hat b_{R_4 }  = \hat b_{S_2 R_4 }$, $\hat b_{R_5 }  = \hat b_{S_2 R_5 }$. The Separation Vector is ${\rm{SV}} = \left[
{{\rm{4}}{\rm{,3}}} \right]$.} \label{Fig6__2S_5R__AllFW} \vspace{-0.75cm}
\end{figure}

\begin{figure}
\centering
\includegraphics [width=\columnwidth] {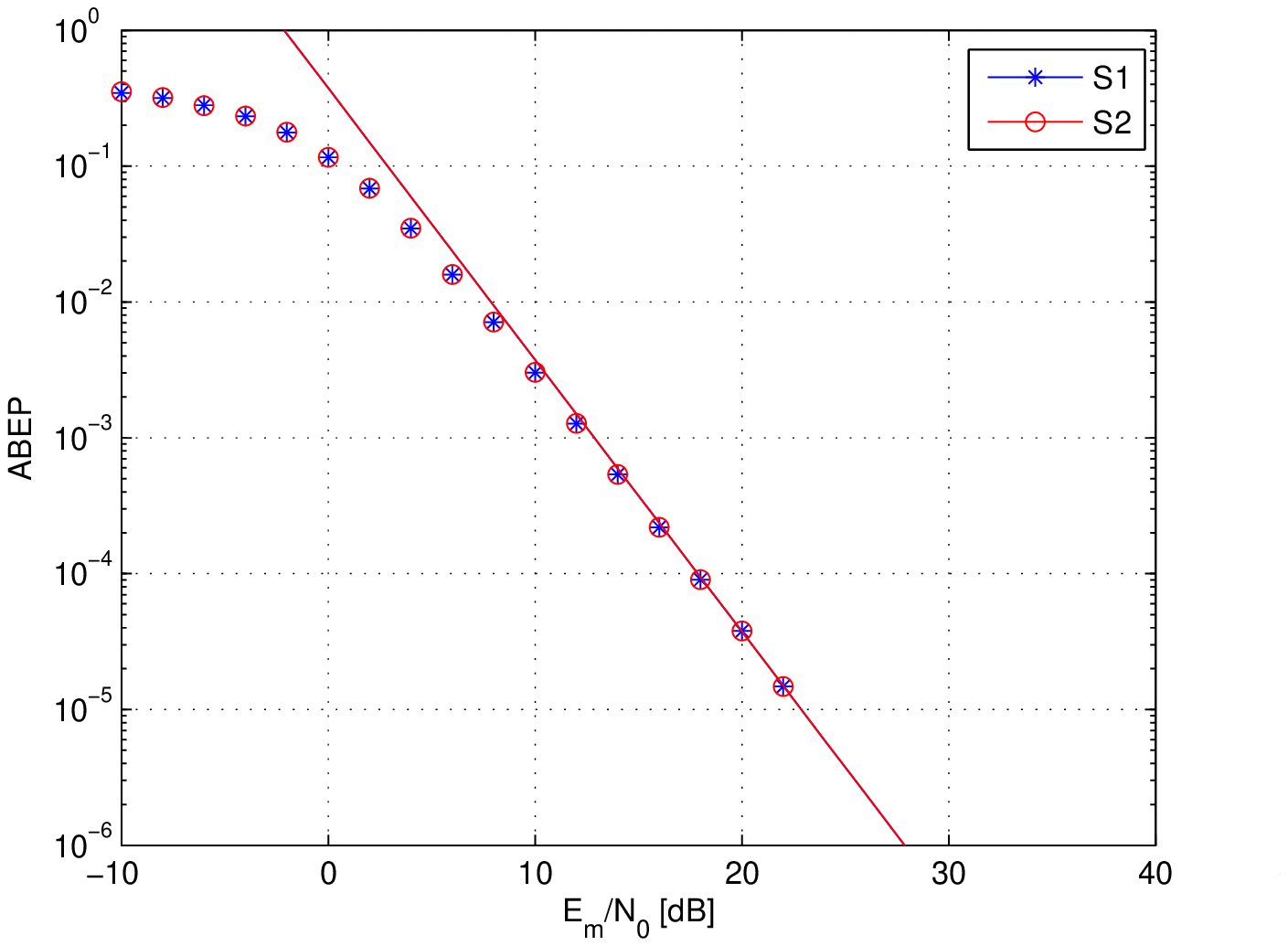}
\vspace{-1.00cm} \caption{\scriptsize ABEP of a 2--source 5--relay network. Markers show Monte Carlo simulations and solid lines show the analytical framework. Setup: i) i.i.d. fading with $\sigma _0^2  = 1$; and ii) $b_{R_1 }  = \hat b_{S_1 R_1 } \oplus \hat b_{S_2 R_1 }$, $b_{R_2 }  = \hat b_{S_1 R_2 } \oplus
\hat b_{S_2 R_2 }$, $b_{R_3 }  = \hat b_{S_1 R_3 } \oplus \hat b_{S_2 R_3 }$, $b_{R_4 }  = \hat b_{S_1 R_4 } \oplus \hat b_{S_2 R_4 }$, $b_{R_5
}  = \hat b_{S_1 R_5 } \oplus \hat b_{S_2 R_5 }$. The Separation Vector is ${\rm{SV}} = \left[ {{\rm{2}}{\rm{,2}}} \right]$.}
\label{Fig7__2S_5R__AllNC} \vspace{-0.75cm}
\end{figure}

\begin{figure}
\centering
\includegraphics [width=\columnwidth] {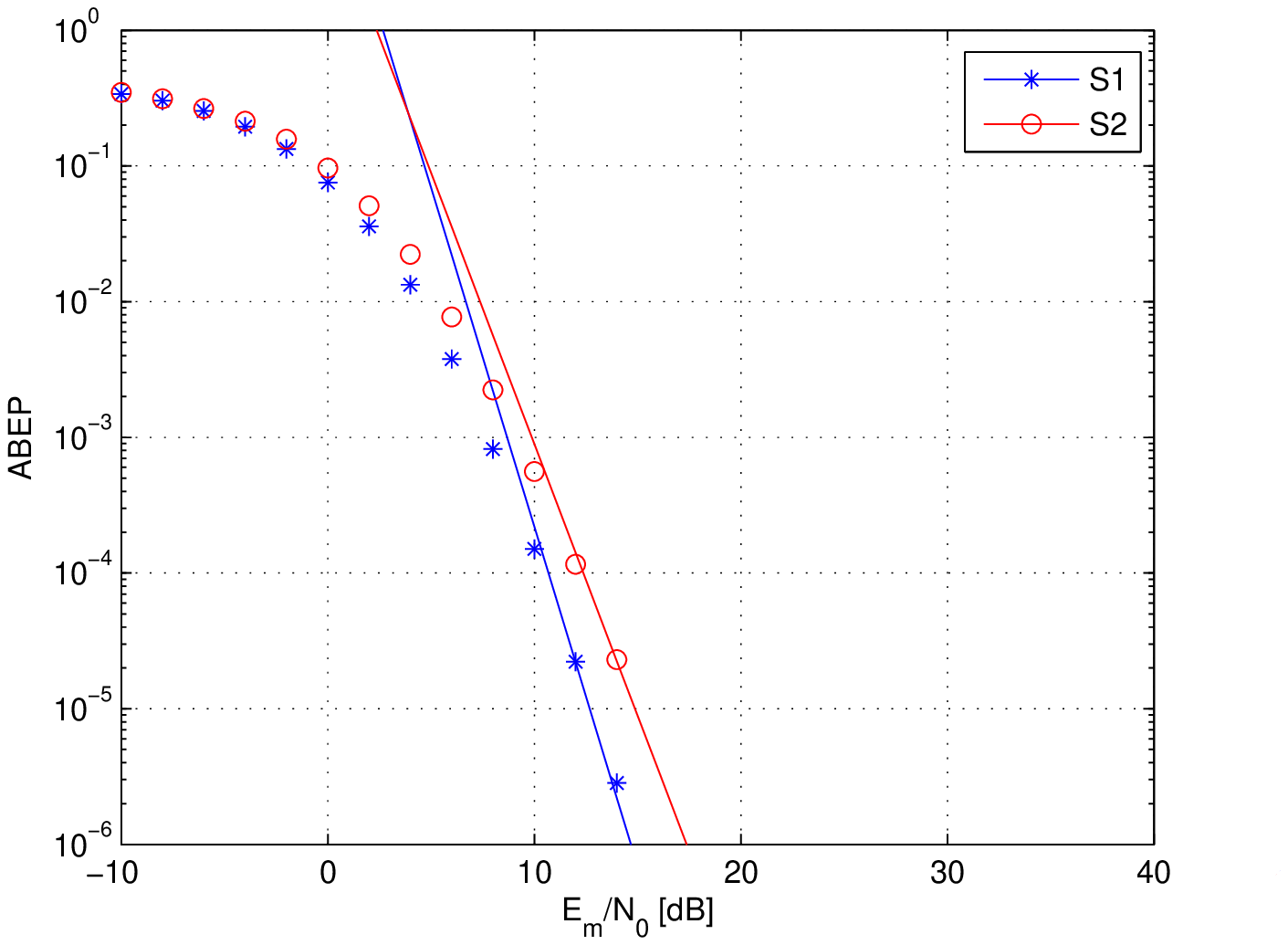}
\vspace{-1.00cm} \caption{\scriptsize ABEP of a 2--source 5--relay network. Markers show Monte Carlo simulations and solid lines show the analytical framework. Setup: i) i.i.d. fading with $\sigma _0^2  = 1$; and ii) $b_{R_1 }  = \hat b_{S_1 R_1 }$, $b_{R_2 }  = \hat b_{S_1 R_2 }$, $b_{R_3 } = \hat b_{S_1 R_3 }
\oplus \hat b_{S_2 R_3 }$, $b_{R_4 }  = \hat b_{S_1 R_4 } \oplus \hat b_{S_2 R_4 }$, $b_{R_5 }  = \hat b_{S_2 R_5 }$. The Separation Vector is
${\rm{SV}} = \left[ {{\rm{5}}{\rm{,4}}} \right]$.} \label{Fig8__2S_5R__MixS1} \vspace{-0.75cm}
\end{figure}

\begin{figure}
\centering
\includegraphics [width=\columnwidth] {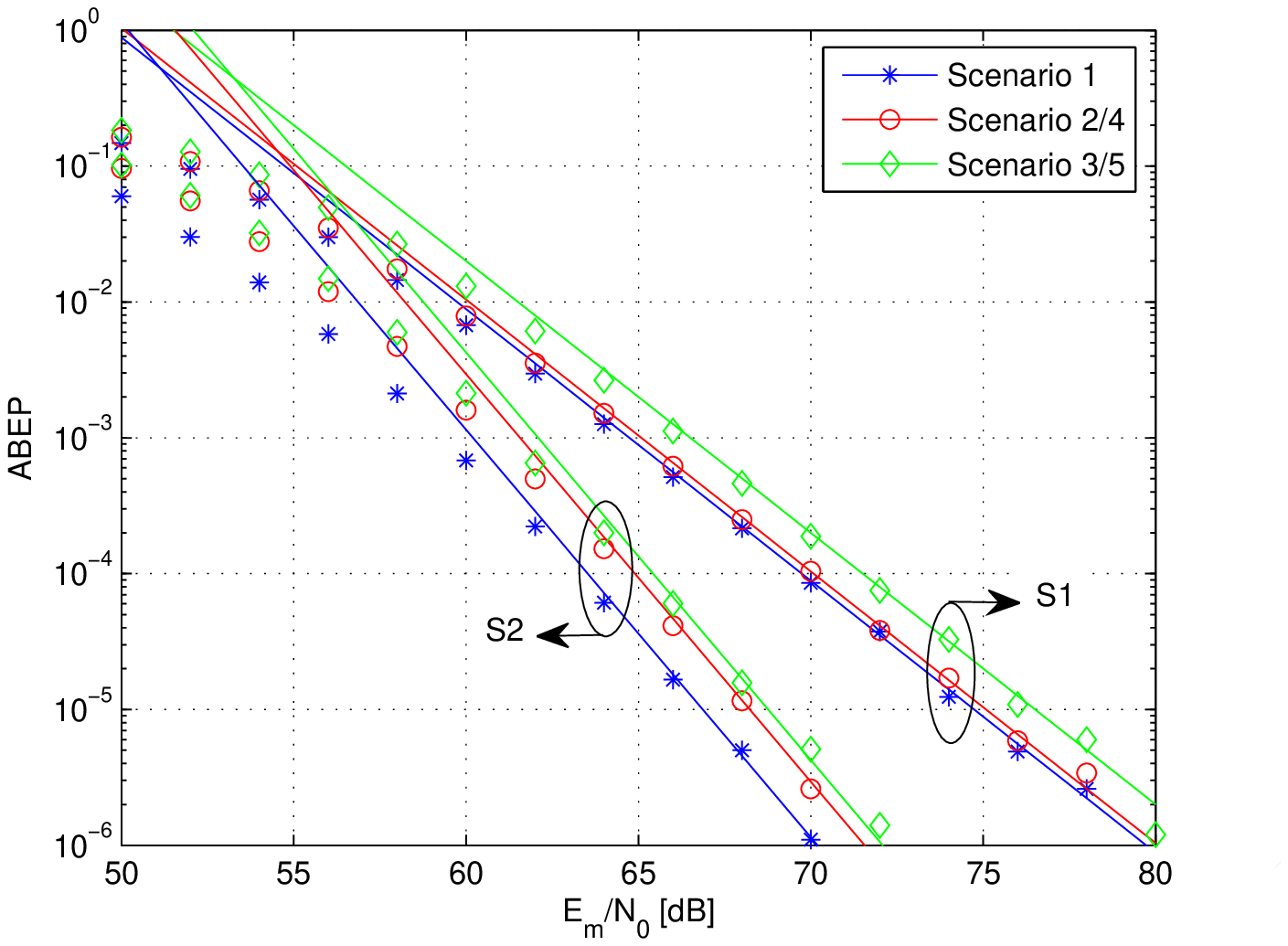}
\vspace{-1.00cm} \caption{\scriptsize ABEP of a 2--source 2--relay network. Markers show Monte Carlo simulations and solid lines show the analytical framework. Setup. i) i.n.i.d. scenario with $\alpha=3$, $\sigma _{XY}^2  = d_{XY}^{ - \alpha}$, ii) the nodes are located at positions (in meters): $S_1  = \left(
{0,25} \right)$, $S_2  = \left( {0,-25} \right)$, $D = \left( {50,0} \right)$, $R_1 = \left( {x_{R_1 } ,12.5} \right)$, $R_2  = \left( {x_{R_2
} ,-12.5} \right)$; and iii) $b_{R_1 } = \hat b_{S_1 R_1 } \oplus \hat b_{S_2 R_1 }$, $b_{R_2 } = \hat b_{S_2 R_2 }$. Furthermore, we have: i)
$x_{R_1 }  = 25$ and $x_{R_2 }  = 25$ in Scenario 1; ii) $x_{R_1 } = 5$ and $x_{R_2 }  = 5$ in Scenario 2; iii) $x_{R_1 }  = 45$ and $x_{R_2 }
= 45$ in Scenario 3; iv) $x_{R_1 }  = 5$ and $x_{R_2 }  = 45$ in Scenario 4; v) $x_{R_1 }  = 45$ and $x_{R_2 }  = 5$ in Scenario 5.}
\label{Fig9__2S_2R__MixS2} \vspace{-0.75cm}
\end{figure}

\begin{figure}
\centering
\includegraphics [width=\columnwidth] {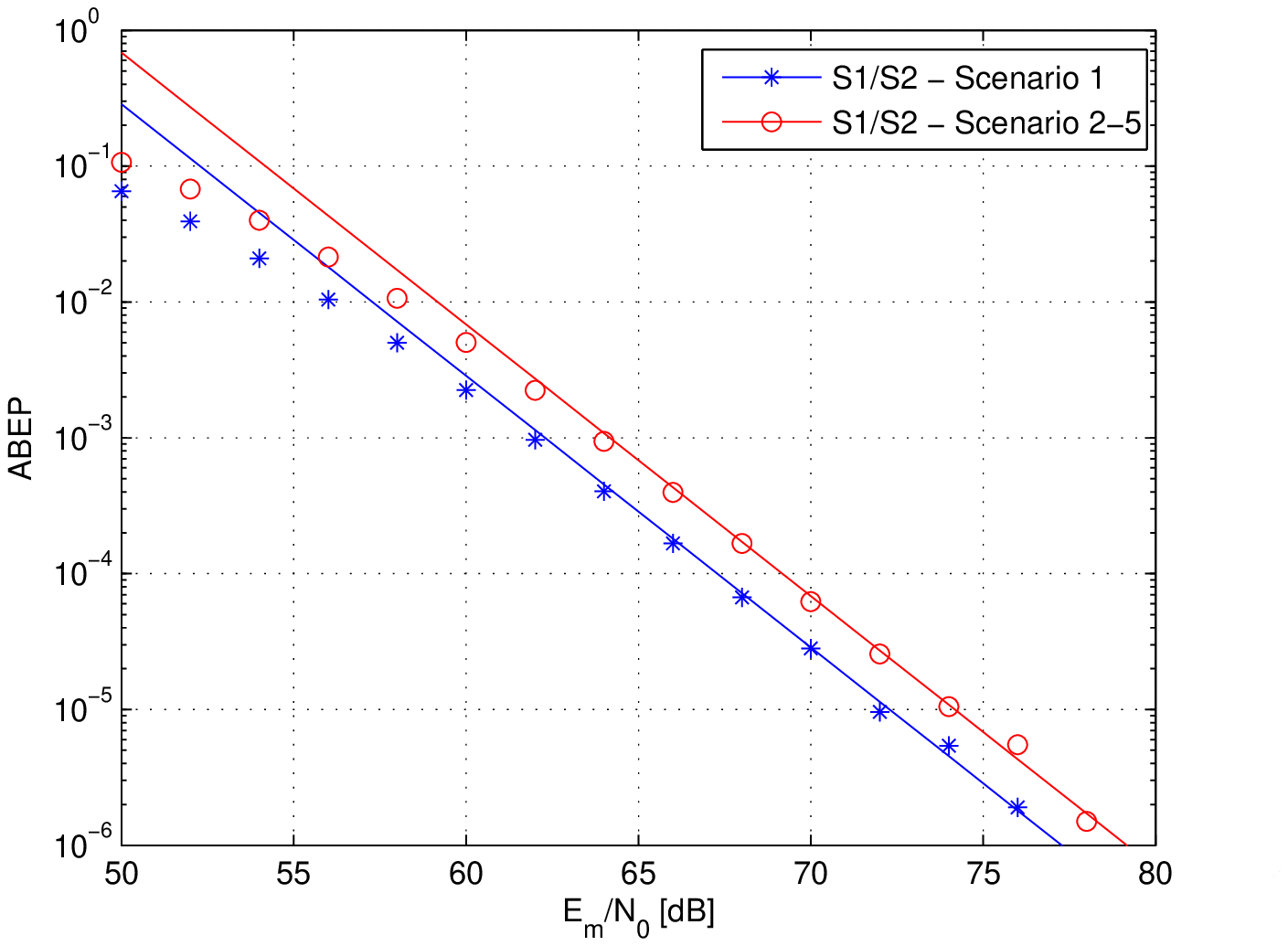}
\vspace{-1.00cm} \caption{\scriptsize ABEP of a 2--source 2--relay network. Markers show Monte Carlo simulations and solid lines show the analytical framework. Setup. i) i.n.i.d. scenario with $\alpha=3$, $\sigma _{XY}^2  = d_{XY}^{ - \alpha}$, ii) the nodes are located at positions (in meters): $S_1  = \left(
{0,25} \right)$, $S_2  = \left( {0,-25} \right)$, $D = \left( {50,0} \right)$, $R_1 = \left( {x_{R_1 } ,12.5} \right)$, $R_2  = \left( {x_{R_2
} ,-12.5} \right)$; and iii) $b_{R_1 } = \hat b_{S_1 R_1 }$, $b_{R_2 } = \hat b_{S_2 R_2 }$. Furthermore, we have: i) $x_{R_1 }  = 25$ and
$x_{R_2 }  = 25$ in Scenario 1; ii) $x_{R_1 } = 5$ and $x_{R_2 }  = 5$ in Scenario 2; iii) $x_{R_1 }  = 45$ and $x_{R_2 }  = 45$ in Scenario 3;
iv) $x_{R_1 }  = 5$ and $x_{R_2 }  = 45$ in Scenario 4; v) $x_{R_1 }  = 45$ and $x_{R_2 }  = 5$ in Scenario 5.} \label{Fig10__2S_2R__AllFW} \vspace{-0.75cm}
\end{figure}

\begin{figure}
\includegraphics [width=\columnwidth] {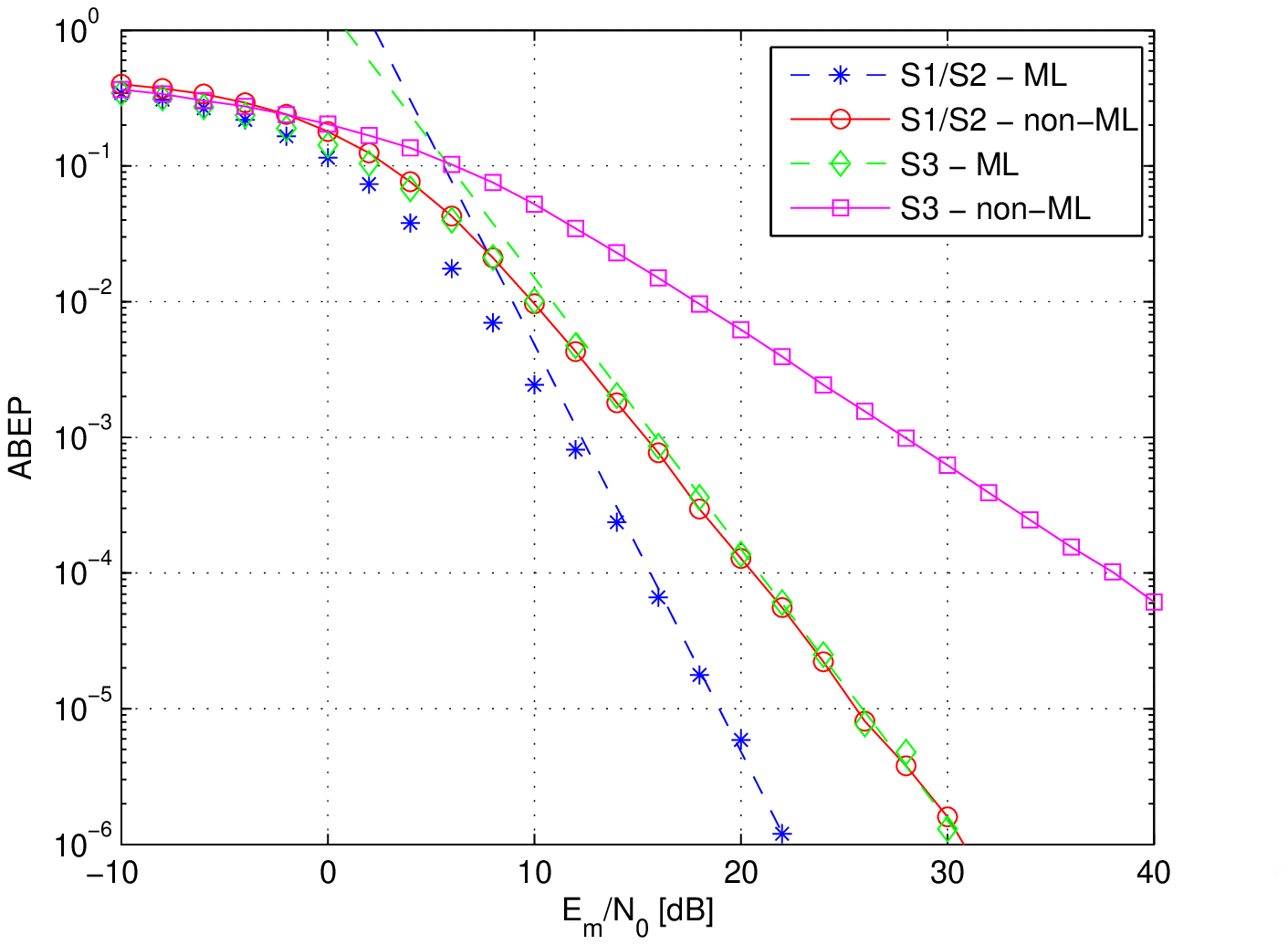}
\vspace{-1.00cm} \caption{\scriptsize ABEP of a 3--source 3--relay network. Markers show Monte Carlo simulations for the ML demodulator, dashed lines show the analytical framework for the ML demodulator, and solid lines with markers show Monte Carlo simulation of the non--ML demodulator. Setup: i) i.i.d. fading with
$\sigma _0^2 = 1$; and ii) $b_{R_1 } = \hat b_{S_1 R_1 }$, $b_{R_2 } = \hat b_{S_2 R_2 }$, $b_{R_3 } = \hat b_{S_1 R_3 } \oplus \hat b_{S_2 R_3
} \oplus \hat b_{S_3 R_3 }$.} \label{Fig11__3S_3R__MixS1S2} \vspace{-0.75cm}
\end{figure}

\begin{figure}
\centering
\includegraphics [width=\columnwidth] {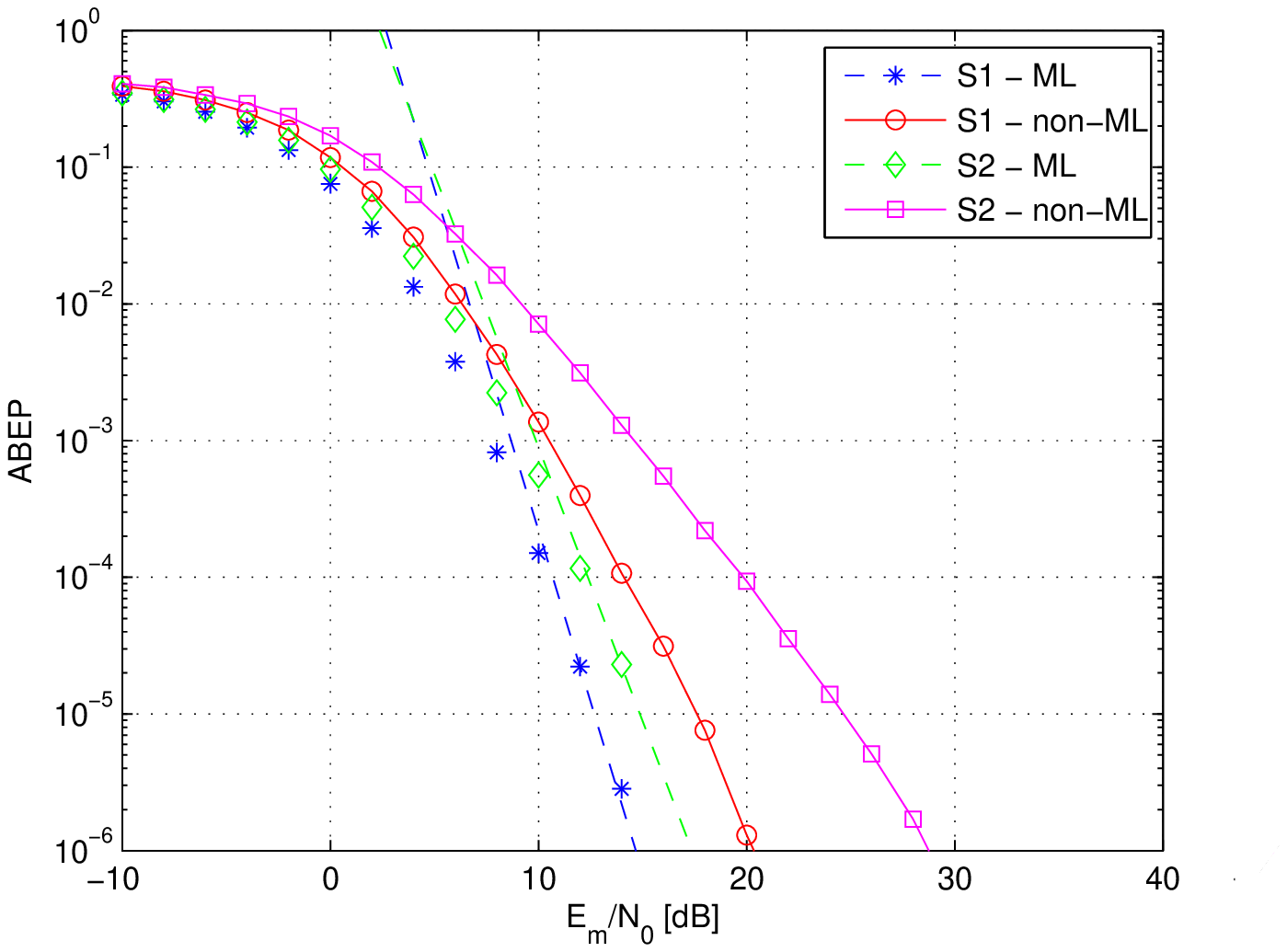}
\vspace{-1.00cm} \caption{\scriptsize ABEP of a 2--source 5--relay network. Markers show Monte Carlo simulations for the ML demodulator, dashed lines show the analytical framework for the ML demodulator, and solid lines with markers show Monte Carlo simulation of the non--ML demodulator. Setup: i) i.i.d. fading with
$\sigma _0^2 = 1$; and ii) $b_{R_1 }  = \hat b_{S_1 R_1 }$, $b_{R_2 }  = \hat b_{S_1 R_2 }$, $b_{R_3 } = \hat b_{S_1 R_3 } \oplus \hat b_{S_2
R_3 }$, $b_{R_4 }  = \hat b_{S_1 R_4 } \oplus \hat b_{S_2 R_4 }$, $b_{R_5 }  = \hat b_{S_2 R_5 }$.} \label{Fig12__2S_5R__MixS1} \vspace{-0.75cm}
\end{figure}

The aim of this section is to show some numerical examples to substantiate analytical derivations, claims, and conclusions of the paper. More
specifically, we are interested in: i) showing the accuracy of the proposed framework for high--SNR, as well as the accuracy of diversity order and
coding gain analysis; ii) understanding the impact of assuming ideal source--to--relay links, as it is often considered in the literature, and
bringing to the attention of the reader that this might lead to misleading conclusions about the usefulness of NC over fading channels; iii)
studying the impact of the network geometry on the end--to--end performance, and, more specifically, the role played by the positions of the
relays; and iv) verifying the diversity reduction caused when the reliability of the source--to relay links is not properly taken into account
at the destination. The analytical frameworks are compared to Monte Carlo simulations, which implements (\ref{Eq_6}) and (\ref{Eq_38}) with no high--SNR approximations. Simulation parameters are summarized in the caption of each figure.
\setcounter{paragraph}{0}
\paragraph{Accuracy of the Framework for i.i.d. Fading Channels}
Figs. \ref{Fig1__2S_2R__AllNC}--\ref{Fig8__2S_5R__MixS1} show the end--to--end ABEP for three network topologies ($N_S=2$ and $N_R=2$; $N_S=3$
and $N_R=3$; $N_S=2$ and $N_R=5$) and for different network codes. In particular, the network codes are chosen according to three criteria: i)
NC is not used and only cooperation is exploited to improve the performance; ii) all the relay nodes implement binary NC on all the received
data, as it is often assumed in the literature \cite{SchoberGLOBE2010}; and iii) only some relay nodes perform NC on a subset of receiver
packets. The first class of codes provides the reference scenario to understand the benefit of NC over classical cooperative protocols. The
second class of codes represents the baseline scenario for network--coded cooperative networks. Finally, the third class of codes is important
to highlight UEP capabilities, and to show that a non--negligible improvement can be obtained if the network code is properly designed and only some
sources are network--coded. Numerical examples confirm the tightness of our framework for high--SNR, and that both diversity order and
coding gain can be well estimated with our simple framework. Furthermore, the UEP behavior of many network codes can be observed as well. In
particular, by comparing the SVs summarized in the caption of each figure with the slope of each curve, we can notice a perfect match, as
predicted in Section \ref{DiversityCodingGain}. Finally, we note that by comparing the results of the 2--source 2--relay network with the
results of the 2--source 5--relay network, we can notice that if the network code is not properly chosen, having multiple relays does not
necessarily lead to a better diversity order. Since the rate of the system is smaller for larger networks (more relays), we can conclude that small
networks with well--optimized network codes can outperform large networks where the network code is not adequately chosen. What really matters
to optimize the performance of multi--source multi--relay networks is the SV of the network code, and, thus, the way the packets received at
the relays are mixed together.
\paragraph{Impact of the Source--to--Relay Links on the Achievable Performance}
In Table \ref{Tab_1}, we show a comparative study of the performance of three network topologies for realistic source--to--relay links, along
with the scenario where ${\sigma _{S_t R_q }^2  \to \infty }$ for $t=1,2,\ldots,N_S$ and $q=1,2,\ldots,N_R$, which is denoted as ``ideal'' in
the table. The results have been obtained from the analytical models and have been verified through Monte Carlo simulations. The accuracy
between model and simulation for the ``realistic'' scenario can be verified in Figs. \ref{Fig1__2S_2R__AllNC}--\ref{Fig8__2S_5R__MixS1}, since
the same simulation setup is used. On the other hand, due to space limitations, similar curves for the ``ideal'' case are not shown, but
similar accuracy has been obtained. The framework used for this latter scenario is given in \emph{Corollary \ref{APEP_IdealSR}}. As
discussed in Section \ref{ParticularChannels}, Table \ref{Tab_1} confirms that there is no diversity loss between the two scenarios, but only a
coding gain loss can be expected. This is because for both scenarios the ML--optimum demodulator is used. However, the conclusions about the
usefulness of NC for both scenarios can be quite different. Let us consider, for example, the 2--source 2--relay network. In the ``ideal''
setting, there is no doubt that NC--3 and NC--4 should be preferred to NC--1 (no NC) and to NC--2 (all received data packets are
network--coded), as one user achieves a higher diversity order while the other has the same ABEP as NC--1 and NC--2. On the other hand, the
conclusion in the ``realistic'' setting is different. In this case, we observe that the higher diversity order achieved by one user is
compensated by a coding gain loss for the second user. In other words, a coding/diversity gain tradeoff exists. However, this behavior is in
the spirit of cooperative networking: one user might tolerate a performance degradation in a given communication round and wait for a reward
during another communication round. Properly choosing the network code enables this possibility. Furthermore, by comparing NC--1 and NC--2, we
can notice that different conclusions can be drawn about the usefulness of NC in the analyzed scenarios. In the ``ideal'' setting, a
cooperative network with NC (NC--2) has the same ABEP as a cooperative network without NC (NC--1). The conclusion is that NC is useless in this
case. On the other hand, the situation changes in the ``realistic'' setting. In this case, we can see that NC--2 is superior to NC--1, and, thus, we conclude that the redundancy introduced by NC can be efficiently exploited at the receiver when it operates in harsh fading
scenarios. In fact, in the ``realistic'' setting, NC--2 can counteract the error propagation due to the dual--hop protocol, even though this
network code is not strong enough to achieve a higher diversity order. Another contradictory behavior can be found when analyzing the 3--source
3--relay network. By comparing NC--1 (no NC) and NC--2 (the relays apply NC to all received packets), we notice that in the ``ideal''
setting NC turns out to be harmful, as NC--2 provides worse performance than NC--1. On the other hand, in the ``realistic'' setting we notice
that NC--1 and NC--2 provides the same ABEP. In other words, NC does not help but at least it is not harmful. These examples, even
though specific to particular networks and codes, clearly illustrate the importance of considering realistic source--to--relay links to draw
sound conclusions about merits and demerits of NC for multi--source multi--relay networks over fading channels. Furthermore, we mention that,
for all the network topologies studied in Table \ref{Tab_1}, NC--2 is representative of a network code that has been designed by keeping
(\ref{Eq_37}) in mind, as it provides the same high--SNR diversity order and coding gain for both ``ideal'' and ``realistic'' settings. Finally, we
emphasize that our conclusions and trends depend on the coding gain of the network, whose study is often neglected due to its analytical
intractability \cite{Laneman2007}, \cite{GiannakisEURASIP2008}, \cite{GiannakisApr2008}, \cite{SchoberGLOBE2010}. In this
paper, we succeeded to provide an accurate estimate of the coding gain as well.
\paragraph{Accuracy of the Framework for i.n.i.d. Fading Channels and Impact of Relay Positions}
In Fig. \ref{Fig9__2S_2R__MixS2} and Fig. \ref{Fig10__2S_2R__AllFW}, we analyze the accuracy of the framework for i.n.i.d. fading channels. We
consider a 2--source 2--relay network with nodes located as described in the caption of the figures. We consider five
network topologies where the relay nodes can occupy different positions with respect to source and destination nodes. We observe a good
accuracy of the framework, and notice that the positions of the relays can affect the end--to--end performance. This example shows that
the proposed framework can be used, for arbitrary fading parameters, for performance optimization via optimal relay placement.
\paragraph{Impact of Receiver CSI on the Diversity Order}
In Fig. \ref{Fig11__3S_3R__MixS1S2} and Fig. \ref{Fig12__2S_5R__MixS1}, we study the impact of using the sub--optimal non--ML demodulator in
(\ref{Eq_38}). In particular, the ABEP of this demodulator is computed by using Monte Carlo simulations, and it is
compared to the analytical investigation in Section \ref{MSD_Decoder}. For comparison, the ABEP (analytical framework and Monte Carlo
simulations) of the ML--optimum demodulator in (\ref{Eq_6}) is shown as well. The non--negligible drop of the diversity order can be observed,
and, by direct inspection, it can be noticed that the curves have the slope predicted in (\ref{Eq_39}). This
confirms the importance of CSI about the source--to--relay links in order to avoid substantial performance degradation.
\section{Conclusion} \label{Conclusion}
In this paper, we have proposed a new analytical framework to study the performance of multi--source multi--relay network--coded cooperative
wireless networks for generic network topologies and binary encoding vectors. Our framework takes into account practical communication
constraints, such as demodulation errors at the relay nodes and fading over all the wireless links. More specifically, closed--form expressions of
the cross--over probability at each relay node are given, and end--to--end closed--form expressions of ABEP and diversity/coding gain are
provided. Our analysis has pointed out that the achievable diversity of each source node coincides with the separation vector of the network
code, which shows that NC can offer unequal diversity capabilities for different sources. Also, the importance of CSI about the
source--to--relay channels has been studied, and it has been proved that half of the diversity might be lost if the
reliability of the source--to--relay links is not properly taken into account at the destination. Monte Carlo simulations have been used
to substantiate analytical modeling and theoretical findings for various network topologies and network codes. In particular, numerical
examples have confirmed that the proposed framework is asymptotically--tight for high SNRs. Finally, by comparing the performance of various
network topologies, with and without taking into account decoding errors at the relays, we have shown that wrong conclusions about the
effectiveness and potential gain of NC for cooperative networks might be drawn when network operations are oversimplified. This highlights the
importance of studying the performance of network--coded cooperative wireless networks with practical communication constraints for a pragmatic
assessment of the end--to--end performance and to enable the efficient optimization of these networks. The framework proposed in this paper
provides an answer to this problem.
\appendices
\section{Proofs of {Lemma \ref{APEP_Lemma1}}, {Lemma \ref{APEP_Lemma2}}, and {Lemma \ref{APEP_Lemma3}}} \label{Appendix_Lemmas}
\subsection{Proof of {Lemma \ref{APEP_Lemma1}}} \label{Lemma1}
\begin{lemma} \label{APEP_Lemma1}
Let $T_1  = {\rm{E}}_{\bf{h}} \left\{ {\prod\nolimits_{k = 1}^{d_H \left( {{\bf{c}},{\bf{\bar c}}} \right)} {{\bf{P}}\left[ {\bar m_{\left(
{{\bf{c}},{\bf{\bar c}}} \right)}^{\left( k \right)} } \right]} } \right\}$, with ${\bf{P}} = \left[ {P_{S_1 D} , \ldots ,P_{S_{N_S } D}
,P_{S_{1:N_S } R_1 D} , \ldots ,P_{S_{1:N_S } R_{N_R } D} } \right]^T$ and ${\bf{P}}\left[ m \right]$, for $m = 1,2, \ldots ,N_S + N_R$, given
in Section \ref{ReceiverDesign} and in \emph{Proposition \ref{CrossProb}}. Then, over i.n.i.d. Rayleigh fading channels and for high--SNR,
$T_1$ has closed--form expression as follows:
\begin{equation} \scriptsize
\label{Eq_1__Lemma1} T_1  \to \left( {4\frac{{E_m }}{{N_0 }}} \right)^{ - d_H \left( {{\bf{c}},{\bf{\bar c}}} \right)} \prod\limits_{m =
1}^{N_S + N_R } {\chi \left\{ {{\bf{\Delta }}_{{\bf{c}},{\bf{\bar c}}} \left[ m \right]{\bf{\bar \Sigma }}_{{\rm{SRD}}}^{\left( {\bf{G}}
\right)} \left[ m \right]} \right\}}
\end{equation}
\noindent where all symbols are defined in \emph{Proposition \ref{APEP}}.

\smallskip \emph{Proof}: Owing to the assumption of independent fading channels, it follows, by direct inspection, that ${\bf{P}}\left[ m \right]$
for $m = 1,2, \ldots ,N_S  + N_R$ are independent RVs, and, thus, $T_1  = \prod\nolimits_{k = 1}^{d_H \left( {{\bf{c}},{\bf{\bar c}}} \right)}
{{\rm{E}}_{\bf{h}} \left\{ {{\bf{P}}\left[ {\bar m_{\left( {{\bf{c}},{\bf{\bar c}}} \right)}^{\left( k \right)} } \right]} \right\}}  =
\prod\nolimits_{k = 1}^{d_H \left( {{\bf{c}},{\bf{\bar c}}} \right)} {{\bf{\bar P}}\left[ {\bar m_{\left( {{\bf{c}},{\bf{\bar c}}}
\right)}^{\left( k \right)} } \right]}$, where ${\bf{\bar P}}\left[ {\bar m_{\left( {{\bf{c}},{\bf{\bar c}}} \right)}^{\left( k \right)} }
\right] = {\rm{E}}_{\bf{h}} \left\{ {{\bf{P}}\left[ {\bar m_{\left( {{\bf{c}},{\bf{\bar c}}} \right)}^{\left( k \right)} } \right]} \right\}$.
Furthermore, from the definition of ${\bf{P}}\left[ m \right]$ in Section \ref{ReceiverDesign} and \emph{Proposition \ref{CrossProb}}, for
high--SNR we have:
\begin{equation} \scriptsize
\label{Eq_2__Lemma1}
 \bar P_{S_t D}  \to \left( {4\frac{{E_m }}{{N_0 }}\sigma _{S_t D}^2 } \right)^{ - 1} \quad {\rm{and}} \quad
 \bar P_{S_{1:N_S } R_q D}  \to \sum\limits_{t = 1}^{N_S } {\left[ {g_{S_t R_q } \left( {4\frac{{E_m }}{{N_0 }}\sigma _{S_t R_q }^2 } \right)^{ - 1} } \right]}  + \left( {4\frac{{E_m }}{{N_0 }}\sigma _{R_q D}^2 } \right)^{ - 1}
\end{equation}

The results in (\ref{Eq_2__Lemma1}) can be obtained from the following chain of equalities and high--SNR approximations:
\begin{equation} \scriptsize
\label{Eq_3__Lemma1} \left\{ \begin{array}{l}
 \bar P_{XY}  = {\rm{E}}_{\bf{h}} \left\{ {Q\left( {\sqrt {2\left( {{{E_m } \mathord{\left/
 {\vphantom {{E_m } {N_0 }}} \right.
 \kern-\nulldelimiterspace} {N_0 }}} \right)\left| {h_{XY} } \right|^2 } } \right)} \right\}\mathop  = \limits^{\left( {a_1 } \right)} \frac{1}{2}\left[ {1 - \sqrt {\frac{{\left( {{{E_m } \mathord{\left/
 {\vphantom {{E_m } {N_0 }}} \right.
 \kern-\nulldelimiterspace} {N_0 }}} \right)\sigma _{XY}^2 }}{{1 + \left( {{{E_m } \mathord{\left/
 {\vphantom {{E_m } {N_0 }}} \right.
 \kern-\nulldelimiterspace} {N_0 }}} \right)\sigma _{XY}^2 }}} } \right]\mathop  \to \limits^{\left( {a_2 } \right)} \left( {4\frac{{E_m }}{{N_0 }}\sigma _{XY}^2 } \right)^{ - 1}  \\
 \bar P_{S_{1:N_S } R_q D}  = {\rm{E}}_{\bf{h}} \left\{ {P_{S_{1:N_S } R_q }  + P_{R_q D}  - 2P_{S_{1:N_S } R_q } P_{R_q D} } \right\}\mathop  = \limits^{\left( {b_1 } \right)} \bar P_{S_{1:N_S } R_q }  + \bar P_{R_q D}  - 2\bar P_{S_{1:N_S } R_q } \bar P_{R_q D} \mathop  \to \limits^{\left( {b_2 } \right)} \bar P_{S_{1:N_S } R_q }  + \bar P_{R_q D}  \\
 \bar P_{S_{1:N_S } R_q }  = \sum\limits_{t = 1}^{N_S } {\left[ {g_{S_t R_q } \bar P_{S_t R_q } \prod\limits_{r = t + 1}^{N_S } {\left( {1 - 2g_{S_r R_q } \bar P_{S_r R_q } } \right)} } \right]} \mathop  \to \limits^{\left( {c_1 } \right)} \sum\limits_{t = 1}^{N_S } {g_{S_t R_q } \bar P_{S_t R_q } } \mathop  \to \limits^{\left( {c_2 } \right)} \sum\limits_{t = 1}^{N_S } {\left[ {g_{S_t R_q } \left( {4\frac{{E_m }}{{N_0 }}\sigma _{S_t R_q }^2 } \right)^{ - 1} } \right]}  \\
 \end{array} \right.
\end{equation}
\noindent where: i) $\mathop  = \limits^{\left( {a_1 } \right)}$ comes from \cite[Eq. (14--3--7)]{Proakis}; ii) $\mathop  \to \limits^{\left(
{a_2 } \right)}$ is the high--SNR approximation of $\mathop  = \limits^{\left( {a_1 } \right)}$ in \cite[Eq. (14--3--13)]{Proakis}; iii)
$\mathop \to \limits^{\left( {b_2 } \right)}$ is the high--SNR approximation of $\mathop  = \limits^{\left( {b_1 } \right)}$, which simply
neglects the term $\bar P_{S_{1:N_S } R_q } \bar P_{R_q D}$, as it decays faster for high--SNR; iv) $\mathop  \to \limits^{\left( {c_1 }
\right)}$ follows by noticing that $1 - 2g_{S_r R_q } \bar P_{S_r R_q }  \to 1$ for high--SNR; and v) $\mathop  \to \limits^{\left( {c_2 }
\right)}$ is, similar to $\mathop  \to \limits^{\left( {a_2 } \right)}$, is the high--SNR approximation of $\mathop \to \limits^{\left( {c_1 }
\right)}$. From (\ref{Eq_3__Lemma1}), (\ref{Eq_1__Lemma1}) follows by using notation and vector representation in
\emph{Proposition \ref{APEP}}. More specifically, the vector ${{\bf{\Delta }}_{{\bf{c}},{\bf{\bar c}}} }$ takes into account that only the
indexes in the set $\Theta \left( {{\bf{c}},{\bf{\bar c}}} \right) = \left\{ {m|{\bf{c}}\left[ m \right] \ne {\bf{\bar c}}\left[ m \right]}
\right\}$ have to be included in $T_1$, and the vector ${{\bf{\bar \Sigma }}_{{\rm{SRD}}}^{\left( {\bf{G}} \right)} }$ accounts for the
dual--hop relaying protocol and the specific network code. This concludes the proof. \hfill $\Box$
\end{lemma}

Two important remarks are worth being made about \emph{Lemma \ref{APEP_Lemma1}}. First, we would like to emphasize that, for ease of
presentation and to stay focused on the most important issues of our analysis, \emph{i.e.}, dual--hop networking and NC, the results in
(\ref{Eq_1__Lemma1}) and (\ref{Eq_2__Lemma1}) are here given for Rayleigh fading only. However, they can be generalized to other fading
distributions for which the high--SNR approximation in \cite{Giannakis} exists. In this paper, Rayleigh fading is studied for illustrative
purposes only. Second, by comparing $\mathop  \to \limits^{\left( {c_2 } \right)}$ and \cite[Eq. (40)]{GiannakisCoopNets}, it follows that, for
high--SNR, the effect, on the error probability at the relays, of performing NC on noisy and faded received data is equivalent to an
Amplify--and--Forward (AF) relay protocol with CSI--assisted relaying \cite{MDR_TCOMSep2009} and with a number of hops equal to the number of
sources that are network--coded at each relay. This conclusion is in agreement with the equivalence between the error probability at the relays
and the error performance of DemF relay protocols already highlighted in Section \ref{CrossoverProbability}. In fact, in \cite{Morgado} it has been shown that, except when the number of hops is very large and the fading severity is very small, the performance of AF and DemF
protocols is very close, for high--SNR, to each other. As the number of sources that can be network--coded is, for practical applications, not
very large, this high--SNR approximation can be very useful to get formulas that provide insights on the system behavior. The high--SNR
equivalency between (\ref{Eq_2__Lemma1}) and AF relaying is exploited in \emph{Lemma \ref{APEP_Lemma2}} to get high--SNR but closed--form
and accurate formulas.
\subsection{Proof of {Lemma \ref{APEP_Lemma2}}} \label{Lemma2}
\begin{lemma} \label{APEP_Lemma2}
Let us consider the term $T_2  = {\rm{E}}_{\bf{h}} \left\{ {\min \left\{ {\prod\nolimits_{k \in {\rm \mathcal{A}}} {{\bf{P}}\left[ {\bar
m_{\left( {{\bf{c}},{\bf{\bar c}}} \right)}^{\left( k \right)} } \right]} ,\prod\nolimits_{k \in \bar {\rm \mathcal{A}}} {{\bf{P}}\left[ {\bar
m_{\left( {{\bf{c}},{\bf{\bar c}}} \right)}^{\left( k \right)} } \right]} } \right\}} \right\}$, with ${\bf{P}} = \left[ {P_{S_1 D} , \ldots
,P_{S_{N_S } D} ,P_{S_{1:N_S } R_1 D} , \ldots ,P_{S_{1:N_S } R_{N_R } D} } \right]^T$ and ${\bf{P}}\left[ m \right]$ for $m = 1,2, \ldots ,N_S
+ N_R$ given in Section \ref{ReceiverDesign} and in \emph{Proposition \ref{CrossProb}}. Then, over i.n.i.d. Rayleigh fading and for
high--SNR, $T_2$ has closed--form expression:
\begin{equation} \scriptsize
\label{Eq_1__Lemma2} T_2  \to \left( {4\frac{{E_m }}{{N_0 }}} \right)^{ - d_H \left( {{\bf{c}},{\bf{\bar c}}} \right)} \left[ {\frac{{2\sqrt
\pi \Gamma \left( {d_H \left( {{\bf{c}},{\bf{\bar c}}} \right) + \frac{1}{2}} \right)}}{{\Gamma \left( {d + \frac{1}{2}} \right)\Gamma \left(
{d_H \left( {{\bf{c}},{\bf{\bar c}}} \right) - d + \frac{1}{2}} \right)}}} \right]\prod\limits_{m = 1}^{N_S  + N_R } {\chi \left\{ {{\bf{\Delta
}}_{{\bf{c}},{\bf{\bar c}}} \left[ m \right]{\bf{\bar \Sigma }}_{{\rm{SRD}}}^{\left( {\bf{G}} \right)} \left[ m \right]} \right\}}
\end{equation}
\noindent where all symbols are defined in \emph{Proposition \ref{APEP}}.

\smallskip \emph{Proof}: The computation of $T_2$ is very analytically involving. To get accurate, but closed--form and insightful formulas
that can shed lights on the network behavior, we exploit some high--SNR approximations. More specifically, the starting point is the following
high--SNR approximation:
\begin{equation} \scriptsize
\label{Eq_2__Lemma2} \left\{ \begin{array}{l}
 P_{S_t D}  = Q\left( {\sqrt {2\left( {{{E_m } \mathord{\left/
 {\vphantom {{E_m } {N_0 }}} \right.
 \kern-\nulldelimiterspace} {N_0 }}} \right)\left| {h_{S_t D} } \right|^2 } } \right) \\
 P_{S_{1:N_S } R_q D}  = P_{S_{1:N_S } R_q }  + P_{R_q D}  - 2P_{S_{1:N_S } R_q } P_{R_q D} \mathop  \to \limits^{\left( a \right)} Q\left( {\sqrt {2\left( {{{E_m } \mathord{\left/
 {\vphantom {{E_m } {N_0 }}} \right.
 \kern-\nulldelimiterspace} {N_0 }}} \right)\left[ {\left| {h_{R_q D} } \right|^{ - 2}  + \sum\limits_{t = 1}^{N_S } {g_{S_t R_q } \left| {h_{S_t R_q } } \right|^{ - 2} } } \right]^{ - 1} } } \right) \\
 \end{array} \right.
\end{equation}

The approximation in $\mathop  \to \limits^{\left( a \right)}$ follows from the closing comment in \emph{Lemma
\ref{APEP_Lemma1}}, where we have proved that for high--SNR the cumulative error due to performing NC on wrong demodulated bits at the relays can
be well--approximated by an equivalent AF multi--hop relay network with a number of hops that is equal to the number of network--coded sources.
In particular, from \cite{MDR_TCOMSep2009} we can recognize that the argument of the Q--function in $\mathop \to \limits^{\left( a \right)}$ is
the end--to--end SNR of an AF relay network, which takes into account the relay--to--destination link and the cumulative error due to
combining, at the most, $N_S$ source. The number of sources that are actually network--coded depends on the number of non--zero NC coefficients
${g_{S_t R_q } }$. Let us emphasize that the formula for the direct source--to--destination link is exact, but we have decided to re--write
it to better understand that high--SNR approximation applies only to the signals forwarded from the relays. Thus, in $T_2$, we have ${\bf{P}}\left[ {\bar m_{\left( {{\bf{c}},{\bf{\bar c}}} \right)}^{\left( k \right)} } \right] \to Q\left( {\sqrt {2\left( {{{E_m } \mathord{\left/ {\vphantom {{E_m } {N_0 }}} \right. \kern-\nulldelimiterspace} {N_0 }}} \right){\rm{SNR}}_{\left( {{\bf{c}},{\bf{\bar c}}} \right)}^{\left( k \right)} } } \right)$, where: i) ${\rm{SNR}}_{\left( {{\bf{c}},{\bf{\bar c}}} \right)}^{\left( k \right)}  = \left| {h_{S_t D} } \right|^2$ for the source--to--destination links, and by bearing in mind that in this case we have a true equality; and ii) ${\rm{SNR}}_{\left( {{\bf{c}},{\bf{\bar c}}} \right)}^{\left( k \right)}  = \left[ {\left| {h_{R_q D} } \right|^{ - 2}  + \sum\nolimits_{t = 1}^{N_S } {g_{S_t R_q } \left| {h_{S_t R_q } } \right|^{ - 2} } } \right]^{ - 1}$ for the relay--to--destination links. Thus, $T_2$ simplifies:
\begin{equation} \scriptsize
\label{Eq_4__Lemma2}
\begin{split}
 T_2  &\to {\rm{E}}_{\bf{h}} \left\{ {\min \left\{ {\prod\limits_{k \in {\rm \mathcal{A}}} {Q\left( {\sqrt {2\left( {{{E_m } \mathord{\left/
 {\vphantom {{E_m } {N_0 }}} \right.
 \kern-\nulldelimiterspace} {N_0 }}} \right){\rm{SNR}}_{\left( {{\bf{c}},{\bf{\bar c}}} \right)}^{\left( k \right)} } } \right)} ,\prod\limits_{k \in \bar {\rm \mathcal{A}}} {Q\left( {\sqrt {2\left( {{{E_m } \mathord{\left/
 {\vphantom {{E_m } {N_0 }}} \right.
 \kern-\nulldelimiterspace} {N_0 }}} \right){\rm{SNR}}_{\left( {{\bf{c}},{\bf{\bar c}}} \right)}^{\left( k \right)} } } \right)} } \right\}} \right\} \\
 & \mathop  \to \limits^{\left( a \right)} {\rm{E}}_{\bf{h}} \left\{ {\min \left\{ {Q\left( {\sqrt {2\left( {{{E_m } \mathord{\left/
 {\vphantom {{E_m } {N_0 }}} \right.
 \kern-\nulldelimiterspace} {N_0 }}} \right)\Upsilon _1 \sum\limits_{k \in {\rm \mathcal{A}}} {{\rm{SNR}}_{\left( {{\bf{c}},{\bf{\bar c}}} \right)}^{\left( k \right)} } } } \right),Q\left( {\sqrt {2\left( {{{E_m } \mathord{\left/
 {\vphantom {{E_m } {N_0 }}} \right.
 \kern-\nulldelimiterspace} {N_0 }}} \right)\Upsilon _2 \sum\limits_{k \in \bar {\rm \mathcal{A}}} {{\rm{SNR}}_{\left( {{\bf{c}},{\bf{\bar c}}} \right)}^{\left( k \right)} } } } \right)} \right\}} \right\} \\
 & \mathop  = \limits^{\left( b \right)} {\rm{E}}_{\bf{h}} \left\{ {Q\left( {\sqrt {2\left( {{{E_m } \mathord{\left/
 {\vphantom {{E_m } {N_0 }}} \right.
 \kern-\nulldelimiterspace} {N_0 }}} \right)\max \left\{ {\Upsilon _1 \sum\limits_{k \in {\rm \mathcal{A}}} {{\rm{SNR}}_{\left( {{\bf{c}},{\bf{\bar c}}} \right)}^{\left( k \right)} } ,\Upsilon _2 \sum\limits_{k \in \bar {\rm \mathcal{A}}} {{\rm{SNR}}_{\left( {{\bf{c}},{\bf{\bar c}}} \right)}^{\left( k \right)} } } \right\}} } \right)} \right\} \\
 \end{split}
\end{equation}
\noindent where: i) the approximation in $\mathop  \to \limits^{\left( a \right)}$ is proved in \emph{Lemma \ref{APEP_Lemma3}}; ii) ${\Upsilon
_1 }$ and ${\Upsilon _2 }$ are two constant factors whose closed--form expression is given in \emph{Lemma \ref{APEP_Lemma3}}; and iii) the
equality in $\mathop  = \limits^{\left( b \right)}$ comes from the fact that the Q--function is monotonically decreasing for increasing values
of its argument.

The last expression in (\ref{Eq_4__Lemma2}) has a convenient structure that can be averaged over fading channel statistics. To this end, the
following considerations can be made: i) $T_2$ can be seen as the ABEP of a dual--branch Selection Combining (SC) scheme, where the equivalent
SNR of first and second branch is \cite{Simon} ${\rm{SNR}}_1  = 2\left( {{{E_m } \mathord{\left/ {\vphantom {{E_m } {N_0 }}} \right.
\kern-\nulldelimiterspace} {N_0 }}} \right)\Upsilon _1 \sum\nolimits_{k \in {\rm \mathcal{A}}} {{\rm{SNR}}_{\left( {{\bf{c}},{\bf{\bar c}}}
\right)}^{\left( k \right)} }$ and ${\rm{SNR}}_2  = 2\left( {{{E_m } \mathord{\left/ {\vphantom {{E_m } {N_0 }}} \right.
\kern-\nulldelimiterspace} {N_0 }}} \right)\Upsilon _2 \sum\nolimits_{k \in \bar {\rm \mathcal{A}}} {{\rm{SNR}}_{\left( {{\bf{c}},{\bf{\bar
c}}} \right)}^{\left( k \right)} }$, respectively; ii) both ${\rm{SNR}}_1$ and ${\rm{SNR}}_2$ can be seen as the equivalent SNR of a Maximum
Ratio Combining (MRC) scheme with a number of branches given by ${\rm{card}}\left\{ \mathcal{A} \right\} = d$ and ${\rm{card}}\left\{
{\mathcal{ \bar A}} \right\} = d_H \left( {{\bf{c}},{\bf{\bar c}}} \right) - d$, respectively; and iii) the ``virtual'' SC and MRC branches
contain independent RVs, as it can be verified via direct inspection. Thus, a closed--form and high--SNR approximation of $T_2$ in
(\ref{Eq_4__Lemma2}) can be obtained by using the method in \cite{Giannakis}. More specifically, by considering: i) the definition of
${\rm{SNR}}_{\left( {{\bf{c}},{\bf{\bar c}}} \right)}^{\left( k \right)}$ in (\ref{Eq_4__Lemma2}); ii) the closed--form expressions of
${\Upsilon _1 }$ and ${\Upsilon _2 }$ in \emph{Lemma \ref{APEP_Lemma3}}; and iii) the general parametrization in \cite[Prop. 1, Prop.
4]{Giannakis} for systems with receive--diversity, we can obtain, after lengthly algebraic manipulations, the final result shown in
(\ref{Eq_1__Lemma2}). In particular, ${{\bf{\Delta }}_{{\bf{c}},{\bf{\bar c}}} }$ and ${{\bf{\bar \Sigma }}_{{\rm{SRD}}}^{\left( {\bf{G}}
\right)} }$ have the same meaning as in \emph{Lemma \ref{APEP_Lemma1}}, while the term into the square brackets accounts for the SC/MRC
high--SNR approximation of $\mathop  = \limits^{\left( b \right)}$ in (\ref{Eq_4__Lemma2}). This concludes the proof. \hfill $\Box$
\end{lemma}

Finally, similar to \emph{Lemma \ref{APEP_Lemma1}} we emphasize once again that the closed--form solution in (\ref{Eq_1__Lemma2}) can be generalized to other fading channel models by using \cite{Giannakis} and \cite{GiannakisCoopNets}.
\subsection{Proof of {Lemma \ref{APEP_Lemma3}}} \label{Lemma3}
\begin{lemma} \label{APEP_Lemma3}
Let $\lambda  = \prod\nolimits_{k \in {\rm \mathcal{A}}} {Q\left( {\sqrt {2\left( {{{E_m } \mathord{\left/ {\vphantom {{E_m } {N_0 }}} \right.
\kern-\nulldelimiterspace} {N_0 }}} \right){\rm{SNR}}_{\left( {{\bf{c}},{\bf{\bar c}}} \right)}^{\left( k \right)} } } \right)}$ with
${{\rm{SNR}}_{\left( {{\bf{c}},{\bf{\bar c}}} \right)}^{\left( k \right)} }$ defined in \emph{Lemma \ref{APEP_Lemma2}}. Then, for
high--SNR and Rayleigh fading, $\lambda$ can be tightly approximated as follows:
\begin{equation} \scriptsize
\label{Eq_1__Lemma3} \lambda  = \prod\limits_{k \in {\rm \mathcal{A}}} {Q\left( {\sqrt {2\left( {{{E_m } \mathord{\left/
 {\vphantom {{E_m } {N_0 }}} \right.
 \kern-\nulldelimiterspace} {N_0 }}} \right){\rm{SNR}}_{\left( {{\bf{c}},{\bf{\bar c}}} \right)}^{\left( k \right)} } } \right)}  \to Q\left( {\sqrt {2\left( {{{E_m } \mathord{\left/
 {\vphantom {{E_m } {N_0 }}} \right.
 \kern-\nulldelimiterspace} {N_0 }}} \right)\Upsilon \sum\limits_{k \in {\rm \mathcal{A}}} {{\rm{SNR}}_{\left( {{\bf{c}},{\bf{\bar c}}} \right)}^{\left( k \right)} } } } \right)
\end{equation}
\noindent where $\Upsilon  = \left[ {\frac{{2^{d - 1} \pi ^{\frac{{d - 1}}{2}} \Gamma \left( {d + \frac{1}{2}} \right)}}{{\Gamma \left(
{\frac{3}{2}} \right)^d \Gamma \left( {d + 1} \right)}}} \right]^{{1 \mathord{\left/ {\vphantom {1 d}} \right. \kern-\nulldelimiterspace} d}}$ and $d = {\rm{card}}\left\{ \mathcal{A} \right\}$.

\smallskip \emph{Proof}: From the Chernoff bound, \emph{i.e.},
$Q\left( x \right) \le \left( {{1 \mathord{\left/ {\vphantom {1 2}} \right. \kern-\nulldelimiterspace} 2}} \right)\exp \left( { - {{x^2 }
\mathord{\left/ {\vphantom {{x^2 } 2}} \right. \kern-\nulldelimiterspace} 2}} \right) \le \exp \left( { - {{x^2 } \mathord{\left/ {\vphantom
{{x^2 } 2}} \right. \kern-\nulldelimiterspace} 2}} \right)$, which is accurate for $x \gg 1$ that in our case implies high--SNR (${{E_m }
\mathord{\left/ {\vphantom {{E_m } {N_0 }}} \right. \kern-\nulldelimiterspace} {N_0 }} \gg 1$), the following approximation holds:
\begin{equation} \scriptsize
\label{Eq_3__Lemma3} \prod\limits_{k \in {\rm \mathcal{A}}} {Q\left( {\sqrt {x_k } } \right)}  \to Q\left( {\sqrt {\Upsilon \sum\limits_{k \in
{\rm \mathcal{A}}} {x_k } } } \right)
\end{equation}
\noindent where $\Upsilon $ is a constant correction term, which is introduced to recover the coding gain inaccuracy that might arise when
using the Chernoff bound \cite{Laneman2007}. The high--SNR approximation in (\ref{Eq_3__Lemma3}) can be explained as follows. By direct
inspection, left-- and right--hand side terms can be shown to have both diversity order equal to $d={\rm{card}}\left\{ \mathcal{A} \right\}$. In
fact, the left--hand side is the product of $d$ terms each one having diversity one. On the other hand, the right--hand side term is the error
probability of a MRC scheme \cite{Simon} with $d$ diversity branches at the receiver, which is known to have diversity $d$ \cite{Giannakis}.
The constant (correction) factor $\Upsilon$ is introduced only to avoid coding gain inaccuracies, which are always present when using the
Chernoff bound. Since the goal of this paper is to accurately estimate both coding gain and diversity order, the accurate evaluation of $\Upsilon$ is
instrumental to estimate the end--to--end performance of the system.

To get an accurate, but simple and useful for further analysis, approximation we use first--order moment matching to estimate
$\Upsilon$ in (\ref{Eq_3__Lemma3}). The motivation is that, as we will better substantiate at the end of this proof, it allows
us to have a closed--form estimate of $\Upsilon$ that depends only on $d$ in (\ref{Eq_1__Lemma3}), while it is
independent of the fading parameters. In formulas, we seek to find $\Upsilon$ such that the following equality is satisfied:
\begin{equation} \scriptsize
\label{Eq_4__Lemma3} {\rm{E}}_{\bf{h}} \left\{ {\prod\limits_{k \in {\rm \mathcal{A}}} {Q\left( {\sqrt {2\left( {{{E_m } \mathord{\left/
 {\vphantom {{E_m } {N_0 }}} \right.
 \kern-\nulldelimiterspace} {N_0 }}} \right){\rm{SNR}}_{\left( {{\bf{c}},{\bf{\bar c}}} \right)}^{\left( k \right)} } } \right)} } \right\} = {\rm{E}}_{\bf{h}} \left\{ {Q\left( {\sqrt {2\left( {{{E_m } \mathord{\left/
 {\vphantom {{E_m } {N_0 }}} \right.
 \kern-\nulldelimiterspace} {N_0 }}} \right)\Upsilon \sum\limits_{k \in {\rm \mathcal{A}}} {{\rm{SNR}}_{\left( {{\bf{c}},{\bf{\bar c}}} \right)}^{\left( k \right)} } } } \right)} \right\}
\end{equation}

To this end, we need closed--form expressions of both averages in (\ref{Eq_4__Lemma3}). Once again, we use the high--SNR parametrization in
\cite{Giannakis}, which leads to the following result:
\begin{equation} \scriptsize
\label{Eq_5__Lemma3} \left\{ \begin{array}{l}
 {\rm{E}}_{\bf{h}} \left\{ {\prod\limits_{k \in {\rm \mathcal{A}}} {Q\left( {\sqrt {2\left( {{{E_m } \mathord{\left/
 {\vphantom {{E_m } {N_0 }}} \right.
 \kern-\nulldelimiterspace} {N_0 }}} \right){\rm{SNR}}_{\left( {{\bf{c}},{\bf{\bar c}}} \right)}^{\left( k \right)} } } \right)} } \right\}\mathop  \to \limits^{\left( a \right)} \left( {4\frac{{E_m }}{{N_0 }}} \right)^{ - d} \prod\limits_{k \in {\rm \mathcal{A}}} {\left[ {\frac{1}{{\sigma _{R_q D}^2 }} + \sum\limits_{t = 1}^{N_S } {\frac{{g_{S_t R_q } }}{{\sigma _{S_t R_q }^2 }}} } \right]}  \\
 {\rm{E}}_{\bf{h}} \left\{ {Q\left( {\sqrt {2\left( {{{E_m } \mathord{\left/
 {\vphantom {{E_m } {N_0 }}} \right.
 \kern-\nulldelimiterspace} {N_0 }}} \right)\Upsilon \sum\limits_{k \in {\rm \mathcal{A}}} {{\rm{SNR}}_{\left( {{\bf{c}},{\bf{\bar c}}} \right)}^{\left( k \right)} } } } \right)} \right\}\mathop  \to \limits^{\left( b \right)} \left( {4\frac{{E_m }}{{N_0 }}\Upsilon } \right)^{ - d} \left[ {\frac{{2^{d - 1} \pi ^{\frac{{d - 1}}{2}} \Gamma \left( {d + \frac{1}{2}} \right)}}{{\Gamma \left( {\frac{3}{2}} \right)^d \Gamma \left( {d + 1} \right)\prod\limits_{k \in {\rm \mathcal{A}}} {\left[ {\left( {\frac{1}{{\sigma _{R_q D}^2 }} + \sum\limits_{t = 1}^{N_S } {\frac{{g_{S_t R_q } }}{{\sigma _{S_t R_q }^2 }}} } \right)^{ - 1} } \right]} }}} \right] \\
 \end{array} \right.
\end{equation}
\noindent where: 1) $\mathop  \to \limits^{\left( a \right)}$ is obtained by taking into account that (i) ${{\rm{SNR}}_{\left(
{{\bf{c}},{\bf{\bar c}}} \right)}^{\left( k \right)} }$ are statistically independent for $k \in {\rm \mathcal{A}}$; (ii) according to
\emph{Lemma \ref{APEP_Lemma2}}, ${{\rm{SNR}}_{\left( {{\bf{c}},{\bf{\bar c}}} \right)}^{\left( k \right)} }$ can be seen as the end--to--end
SNRs of an equivalent multi--hop AF relay protocol; and (iii) by using asymptotic analysis for multi--hop AF relay networks in
\cite{GiannakisCoopNets}; and 2) $\mathop  \to \limits^{\left( b \right)}$ is obtained from \cite{Giannakis} and \cite{GiannakisCoopNets} by
recognizing that we have to compute the average of an equivalent MRC scheme where each branch is an equivalent multi--hop network that uses the
AF relay protocol. Finally, by equating the two terms in (\ref{Eq_5__Lemma3}), $\Upsilon$ in (\ref{Eq_1__Lemma3}) can be obtained. As mentioned above, $\Upsilon$ is independent of channel statistics. Similar to \emph{Lemma \ref{APEP_Lemma1}} and \emph{Lemma
\ref{APEP_Lemma2}} we mention that the proposed procedure can be applied to any fading channel model, for which the parametrization in
\cite{Giannakis} is available. This concludes the proof. \hfill $\Box$
\end{lemma}
%
%
%
%
%
%

%
%
%
%
%
%
%
\begin{table*}[!b] \scriptsize
\renewcommand{\arraystretch}{1.5}
\caption{\scriptsize ${\rm{ABEP}}_{S_t }  = \left( {G_c^{\left( {S_t } \right)} \bar \gamma _m } \right)^{ - G_d^{\left( {S_t } \right)} }$, where
${G_d^{\left( {S_t } \right)} }$ is the diversity order, ${G_c^{\left( {S_t } \right)} }$ is the coding gain, and we have defined $\gamma_m =
{E_m /N_0 }$. 2--source, 2--relay network. NC--1: $b_{R_1 }  = \hat b_{S_1 R_1 }$, $b_{R_2 } = \hat b_{S_2 R_2 }$; NC--2: $b_{R_1 }  = \hat
b_{S_1 R_1 }  \oplus \hat b_{S_2 R_1 }$, $b_{R_2 }  = \hat b_{S_1 R_2 }  \oplus \hat b_{S_2 R_2 }$; NC--3: $b_{R_1 }  = \hat b_{S_1 R_1 }
\oplus \hat b_{S_2 R_1 }$, $b_{R_2 }  = \hat b_{S_2 R_2 }$; NC--4: $b_{R_1 }  = \hat b_{S_1 R_1 }$, $b_{R_2 }  = \hat b_{S_1 R_2 }  \oplus \hat
b_{S_2 R_2 }$. 3--source, 3--relay network. NC--1: $b_{R_1 }  = \hat b_{S_1 R_1 }$, $b_{R_2 }  = \hat b_{S_2 R_2 }$, $b_{R_3 }  = \hat b_{S_3
R_3 }$; NC--2: $b_{R_1 }  = \hat b_{S_1 R_1 }  \oplus \hat b_{S_2 R_1 }  \oplus \hat b_{S_3 R_1 }$, $b_{R_2 }  = \hat b_{S_1 R_2 }  \oplus \hat
b_{S_2 R_2 }  \oplus \hat b_{S_3 R_2 }$, $b_{R_3 }  = \hat b_{S_1 R_3 }  \oplus \hat b_{S_2 R_3 }  \oplus \hat b_{S_3 R_3 }$; NC--3: $b_{R_1 }
= \hat b_{S_1 R_1 }$, $b_{R_2 }  = \hat b_{S_2 R_2 }$, $b_{R_3 }  = \hat b_{S_1 R_3 }  \oplus \hat b_{S_2 R_3 }  \oplus \hat b_{S_3 R_3 }$.
2--source, 5--relay network. NC--1: $b_{R_1 }  = \hat b_{S_1 R_1 }$, $b_{R_2 }  = \hat b_{S_1 R_2 }$, $b_{R_3 }  = \hat b_{S_1 R_3 }$, $b_{R_4
} = \hat b_{S_2 R_4 }$, $b_{R_4 }  = \hat b_{S_2 R_5 }$; NC--2: $b_{R_1 }  = \hat b_{S_1 R_1 } \oplus \hat b_{S_2 R_1 }$, $b_{R_2 }  = \hat
b_{S_1 R_2 }  \oplus \hat b_{S_2 R_2 }$, $b_{R_3 }  = \hat b_{S_1 R_3 }  \oplus \hat b_{S_2 R_3 }$, $b_{R_4 }  = \hat b_{S_1 R_4 }  \oplus \hat
b_{S_2 R_4 }$, $b_{R_5 }  = \hat b_{S_1 R_5 }  \oplus \hat b_{S_2 R_5 }$; NC--3: $b_{R_1 }  = \hat b_{S_1 R_1 }$, $b_{R_2 }  = \hat b_{S_1 R_2
}$, $b_{R_3 }  = \hat b_{S_1 R_3 }  \oplus \hat b_{S_2 R_3 }$, $b_{R_4 }  = \hat b_{S_1 R_4 } \oplus \hat b_{S_2 R_4 }$, $b_{R_5 }  = \hat
b_{S_2 R_5 }$. Finally, i.i.d. fading with $\sigma _0^2  = 1$ is considered. \vspace{-0.5cm}} \label{Tab_1}
\begin{center}
\begin{tabular}{|c||c|c|c||c|c|c|}
\hline
\multicolumn{7}{|c|} {Network: 2--source, 2--relay} \\
\hline
 & \multicolumn{3}{|c||} {Ideal source--to--relay channels} & \multicolumn{3}{|c|} {Realistic source--to--relay channels} \\
\hline
 & ${\rm{ABEP}}_\infty ^{\left( {S_1 } \right)}$ & ${\rm{ABEP}}_\infty ^{\left( {S_2 } \right)}$ & ${\rm{ABEP}}_\infty ^{\left( {S_3 } \right)}$ & ${\rm{ABEP}}_\infty ^{\left( {S_1 } \right)}$ & ${\rm{ABEP}}_\infty ^{\left( {S_2 } \right)}$ & ${\rm{ABEP}}_\infty ^{\left( {S_3 } \right)}$ \\
\hline
NC-1 & $0.3750\gamma_m^{ - 2}$ & $0.3750\gamma_m^{ - 2}$ & - & $0.7500\gamma_m^{ - 2}$ & $0.7500\gamma_m^{ - 2}$ & - \\
\hline
NC-2 & $0.3750\gamma_m^{ - 2}$ & $0.3750\gamma_m^{ - 2}$ & - & $0.3750\gamma_m^{ - 2}$ & $0.3750\gamma_m^{ - 2}$ & - \\
\hline
NC-3 & $0.3750\gamma_m^{ - 2}$ & $0.9688\gamma_m^{ - 3}$ & - & $1.1250\gamma_m^{ - 2}$ & $3.8750\gamma_m^{ - 3}$ & - \\
\hline
NC-4 & $0.9688\gamma_m^{ - 3}$ & $0.3750\gamma_m^{ - 2}$ & - & $3.8750\gamma_m^{ - 3}$ & $1.1250\gamma_m^{ - 2}$ & - \\
\hline
\hline
\multicolumn{7}{|c|} {Network: 3--source, 3--relay} \\
\hline
 & \multicolumn{3}{|c||} {Ideal source--to--relay channels} & \multicolumn{3}{|c|} {Realistic source--to--relay channels} \\
\hline
 & ${\rm{ABEP}}_\infty ^{\left( {S_1 } \right)}$ & ${\rm{ABEP}}_\infty ^{\left( {S_2 } \right)}$ & ${\rm{ABEP}}_\infty ^{\left( {S_3 } \right)}$ & ${\rm{ABEP}}_\infty ^{\left( {S_1 } \right)}$ & ${\rm{ABEP}}_\infty ^{\left( {S_2 } \right)}$ & ${\rm{ABEP}}_\infty ^{\left( {S_3 } \right)}$ \\
\hline
NC-1 & $0.3750\gamma_m^{ - 2}$ & $0.3750\gamma_m^{ - 2}$ & $0.3750\gamma_m^{ - 2}$ & $0.7500\gamma_m^{ - 2}$ & $0.7500\gamma_m^{ - 2}$ & $0.7500\gamma_m^{ - 2}$ \\
\hline
NC-2 & $0.7500\gamma_m^{ - 2}$ & $0.7500\gamma_m^{ - 2}$ & $0.7500\gamma_m^{ - 2}$ & $0.7500\gamma_m^{ - 2}$ & $0.7500\gamma_m^{ - 2}$ & $0.7500\gamma_m^{ - 2}$ \\
\hline
NC-3 & $0.9688\gamma_m^{ - 3}$ & $0.9688\gamma_m^{ - 3}$ & $0.3750\gamma_m^{ - 2}$ & $4.8438\gamma_m^{ - 3}$ & $4.8438\gamma_m^{ - 3}$ & $1.5000\gamma_m^{ - 2}$ \\
\hline
\hline
\multicolumn{7}{|c|} {Network: 2--source, 5--relay} \\
\hline
 & \multicolumn{3}{|c||} {Ideal source--to--relay channels} & \multicolumn{3}{|c|} {Realistic source--to--relay channels} \\
\hline
 & ${\rm{ABEP}}_\infty ^{\left( {S_1 } \right)}$ & ${\rm{ABEP}}_\infty ^{\left( {S_2 } \right)}$ & ${\rm{ABEP}}_\infty ^{\left( {S_3 } \right)}$ & ${\rm{ABEP}}_\infty ^{\left( {S_1 } \right)}$ & ${\rm{ABEP}}_\infty ^{\left( {S_2 } \right)}$ & ${\rm{ABEP}}_\infty ^{\left( {S_3 } \right)}$ \\
\hline
NC-1 & $0.4961\gamma_m^{ - 4}$ & $0.4844\gamma_m^{ - 3}$ & - & $3.9688\gamma_m^{ - 4}$ & $1.9375\gamma_m^{ - 3}$ & - \\
\hline
NC-2 & $0.3750\gamma_m^{ - 2}$ & $0.3750\gamma_m^{ - 2}$ & - & $0.3750\gamma_m^{ - 2}$ & $0.3750\gamma_m^{ - 2}$ & - \\
\hline
NC-3 & $0.9980\gamma_m^{ - 5}$ & $0.4961\gamma_m^{ - 4}$ & - & $21.9570\gamma_m^{ - 5}$ & $8.9297\gamma_m^{ - 4}$ & - \\
\hline
\end{tabular}
\end{center}
\end{table*}
%
%
\twocolumn
\clearpage
%
%
%
%
%
%
%
%
%
%
%
%
\clearpage
%
%
%
%
%
%
%
%
\newpage
%
%
%
%
%
%
%
%
%
%
%
%
\end{document}